\documentclass[letterpaper,12pt]{article}
\usepackage{amsmath}
\usepackage{amssymb}
\usepackage{amsfonts}
\usepackage{setspace}
\usepackage{bbm}
\usepackage{color}
\usepackage{enumerate}
\usepackage{mathrsfs}
\usepackage[longnamesfirst]{natbib}
\usepackage[bottom]{footmisc}
\usepackage{graphicx}
\usepackage{subcaption}
\usepackage{hyperref}
\usepackage[shortlabels]{enumitem}
\usepackage[usenames,dvipsnames,svgnames,table]{xcolor}
\usepackage{multirow}
\usepackage[margin=1in]{geometry}
\usepackage{tikz}
\usetikzlibrary{topaths,calc}
\usepackage{enumitem}
\usepackage{cancel}
\usepackage[normalem]{ulem}

\newtheorem{theorem}{Theorem}
\newtheorem{lemma}{Lemma}
\newtheorem{proposition}{Proposition}
\newtheorem{assumption}{Assumption}

\newtheorem{corollary}{Corollary}
\newtheorem{definition}{Definition}

\newenvironment{proof}[1][Proof]{\noindent \textbf{#1.} }{\  \rule{0.5em}{0.5em}}
\newcommand{\mb}[1]{\mathbb{#1}}

\newcommand{\mr}[1]{\mathrm{#1}}

\newcommand{\mc}[1]{\mathcal{#1}}

\newcommand{\ul}[1]{\underline{#1}}
\newcommand{\ol}[1]{\overline{#1}}

\hypersetup{
     colorlinks   = true,
     citecolor    = DarkBlue,
     urlcolor     = Black, 
     linkcolor    = DarkBlue
}

\makeatletter
\renewcommand\paragraph{\@startsection{paragraph}{4}{\z@}%
                                    {0pt \@plus1ex \@minus.2ex}%
                                    {-1em}%
                                    {\normalfont\normalsize\bfseries}}
\makeatother

\usepackage{float}
\DeclareMathOperator*{\argmax}{argmax}
\DeclareMathOperator*{\argmin}{argmin} 
\newcommand{\E}{\mathbb{E}}
\begin{document}

\onehalfspacing

\author{%
Timothy Christensen\thanks{%
 Yale University. \texttt{timothy.christensen@yale.edu}}
 \quad
Hyungsik Roger Moon\thanks{%
 University of Southern California. \texttt{moonr@usc.edu}}
 \quad
Frank Schorfheide\thanks{%
 University of Pennsylvania, CEPR, NBER, and PIER. \texttt{schorf@ssc.upenn.edu}}
}

\title{%
Optimal Decision Rules \\when Payoffs are Partially Identified\footnote{
We are grateful to X.~Chen, L.~Hansen, C.~Manski, F. Molinari, J.~Porter, Q.~Vuong,  E.~Vytlacil, three anonymous referees, and participants in various seminars and conferences
for helpful comments and suggestions. This paper supersedes the preprint \texttt{arXiv:2011.03153} \citep{CMS2020forecast}. This material is based upon work supported by the National Science Foundation under Grants No. SES-1919034 (Christensen), SES-1625586 (Moon), and SES-1851634 (Schorfheide).}
}

\date{December 18, 2025}

\maketitle

\begin{abstract}  
\singlespacing
\noindent 
We derive asymptotically optimal statistical decision rules for discrete choice problems when payoffs depend on a partially-identified parameter $\theta$ and the decision maker can use a point-identified parameter $\mu$ to deduce restrictions on $\theta$. Examples include treatment choice under partial identification and pricing with rich unobserved heterogeneity. 
Our notion of optimality combines a minimax approach to handle the ambiguity from partial identification of $\theta$ given $\mu$ with an average risk minimization approach for $\mu$.
We show how to implement optimal decision rules using the bootstrap and (quasi-)Bayesian methods in both parametric and semiparametric settings. We provide detailed applications to treatment choice and optimal pricing. Our asymptotic approach is well suited for realistic empirical settings in which the derivation of finite-sample optimal rules is intractable. 

\medskip 


\medskip

\end{abstract}

\thispagestyle{empty}

\newpage

\setcounter{page}{1}

\section{Introduction}

Many important policy decisions involve discrete choices. Examples include whether or not to treat an aggregate population or large sub-population, firm or worker decisions at the extensive margin, and pricing policies when, in practice, prices must be expressed in whole currency units. Suppose a decision maker must choose a policy from a discrete choice set. The decision maker has data that may be used to bound, but not point identify, the payoffs associated with some choices. 
How should they proceed?

In this paper, we propose an approach for making asymptotically optimal discrete statistical (i.e., data-driven) decisions when the payoffs associated with some choices are only partially identified. 
We assume the decision maker observes data which may be used to learn about a vector of parameters $\mu$. The decision maker then chooses a policy from a finite set. The distribution of payoffs associated with the different policies depends on a parameter $\theta$. A key assumption underlying the analysis in this paper is that $\theta$ is possibly set-identified, but the parameters $\mu$ may be used to deduce restrictions on $\theta$. The decision maker therefore confronts both ambiguity (the payoff distribution is not point-identified) and statistical uncertainty ($\mu$ must be estimated from the data).

We propose a theory of optimal statistical decision making in this setting, building on a line of research going back to \cite{Manski2000}. We depart from this body of work by combining a minimax approach to handle the ambiguity that arises from partial identification of $\theta$ given $\mu$ and average (or integrated) risk minimization for $\mu$. This asymmetric treatment of parameters is in the spirit of the generalized Bayes-minimax principle of \cite{Hurwicz1951}. The resulting optimal decision rules are ``robust'' in the sense that they minimize maximum risk or regret over the identified set $\Theta_0(\mu)$ for $\theta$ conditional on $\mu$, and use the data to learn efficiently about features of $\mu$ germane to the choice problem. 

Our optimal decision rules can be implemented very easily in realistic empirical settings. All we require is that, for every choice, the maximum risk (or regret) over $\theta \in \Theta_0(\mu)$ conditional on $\mu$ can be computed. The maximum risk is averaged across a bootstrap distribution for an efficient estimator $\hat \mu$ of $\mu$, a posterior distribution for $\mu$ in parametric models, or a quasi-posterior based on a limited-information criterion for $\mu$ in semiparametric models. Our optimal decision is then simply to choose whatever choice has smallest average maximum risk. We refer to this rule as a \emph{bootstrap decision} when the maximum risk is averaged over a bootstrap distribution, a \emph{Bayes decision} when averaged over a posterior,\footnote{These are not Bayes decision rules in the usual sense, but rather hybrid Bayes-minimax rules which, as we explain below, also have an interpretation as a type of robust Bayes (or conditional $\Gamma$-minimax) rules. For simplicity, we refer to them as Bayes decisions and drop the robust/minimax qualifiers.} and a \emph{quasi-Bayes} decision when averaged over a quasi-posterior. Despite its simplicity, we provide a formal asymptotic optimality theory to justify our approach. 

As running example, we show how to implement optimal decisions in the context of a treatment assignment problem under partial identification of the average treatment effect (ATE). In an empirical illustration, taken from \cite{IK}, a decision maker decides whether or not to adopt a job-training program based on several RCT estimates from other studies and their standard errors. The identified set $\Theta_0(\mu)$ is defined through intersection bounds derived by extrapolating multiple studies. In addition to treatment assignment problems, we discuss how our framework can be used for optimal pricing decisions in an environment with rich unobserved heterogeneity, where revealed preference arguments may be used to derive bounds on demand responses under counterfactual prices. In both contexts, the maximum risk of different choices is available in closed form or can be computed by solving a standard optimization problem, e.g., a linear program. 
 
To elaborate on practicality a little, consider the two competing paradigms: Bayes and minimax. In our setting, the usual Bayes decision rule requires specifying a prior on the parameter space for $(\theta,\mu)$, computing the posterior for $(\theta,\mu)$ having observed the data, then choosing the decision that minimizes posterior risk. A common criticism of Bayes decisions (see, e.g., \cite{Manski2021Haavelmo}) is that they are only justified if one can elicit a credible subjective prior, which can be difficult in practice, and that the resulting decision will depend, to some extent, on the decision maker's choice of prior. While this is true under both point- and partial identification, the problem is more severe under the latter because the prior for $\theta$ is not updated by the data (e.g., \cite{MoonSchorfheide2012}). Thus, even asymptotically, the decision will depend on the prior for $\theta$. By contrast, our (quasi-)Bayesian rules only require specifying a prior for $\mu$ and our decisions are asymptotically independent of this choice of prior. Moreover, our bootstrap rules sidestep the choice of a prior altogether.

Minimax decisions minimize the maximum risk over the parameter space for $(\theta,\mu)$. While this may be desirable, it is often not feasible. To derive minimax decisions one usually has to make strong assumptions on the data-generating process and restrict the dependence of payoffs on parameters to be of a very simple form. Indeed, we are not aware of any work deriving a minimax treatment rule under partial identification even in the simplest case of a binary outcome and binary treatment when randomization is not permitted and bounds on the ATE must be estimated from data. Algorithms such as that of \cite{Chamberlain2000} may be used to compute approximate minimax decisions, but the performance gap between these and the true minimax decision can be difficult to quantify. Adopting a minimax approach with respect to $\theta$ and a Bayes approach with respect to $\mu$ lends a great deal of tractability, allowing us to derive asymptotically optimal rules for a very broad class of empirically relevant settings where minimax rules are intractable.

We develop an asymptotic optimality theory for decisions based on parametric and semiparametric models. For parametric models, our optimality criterion extends the asymptotic average risk criterion introduced by \cite{HiranoPorter2009} for point-identified settings to partially identified settings. We further extend this notion to semiparametric models via a least favorable parametric submodel. Both of these extensions represent new contributions to the literature on asymptotic optimality for statistical decision rules. Our main results show formally that the proposed (quasi-)Bayesian rule is asymptotically optimal. Moreover, any decision rule that is asymptotically equivalent to the (quasi-)Bayes rule is optimal as well. This includes an implementation that replaces averaging under a (quasi-)posterior by averaging under a bootstrap approximation of the sampling distribution of an efficient estimator $\hat \mu$ of $\mu$, in cases in which these two distributions are asymptotically equivalent. 

Importantly, we show asymptotic equivalence to the (quasi-)Bayes or bootstrap rules is \emph{necessary} for optimality: any decision whose asymptotic behavior is different from these is sub-optimal. It follows that ``plug-in'' rules, which plug an efficient estimator $\hat \mu$ into the oracle decision rule if $\mu$ were known, can perform sub-optimally.\footnote{\cite{Manski2021Haavelmo,Manski2021} refers to plug-in rules as ``as-if'' optimization, because estimators $\hat \mu$ are treated as if they are the true parameters.} \cite{Manski2021Haavelmo,Manski2021} shows in numerical experiments that plug-in rules may perform poorly under finite-sample minimax regret criteria. Our necessity result provides a complementary and quite general theoretical explanation for why plug-in rules may perform poorly, albeit under a different (but related) optimality criterion.\footnote{For the intuition, consider a treatment assignment problem under partial identification of the ATE. Oracle rules depend on a robust welfare contrast $b(\mu)$ formed from bounds on the ATE. The bounds are non-smooth functions of $\mu$ in many empirically relevant settings reviewed in Section~\ref{sec:treatment}. This non-smoothness leads to a failure of the $\delta$-method that breaks the asymptotic equivalence between plug-in rules, which depend on $b(\hat \mu)$, and optimal rules, which depend on the average of $b(\cdot)$ across a bootstrap or posterior distribution.} In our empirical application to the adoption of a job-training program, our optimal decision produces different treatment recommendations than the plug-in rule for some sub-populations.

On the technical side, partial identification often leads to non-differentiability of key components of the maximum risk function in $\mu$. We address this challenge by extending arguments in \cite{HiranoPorter2009} for models with smooth welfare contrasts to settings with directional (but not full) differentiability. An important technical building-block is the derivation of the asymptotic distribution of the (quasi-)posterior mean of directionally differentiable functions in both parametric and semiparametric models (see Propositions~\ref{prop:BVM} and~\ref{prop:BVM.semiparametric} in Appendix~\ref{appsec:BVM}). These results are of independent interest. While \citet{KitagawaOleaPayneVelez2020} derived the asymptotic behavior of the posterior distribution of directionally differentiable functions, we instead characterize the large-sample (frequentist) distribution of the posterior mean of such functions. Our results offer a novel contribution to the literature on asymptotics for non-smooth functions \citep{Dumbgen1993,FangSantos2019}. We also introduce the concept of $\sigma$-optimality to handle settings in which the average (under Lebesgue measure) maximum risk is infinite at the optimal decision.

Our paper complements prior work on treatment assignment under partial identification, including  \cite{Manski2000,Manski2007treatment,Manski2020Meta,Manski2021Haavelmo,Manski2021}, \cite{Chamberlain2011}, \cite{Stoye2012}, \cite{Russell2020}, \cite{IK}, and \cite{Yata2021}. Except for \cite{Chamberlain2011}, these works seek decision rules that are optimal under finite-sample minimax regret criteria. We depart from these works in two respects. First, our asymptotic framework allows us to relax restrictive parametric assumptions and accommodate a much broader class of data-generating processes, including semiparametric models. Our optimality results apply to settings where the decision maker cannot confidently assert that the data are drawn from a given parametric model, or where bounds on the ATE are estimated using a vector of moments or summary statistics (e.g. regression or IV estimates from observational studies) whose  finite-sample distribution is unknown.\footnote{Finite-sample results are sometimes developed for this case assuming Gaussianity of the statistics,  by arguing that the studies are sufficiently large that the sampling distribution of the statistics is approximately normal. In these cases, it seems logically consistent to use a large-sample optimality criterion.} Second, except for \cite{Chamberlain2011}, these works use optimality criteria that (in our notation) are minimax over $(\theta,\mu)$, whereas our criterion is minimax over the partially-identified parameter $\theta$ and averages over the point-identified parameter $\mu$, reflecting the asymmetric parameterization of the problem. 

Our approach is related to  the multiple priors framework of \cite{GilboaSchmeidler1989} and a particular form of robust Bayes decision making, called conditional $\Gamma$-minimax; see \cite{DasGuptaStudden1989}, \cite{BetroRuggeri1992}, and the recent survey by \cite{GiacominiKitagawaRead2021}. We show how, under specific informational assumptions, our decision rule arises from a two-player zero-sum game between a decision maker and an adversarial nature. In the conditional $\Gamma$-minimax version of this game, nature chooses a prior for $\theta \in \Theta_0(\mu)$ conditional on the reduced-form parameter $\mu$ and the data that are available for estimation of $\mu$.

In the optimal pricing application we consider an environment in which a monopolist is choosing to set a price but faces ambiguity about the true distribution of demand. As in \cite{BergemannSchlag2011}, we assume the decision maker has a preference for robustness. However, we  use revealed preference demand theory to derive bounds on demand, rather than assuming true demand is in a neighborhood of a pre-specified distribution. Moreover, in our setting the decision maker must estimate the bounds from data.  This application builds on prior work on revealed-preference demand theory, including \cite{Blundelletal2007,BlundellBrowningCrawford2008}, \cite{Blundelletal2014,Blundelletal2017}, \cite{HoderleinStoye2015}, \cite{Manski2007,Manski2014}, and \cite{KitamuraStoye2018,KitamuraStoye2019}. These works are primarily concerned with testing rationality or deriving bounds on demand. We instead focus on using bounds to solve an optimal pricing problem, and using  demand data efficiently in that context. 

The remainder of the paper is structured as follows. Section~\ref{sec:treatment.empirical} presents an application to treatment assignment based on extrapolating meta-analyses. We first discuss the oracle decision based on a known $\mu$, then describe our proposed optimal decision rule in this context, and finally provide an empirical illustration. Section~\ref{sec:decisions} outlines our general framework for optimal decision making, discusses Bayesian and bootstrap implementations, and provides comparisons to minimax and conditional $\Gamma$-minimax decisions. The large sample optimality theory is presented in Section~\ref{sec:optimality} for parametric models  and Section~\ref{sec:semiparametric} for semiparametric models. Further applications to other treatment assignment problems and optimal pricing are discussed in Sections~\ref{sec:treatment} and~\ref{sec:pricing}, respectively. Finally, Section~\ref{sec:conclusion} concludes. Proofs and additional results and lemmas are relegated to the Appendix.

\section{Treatment Assignment}
\label{sec:treatment.empirical}

To set the stage, we will present a specific treatment assignment application in Section~\ref{subsec:treatment.empirical.problem}, describe the proposed decision rule in Sections~\ref{subsec:treatment.empirical.oracle} and~\ref{subsec:treatment.empirical.implementation}, and implement it numerically for the application in Section~\ref{subsec:treatment.empirical.numerics}.

\subsection{The Basic Problem and an Application} 
\label{subsec:treatment.empirical.problem}

Suppose a social planner is choosing whether or not to introduce a treatment in a target population. The social planner does not know the target population's ATE, defined as $\theta = \mathbb{E}[Y]$, where $Y:=Y_1-Y_0$ is the treatment effect associated with any individual, $Y_1$ is the outcome under treatment and $Y_0$ the outcome in the absence of treatment. Instead, the social planner observes a meta-analysis consisting of estimates $\hat{\mu}_k$ of the ATE $\mu_k$ in several related populations $k=1,\ldots,K$, their standard errors $s_k$, characteristics $x_k$ of the related populations, and characteristics $x_0$ of the target population. 

For concreteness, we revisit Example~2 of \cite{IK}. The authors consider a subset of $K = 14$ studies form the database of \cite{CardKluveWeber2017}, each of which is an RCT looking at the impact of job training programs on employment. Studies are implemented in a number of different countries and in groups that differ by characteristics $x_k$ consisting of gender (males, females, or both), age (youths, adults, or both), OECD membership status, GDP growth (standardized) and unemployment (standardized). We consider the hypothetical question of whether to roll out a job-training program in two populations: German male youths in 2010 and German female youths in 2010 (with GDP growth of 3.48\% and an unemployment rate of 9.45\%).

To connect the RCT studies to the ATE $\theta$ associated with the target population, suppose that treatment effects are Lipschitz in the distance between size-specific covariates:
\begin{equation}
	|\mu_k - \theta| \leq C \|x_k - x_0\| , \quad k = 1,\ldots,K,
	\label{eq:lipschitz}
\end{equation}
where $C$ is a pre-specified constant. We refer to $\mu := (\mu_1,\ldots,\mu_k)$ as the reduced-form parameter. It is assumed to be consistently estimable from the data and its estimate is denoted by $\hat{\mu}$. We obtain the following lower ($L$) and upper ($U$) bounds for $\theta$: 
\begin{equation} \label{eq:bounds.treatment.empirical}
	b_L(\mu) :=  \max_{1 \leq k \leq K} (\mu_k - C \|x_0 - x_k\|) \leq \theta \leq \min_{1 \leq k \leq K} (\mu_k + C \|x_0 - x_k\|) =: b_U(\mu) \,.
\end{equation}
The bounds in (\ref{eq:bounds.treatment.empirical}) define an identified set for $\theta$ as a function of $\mu$: 
\begin{equation}
	\Theta_0(\mu) := [b_L(\mu), b_U(\mu)].
\end{equation}

\subsection{Oracle Decision} 
\label{subsec:treatment.empirical.oracle}

For now we will assume that $\mu$ is known. Following \cite{Manski2000,Manski2004}, it is common to derive treatment rules under a utilitarian social welfare function that is linear in the target population's ATE. Interpreting negative utility as loss, we write
\[
  l(d,\theta) = -d \theta,
\]
where $d \in \{0,1\}$ indicates treatment. The loss function is minimized by the treatment decision $\mathbb{I}[ \theta \ge 0]$, where 
$\mathbb{I}[\cdot]$ is the indicator function that is equal to one if its argument is true and zero otherwise. Following \cite{Manski2000,Manski2004}, we 
subsequently use the 
regret criterion
\begin{equation}
  r(d,\theta) = l(d,\theta) - \min_{d'} l(d',\theta)  = \big( \mathbb{I}[ \theta \geq 0] - d \big) \theta. \label{eq:regret.loss}
\end{equation}
Failure to treat ($d=0$) incurs zero regret when $\theta <0 $, otherwise the regret is $\theta$. Similarly, treating ($d=1$) incurs zero regret when $\theta \geq 0$, otherwise the regret is $-\theta$. 

If the social planner knows $\mu$, they can choose the decision that minimizes maximum regret over $\Theta_0(\mu)$, defined as
\[
	R(d,\mu) := \sup_{\theta \in \Theta_0(\mu)} r(d,\theta).
\]
In our specific application we obtain
\begin{equation}
	R(0,\mu) = \left( b_U(\mu) \right)_+ \quad \mbox{and} \quad R(1,\mu) = -\left( b_L(\mu) \right)_-\,,
    \label{eq:regret.maximum}	
\end{equation}
where $(a)_+ := \max\{a,0\}$ and $(a)_- := \min\{a,0\}$. 
This leads to the \emph{oracle decision}
\begin{equation}\label{eq:oracle_treatment}
\delta^o = \mb I\Big[ b(\mu) \geq 0 \Big]
\end{equation}
that assigns treatment if the \emph{robust welfare contrast}
\[
b(\mu) :=  \left( b_U(\mu) \right)_+ + \left( b_L(\mu) \right)_-
\]
is non-negative, and non-treatment otherwise.

Notice from plugging  (\ref{eq:bounds.treatment.empirical}) into  (\ref{eq:regret.maximum}) that the max and min operators make $R(0,\mu)$ and $R(1,\mu)$ only directionally differentiable with respect to $\mu$. Directional differentiability of the maximum risk function is a generic feature of all the applications discussed in this paper. It generates technical challenges that we tackle in our large sample analysis and distinguishes our analysis from that in \cite{HiranoPorter2009}.

\subsection{Optimal Decision Rule}
\label{subsec:treatment.empirical.implementation}

The oracle rule is an infeasible first-best as it requires knowledge of the true $\mu$. In practice $\mu$ is unknown and any practical rule must also confront sampling uncertainty about $\mu$. One option is to simply plug $\hat \mu$ into (\ref{eq:oracle_treatment}), which leads to the plug-in rule
\begin{equation} \label{eq:example.treatment.plug}
 \delta^{plug}_n  = \mb I\Big[ b(\hat \mu) \geq 0 \Big]. 
\end{equation}
However, this rule does not account for the precision of $\hat \mu$. \cite{Manski2021} explores how different estimates of parameters affect the performance of plug-in rules.

We develop an optimality theory in Sections~\ref{sec:decisions} to~\ref{sec:semiparametric} below to guide decision-making in situations such as these where a payoff-relevant parameter $\theta$ is partially identified and its identified set $\Theta_0(\mu)$ is indexed by a parameter $\mu$ which can be estimated from data. In the context of the treatment assignment application, the procedure amounts to replacing $b(\mu)$ in (\ref{eq:oracle_treatment}) by its average $\bar b_n$ under a posterior for $\mu$ (in parametric models) or quasi-posterior for $\mu$ (in semiparametric models):
\begin{equation} \label{eq:example.treatment.optimal}
\delta^*_n = \mb I\Big[\bar b_n \geq 0\Big].
\end{equation}
We show formally in Sections~\ref{sec:optimality} and~\ref{sec:semiparametric} that this decision is more efficient than the plug-in rule, as it takes into account the sampling uncertainty in  $\hat \mu$.

\subsection{Empirical Illustration} 
\label{subsec:treatment.empirical.numerics}

We use a  limited-information quasi-posterior distribution for $\mu$ that interprets the sampling distribution  $\hat{\mu}_k \stackrel{\mbox{\tiny approx}}{\sim} N(\mu_k,s_k^2)$ as a quasi-likelihood. The estimates $\hat \mu_k$ are independent as they come from independent RCTs. Following \cite{DoksumLo1990}, \cite{Kim2002}, and \cite{Mueller2013}, we combine the $N(\mu, \hat \Sigma)$ quasi-likelihood for $\hat \mu$ with a flat prior for $\mu$ to obtain a $N(\hat \mu, \hat \Sigma)$ quasi-posterior for $\mu$, where $\hat \Sigma$ denotes a diagonal matrix with $(s_k^2)_{k=1}^K$ down its diagonal. The quasi-posterior mean of $b(\mu)$, denoted by $\bar{b}_n$, is computed by drawing independent  $\mu \sim N(\hat \mu,\hat \Sigma)$, computing $b(\mu)$, then averaging across a large number of draws. The efficient decision from (\ref{eq:example.treatment.optimal}) is then to treat if $\bar b_n$ is non-negative, and not treat otherwise.

\begin{figure}[t!]
	\noindent
	\makebox[\textwidth]{
		\begin{subfigure}{.5\textwidth}
			\centering
			\includegraphics[width=\linewidth]{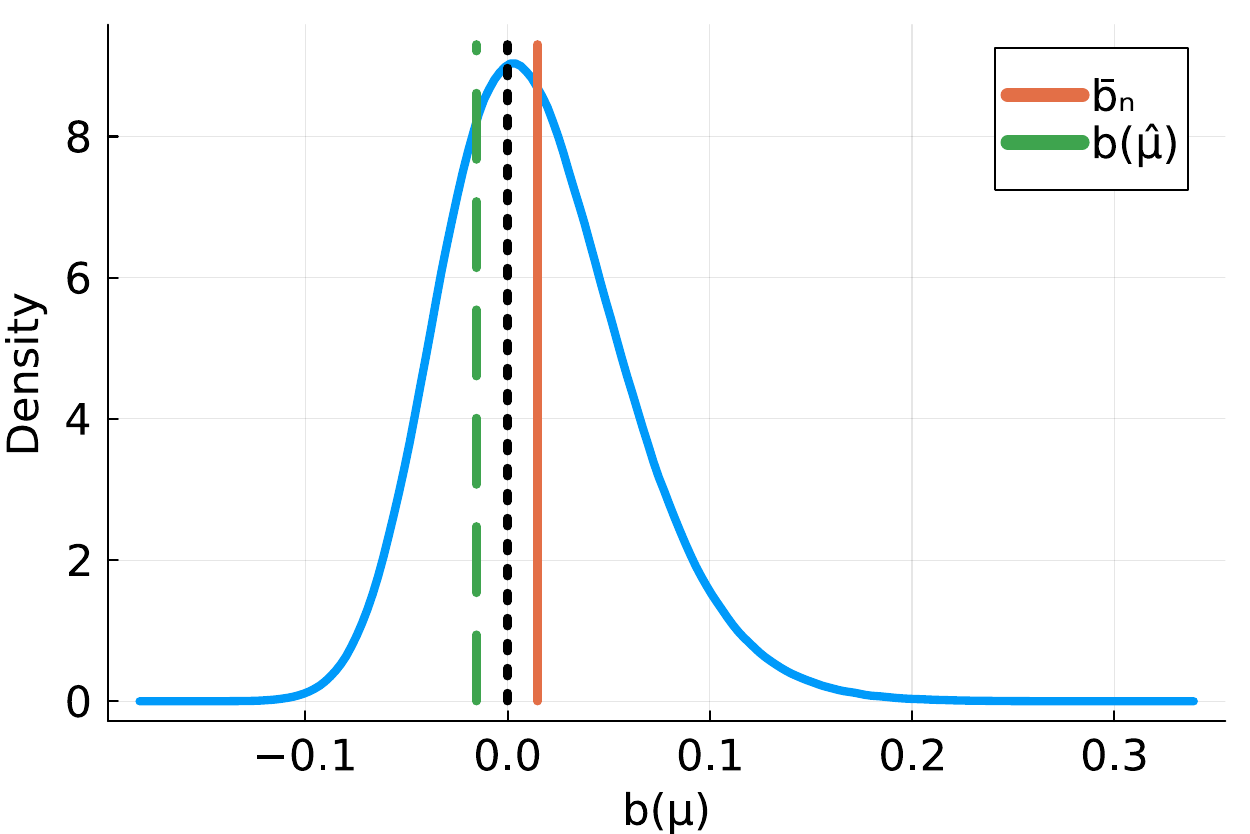}
			\caption{Male youths.} 
			\label{f:ik_de_male}
		\end{subfigure}%
		\begin{subfigure}{.5\textwidth}
			\centering
			\includegraphics[width=\linewidth]{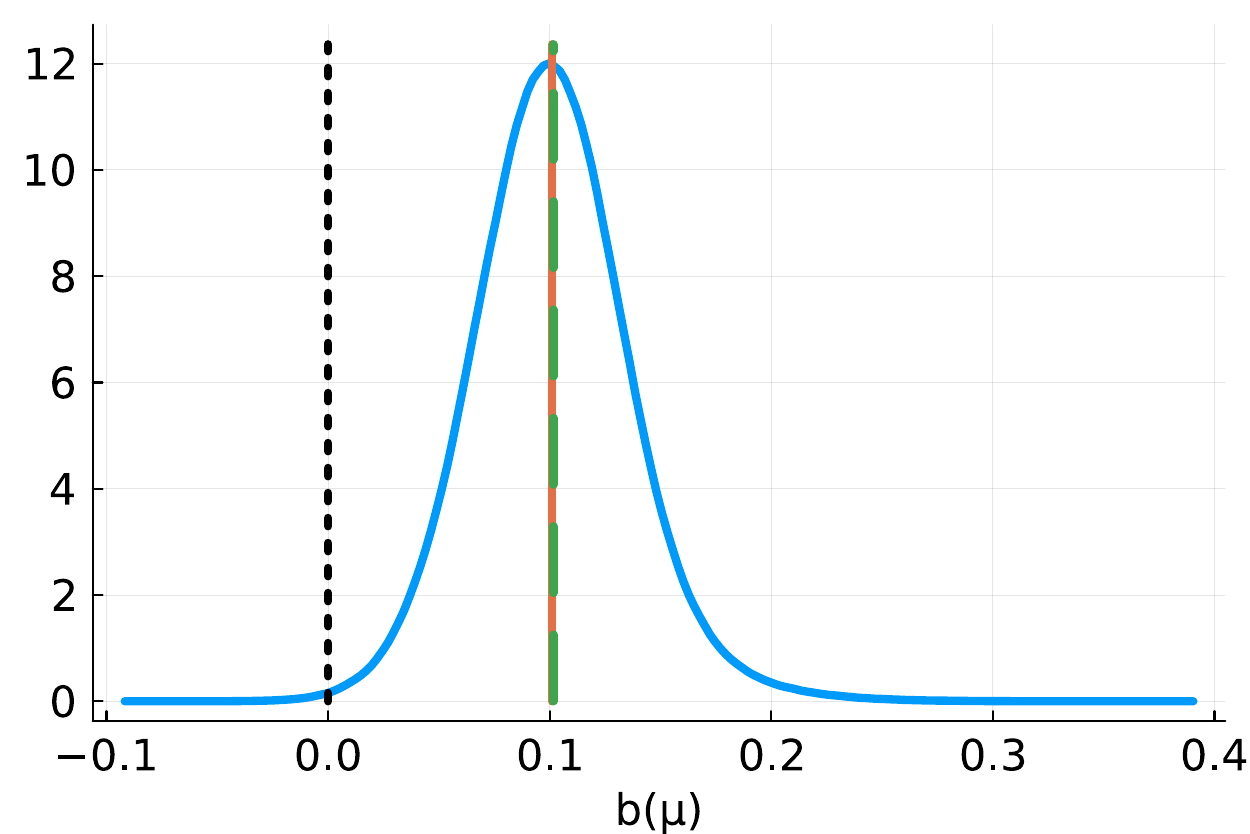}
			\caption{Female youths.} 
			\label{f:ik_de_female}
		\end{subfigure}
	}
	\centering
	
	\vskip 4pt
	
	\parbox{\textwidth}{\caption{\label{f:ik_de} Distribution of the robust welfare contrast $b(\mu)$ for $\mu \sim N(\hat \mu,\hat \Sigma)$, its quasi-posterior mean $\bar b_n$, and the plug-in value $b(\hat \mu)$.}}
\end{figure}

We compute the optimal decision as described above using data from \cite{IK}.  We increase the Lipschitz constant in (\ref{eq:lipschitz}) from $C = 0.025$ in their paper to $C = 0.25$ so that the bounds on $\theta$ are non-empty. Figures \ref{f:ik_de_male} and \ref{f:ik_de_female} plot the quasi-posterior distribution of $b(\mu)$ under $\mu \sim N(\hat \mu,\hat \Sigma)$ for males and females, respectively. Both figures display the mean $\bar b_n$ whose sign determines the optimal treatment decision $\delta^*_n$ in (\ref{eq:example.treatment.optimal}). We also display the plug-in value $b(\hat \mu)$ whose sign determines the plug-in rule $\delta^{plug}_n$ in (\ref{eq:example.treatment.plug}).

In Figure \ref{f:ik_de_male} (male youths) we see that the distribution has a pronounced right-skew. Recall that the lower bound is $\max_{1 \leq k \leq K} (\mu_k - C \|x_0 - x_k\|)$. When evaluated at $\hat \mu$, the largest value of $\hat \mu_k - C \|x_0 - x_k\|$ is $-0.3190$ (corresponding to a US study) and the second-largest is $-0.3298$ (corresponding to a Brazilian study). As the maximum is not well separated relative to sampling uncertainty (the average $s_k$ across the studies is $0.034$), the distribution of the lower bound is right-skewed because it behaves like a maximum of two Gaussians. The upper bound is less skewed because the minimum (corresponding to a Colombian study) is better separated. As a consequence of this asymmetry, the  mean $\bar b_n$ is positive and the optimal decision is to treat. By contrast,  the plug-in value $b(\hat \mu)$ is negative so the plug-in decision is to not treat. In Section~\ref{sec:optimality} we show this difference between the two rules, where the optimal rule predicts treatment more aggressively than the plug-in rule, is as predicted by our optimality theory.

The situation is different for female youths (Figure \ref{f:ik_de_female}). Here the minima and maxima are relatively better separated, so the distribution of $b(\mu)$ for $\mu \sim N(\hat \mu,\hat \Sigma)$ is close to Gaussian. In consequence, the mean $\bar b_n$ and the plug-in value $b(\hat \mu)$ are almost identical, and the optimal and plug-in decisions are both to treat.

\section{Optimal Decisions and Their Implementation}
\label{sec:decisions}

The treatment assignment application in the previous section imposed a lot of structure which we will now relax. We begin by stating the general decision problem in Section~\ref{subsec:decisions.problem}. Many empirical examples are covered by our framework: see Sections~\ref{sec:treatment} and~\ref{sec:pricing} for applications to treatment assignment and optimal pricing and \cite{CMS2020forecast} for an application to forecasting. We develop our concept of optimal decision rules in Section~\ref{subsec:decisions.efficient.robust}. Bayesian and bootstrap implementations are presented in Sections~\ref{subsec:decisions.bayesian} and~\ref{subsec:decisions.bootstrap}. Finally, in Section~\ref{sec:decisions.discussion} we discuss a two-player zero-sum game representation of our setup and its relationship to minimax and robust Bayesian decision making.

\subsection{Decision Problem}
\label{subsec:decisions.problem}

A decision maker (DM) observes $X^n \sim P_{n,\mu}$ taking values in a sample space $\mc X^n$. Often $X^n$ may be a sample $(X_1,\ldots,X_n)$ of size $n$ with joint distribution $P_{n,\mu}$. But $X^n$ may also be a vector of estimators (e.g., OLS or IV estimators) with sampling distribution $P_{n,\mu}$. The distribution $P_{n,\mu}$ is indexed by a  \emph{reduced-form parameter} $\mu \in \mc M$. 

After observing $X^n$, the DM chooses an action $d$ in the action space $\mc D = \{0,1,\ldots,D\}$. A \emph{statistical decision rule} $\delta_n : \mc X^n \to  \mc D$ maps realizations of the data into actions.\footnote{We consider nonrandomized rules as opposed to randomized rules where the action space is the set of all distributions over $\{0,1,\ldots,D\}$.}
The DM's utility $u(d,Y,\theta,\mu)$ from choosing $d$ depends on a random variable $Y \sim G_\theta$ independent of $X^n$, a \emph{structural parameter} $\theta \in \Theta$, and $\mu$. In a treatment assignment problem, $X^n$ may represent data or summary statistics from an experiment, $Y$ may represent the treatment effect for an individual sampled at random from the remaining population (hence independent of $X^n$), and $\E_\theta[Y]$ may represent the average treatment effect. Both $G_\theta$ and $\E_\theta[\,\cdot\,]$ can be conditional on covariates but we suppress this to simplify notation.

We interpret negative utility as loss and write $l(d,Y,\theta,\mu) \equiv -u(d,Y,\theta,\mu)$. The DM incurs this loss once $Y$ is realized. Here it is helpful to draw a distinction between the \emph{ex-post problem} that the DM faces after observing $X^n$ but before $Y$ is realized, and the \emph{ex-ante problem} the DM faces before $X^n$ is realized. In the ex-post problem, the DM faces risk
\[
 r(d,\theta,\mu) := \E_\theta[l(d,Y,\theta,\mu)]
\] 
from choosing $d \in \mc D$, where the expectation is taken with respect to $Y \sim G_\theta$.\footnote{This notation nests welfare regret by setting $l(d,y,\theta,\mu) = \max_{d' \in \mc D} \; \E_\theta [W(d',Y,\theta,\mu)] - W(d,y,\theta,\mu)$
for a welfare function $W$. Hence, we do not distinguish between ``risk'' and ``regret'' in what follows.} We refer to $r$ as ``risk'' because it involves taking an expectation with respect to the random outcome $Y$ in the ex-post problem. One can equally view $r$ as a loss function in the DM's ex-ante problem of choosing a decision rule $\delta_n$ before $X^n$ is observed.

An important aspect of this paper is that we allow the payoff-relevant parameter $\theta$ to be set-identified. That is, the most that the DM can infer about $\theta$ is that $\theta \in \Theta_0(\mu) \subseteq \Theta$, where $\Theta_0(\cdot)$ is a known set-valued mapping from $\mathcal M$ to $\Theta$. While $X^n$ can be used to learn about $\mu$, it contains no information about the identity of the true $\theta$ within $\Theta_0(\mu)$.

\bigskip

\paragraph{Example: Treatment Assignment.}

Consider the setup in Section~\ref{sec:treatment.empirical}: $d \in \{0,1\}$ is a binary treatment indicator, $\theta$ is the ATE, and $r(d,\theta,\mu) \equiv r(d,\theta)$ is given in display~(\ref{eq:regret.loss}). 
The DM has bounds $b_L(\mu)$ and $b_U(\mu)$ on $\theta$ as a function of $\mu$ and observes data or summary statistics $X^n \sim P_{n,\mu}$ that may be used to learn $\mu$. In the application from Section~\ref{sec:treatment.empirical}, the statistics were the vector of ATE estimators $X^n = (\hat \mu_1,\ldots,\hat \mu_k)$, $P_{n,\mu}$ is their sampling distribution, the reduced-form parameter is the vector of ATEs $\mu = (\mu_1,\ldots,\mu_k)$, and $b_L(\mu)$ and $b_U(\mu)$ are given in (\ref{eq:bounds.treatment.empirical}). We review other methods for constructing bounds in different empirical settings in Section \ref{sec:treatment}. The identified set is $\Theta_0(\mu) = [b_L(\mu) ,b_U(\mu)]$. $\square$

\subsection{Optimality}
\label{subsec:decisions.efficient.robust}

Our setup features two types of parameters: a point-identified parameter $\mu$ and a set-identified parameter $\theta$. We adopt a minimax approach to handle the ambiguity that arises from the partial identification of $\theta$ conditional on $\mu$ and average (or integrated) risk minimization to estimate $\mu$. Our asymmetric treatment of parameters is in the spirit of \cite{Hurwicz1951}; we discuss the connection in more detail in Section~\ref{subsec:decisions.bayesian}. We first consider the case in which $\mu$ is known and then extend the analysis to the case of unknown $\mu$, which will lead us to the definition of our optimality concept.

\bigskip

\paragraph{Known $\boldsymbol \mu$.} If the DM knows $\mu$, then the data are irrelevant and the germane problem is the ex-post problem in which $d$ is chosen. To handle the ambiguity about $\theta \in \Theta_0(\mu)$, the DM evaluates actions by their maximum risk
\begin{equation}\label{eq:maxrisk}
 R(d,\mu) := \sup_{\theta \in \Theta_0(\mu)} r(d,\theta,\mu) .
\end{equation}
Define
\[
 \delta^o(\mu) = \argmin_{d \in \mc D} R(d,\mu),
\]
if the argmin is unique, otherwise $\delta^o(\mu)$ is chosen randomly from $\argmin_{d \in \mc D} R(d,\mu)$. This choice can be interpreted as the equilibrium in a two-player zero-sum game in which an adversarial nature chooses $\theta$ in response to the DM's choice of $d$ to minimize $\mathbb{E}_\theta[u(d,Y,\theta,\mu)]$. We call $\delta^o(\mu)$ the \emph{oracle} decision: it represents the DM's optimal choice if $\mu$ was known. The oracle decision is infeasible in any practical application because $\mu$ will need to be estimated.  Nevertheless, it serves as a useful benchmark because $R(\delta_n(X^n),\mu) \geq R(\delta^o(\mu), \mu) \equiv \min_{d \in \mc D} R(d,\mu)$ for any data-dependent decision rule $\delta_n(X^n)$.

\bigskip

\paragraph{Unknown $\boldsymbol \mu$.} Here the DM uses the data $X^n$ to learn about $\mu$ before choosing $d \in \mc D$. We treat $R(d,\mu)$ as a loss function for the DM's ex-ante problem. For a decision rule $\delta_n$, the DM incurs ex-ante risk
\begin{equation} \label{eq:amr}
 \mb E_\mu [R(\delta_n(X^n), \mu)],
\end{equation}
where $\E_\mu[\,\cdot\,]$ denotes expectation with respect to $X^n \sim P_{n,\mu}$. Our goal is to construct sequences $\{\delta_n\}$ of decision rules that use the data efficiently, so that criterion (\ref{eq:amr}) is minimized in large samples over a range of data-generating processes. 

To simplify exposition, we consider parametric models in the rest of this section and defer discussion of semiparametric models to Section~\ref{sec:semiparametric}.  Following \cite{HiranoPorter2009}, we work in a local asymptotic framework centered at a fixed $\mu_0 \in \mathcal M \subseteq \mathbb R^K$ and parameterized by a local parameter $h = \sqrt n(\mu - \mu_0)$ ranging over $\mathbb R^K$. As $h$ is not consistently estimable, this asymptotic framework is designed to mimic the finite-sample problem faced by the researcher, where $\mu$ is not known with certainty. We let $P_{n,h}$ denote the distribution of $X^n$ when $\mu = \mu_0 + h/\sqrt n$ and let $\E_{n,h}$ denote expectation under $P_{n,h}$. 

We rank decision rules by their ex-ante risk in excess of the oracle. We scale excess risk by $\sqrt n$ so that the large-sample limit is not degenerate. This leads to the criterion
\begin{equation}  \label{eq:criterion_asymptotic_h}
  \mc R(\{\delta_n\};\mu_0, h) =  \limsup_{n \to \infty} \sqrt n \left( \E_{n,h}\left[R\big(\delta_n(X^n), \mu_0 + h/\sqrt n \big)\right] - \min_{d\in \mc D} R\big(d,\mu_0 + h/\sqrt n \big) \right).
\end{equation}
As there is no a priori reason to assign more weight to one local parameter than another, we integrate with respect to Lebesgue measure over $h$ to arrive at the criterion
\begin{equation} \label{eq:criterion_asymptotic}
 \mc R(\{\delta_n\};\mu_0) = \int \mc R(\{\delta_n\};\mu_0, h) \, d h.
\end{equation}
We rank sequences of decision rules by criterion (\ref{eq:criterion_asymptotic}). For now we assume $\mc R(\{\delta_n\};\mu_0)$ is finite for at least one sequence of decisions $\{\delta_n\}$, so that criterion (\ref{eq:criterion_asymptotic}) provides a meaningful ranking. Section~\ref{subsec:finiteness} discusses extensions to settings where this criterion is infinite and the improper Lebesgue prior on $h$ is approximated by a sequence of proper priors. 

Before introducing our definition of optimality, we first define a class $\mathbb D$ of sequences of decision rules for which the above criteria are well defined. Let 
\[
 \underline{\mathcal D}_{\mu_0} = \argmin_{d \in \mathcal D} R(d,\mu_0)
\]
denote the set of optimal choices at $\mu_0$. Note that $\underline{\cal D}_{\mu_0}$ will be a non-singleton at values of $\mu_0$ where there are multiple choices that minimize $R(d,\mu_0)$.

\begin{definition} \label{def:D}
$\mathbb D$ consists of all sequences $\{\delta_n\}$ such that, for all $\mu_0 \in \mathcal M$,
\begin{enumerate}[label={(\roman*)}, nosep]
\item\label{D.1} $\lim_{n \to \infty} P_{n,h}(\delta_n(X^n) = d)$ exists for all $d \in \mathcal D$ and $h \in \mathbb R^K$;
\item\label{D.2} $\lim_{n \to \infty} \sqrt n \, P_{n,h} \left( \delta_n(X^n) \not \in \underline{\mathcal D}_{\mu_0} \right) = 0$
for all $h \in \mathbb R^K$.
\end{enumerate}
\end{definition}

Definition~\ref{def:D}\ref{D.1} requires that the sequence of random variables $\{\delta_n(X^n)\}$ converges in distribution along $\{P_{n,h}\}$. Definition~\ref{def:D}\ref{D.2} is a minor technical condition requiring $\delta_n$ to choose among $\underline{\mc D}_{\mu_0}$ with high probability under small perturbations of $\mu$ around $\mu_0$. For the intuition, by the maximum theorem we know that  optimal actions under $\mu$ near $\mu_0$ should belong to $\underline{\mc D}_{\mu_0}$. If $d_n$ chooses actions outside this set, then it makes mistakes: it chooses actions that should be easy to identify as sub-optimal. Condition~\ref{D.2} requires the probability of these mistakes vanishes sufficiently fast. Bayes decisions satisfy this condition: their mistake probabilities vanish at rate $n^{-1}$ (see Lemma~\ref{lem:mistake_prob_bayes} in Appendix~\ref{appsec:mistake}). 

We refer to sequences $\{\delta_n\}$ as optimal if they satisfy the following definition:

\begin{definition}
	\label{def:efficient.robust}
	Say $\{\delta_n\} \in \mb D$ is \textit{\textbf{optimal}} if $\mc R(\{\delta_n\};\mu_0) = \inf_{\{\delta_n'\} \in \mathbb D} \mc R(\{\delta_n'\}; \mu_0) $ for all $\mu_0 \in \mc M$.
\end{definition}

This is a non-standard problem and there are different notions of optimality that may lead to different rankings over sequences of decision rules, as we discuss in Section~\ref{sec:decisions.discussion} below. An attractive feature of our optimality criterion is that the DM only needs to be able to compute $R(d,\mu)$ in order to implement optimal decisions. This makes our approach broadly applicable including in scenarios when $R(d,\mu)$ has no closed-form expression, such as those in Sections~\ref{sec:treatment} and~\ref{sec:pricing}. In the next subsections we present Bayesian and bootstrap implementations.

\subsection{Bayesian Interpretation and Implementation}
\label{subsec:decisions.bayesian}

Integrating criterion (\ref{eq:amr}) with respect to a prior $\pi$ yields the \emph{integrated maximum risk} criterion 
\begin{equation}\label{eq:imr}
	\int \mb E_\mu [R(\delta_n(X^n), \mu)] \, d\pi(\mu) .
\end{equation}
After observing $X^n$, the DM can form a posterior $\pi_n(\mu) = \pi_n(\mu|X^n)$ for $\mu$. Standard arguments (e.g. \citeauthor{Wald1950}, \citeyear{Wald1950}, Chapter 5.1) imply that (\ref{eq:imr}) can be minimized by minimizing the posterior maximum risk
\begin{equation} \label{eq:pmr}
	\bar R_n(d) : = \int R(d,\mu) \, d \pi_n(\mu)
\end{equation} 
with respect to $d \in \mc D$ for almost every realization of the data. This leads to the Bayes rule
\begin{equation}\label{eq:decision}
	\delta_n^*(X^n;\pi) = \argmin_{d \in \mc D} \bar R_n(d)
\end{equation}
if the argmin is unique, otherwise $\delta_n^*(X^n;\pi)$ is chosen randomly from $\argmin_{d \in \mc D} \bar R_n(d)$. We include $\pi$ as an argument to indicate that the decision depends on $\pi$ in finite samples.

Bayes decisions $\delta_n^*(X^n;\pi)$ are  optimal under criterion (\ref{eq:criterion_asymptotic}) for any prior $\pi$ whose density is positive, bounded, and continuous (Theorem~\ref{t1}). To see the connection between (\ref{eq:criterion_asymptotic}) and~(\ref{eq:imr}), use a change-of-variables to express the prior for $h$ as $\propto \pi(\mu_0+ h/\sqrt{n})$. This prior becomes uniform as $n \to \infty$, which is the weight function underlying (\ref{eq:criterion_asymptotic}).

\medskip

\paragraph{Example: Treatment Assignment.}
Recall from Section~\ref{sec:treatment.empirical} that the maximum risks of $d \in \{0,1\}$ are the positive and negative parts of the bounds on the ATE:
\[
\begin{aligned}
R(0,\mu) &= (b_U(\mu))_+ \,, & & 
R(1,\mu) &= - (b_L(\mu))_- \,.
\end{aligned}
\]
Averaging  $R(1,\mu)$ and $R(0,\mu)$ across $\pi_n$ yields
\[
\begin{aligned}
\bar R_n(0) &= \int (b_U(\mu))_+ \,  d \pi_n(\mu)\,, &
\bar R_n(1) &= \int - (b_L(\mu))_- \,  d \pi_n(\mu) \,.
\end{aligned} 
\]
We have $\bar R_n(1)  \leq \bar R_n(0)$ if and only if $\bar b_n := \int b(\mu) \,  d \pi_n(\mu) \geq 0$ with $b(\mu) := (b_U(\mu))_+ + (b_L(\mu))_-$. Hence, the Bayes rule is  $\delta_n^*(X^n;\pi) = \mb I\left[ \bar b_n \geq 0\right]$.\footnote{This rule deterministically chooses treatment when there is a tie (i.e., when $\bar b_n = 0$). Any (possibly randomized) tie-breaking rule leads to decision which is optimal under our optimality criteria.}
We discuss implementation in a number of different empirical settings in Section \ref{sec:treatment}.
$\square$

\subsection{Bootstrap Implementation}
\label{subsec:decisions.bootstrap}

Let $\hat \mu$ denote an efficient estimator of $\mu$ and let $\E_n^*$ denote expectation with respect to the bootstrap version $\hat \mu^*$ of $\hat \mu$ conditional on $X^n$. Define the bootstrap average maximum risk
\[
 R_n^*(d) = \E_n^* \left[ R(d,\hat \mu^*) \right] .
\]
The bootstrap rule is
\[
 \delta_n^{**}(X^n) = \argmin_{d \in \mc D} R_n^*(d)
\]
if the argmin is unique (with a random selection from the argmin otherwise). We show in Theorem~\ref{t1} below that as long as $\delta_n^{**}$ behaves asymptotically like $\delta_n^*$ it inherits the optimality property of the Bayes decision.

\bigskip

\paragraph{Example: Treatment assignment.}
Here we have
\[
\begin{aligned}
 R_n^*(1) &= \E_n^* \left[ - (b_L(\hat \mu^*))_- \right] , &
 R_n^*(0) &= \E_n^* \left[ (b_U(\hat \mu^*))_+  \right].
\end{aligned} 
\]
Therefore, the bootstrap rule is $\delta_n^{**}(X^n) =  \mb I \left[ \E_n^* \left[  b(\hat \mu^*) \right] \geq 0\right]$. $\square$

\subsection{Further Discussion} 
\label{sec:decisions.discussion}

The first distinguishing feature of our analysis is that we separate two groups of parameters, $\theta$ and $\mu$, and apply the minimax reasoning only to $\theta$ because only it is partially identified. The approach of treating groups of parameters differently dates back to \cite{Hurwicz1951}. He argued that one might consider multiple priors for some parameters and referred to it as a generalized Bayes-minimax principle. He provided a two-parameter example in which the marginal prior  for one of the parameters, $\mu$ in our notation, is fixed, whereas a family of priors is considered for the conditional distribution of the second parameter $\theta$ given $\mu$. 

Recall that $\delta_n^*(\,\cdot\,;\pi)$ minimizes criterion~(\ref{eq:imr}) over all decision rules $\delta_n : \mc X^n \to \mc D$. By the definition of $R(d,\mu)$ and the discussion in Section~\ref{subsec:decisions.bayesian}, we see that $\delta_n^*(x^n;\pi)$ solves
\begin{equation}\label{eq:DM.risk.equations}
 \min_{d \in\mathcal D} \; \int_{\mathcal M} \bigg(\sup_{\theta \in \Theta_0(\mu)}\; r(d,\theta,\mu) \bigg) \, d\pi_n(\mu|x^n)
\end{equation}
for almost every realization $x^n$ of $X^n$, where $r(d,\theta,\mu) = \E_\theta[-u(d,Y,\theta,\mu)]$.
Unlike the usual minimax framework, we allow the adversary (``nature'') to choose $\theta$ conditional on $X^n$. This is the second distinguishing feature of our analysis. It creates a more adversarial setting than usual, but it is disciplined by restricting nature's choice to the identified set $\Theta_0(\mu)$ rather than the whole parameter space $\Theta$. 

In this regard, our approach is closely related to the derivation of conditional $\Gamma$-minimax decision rules in the Bayesian literature, e.g., \cite{DasGuptaStudden1989}, \cite{BetroRuggeri1992}, and \cite{GiacominiKitagawaRead2021}. One difference is that we keep the marginal prior distribution of $\mu$ fixed, which alleviates concerns about the conservativeness of the approach. Suppose one combines the unique prior $\pi$ for $\mu$ with a family $\Lambda$ of conditional priors $\lambda := \{\lambda(\cdot|\mu): \mu \in \mc M\}$ for $\theta$, where $\lambda(\theta|\mu)$ has support contained in $\Theta_0(\mu)$ for each $\mu \in \mc M$. Then one can define $\Gamma = \{ \pi \otimes \lambda : \lambda \in \Lambda\}$, where each $\gamma \in \Gamma$ is a prior over $(\theta,\mu)$.\footnote{We use the notation $\pi \otimes \lambda$ to denote the conditional probability measures $\lambda = \{\lambda(\cdot|\mu): \mu \in \mc M\}$ integrated against a marginal probability measure $\pi$ for $\mu$. Thus, for $\gamma = \pi \otimes \lambda$ and $g : \Theta \times \mc M \to \mb R$, we define $\int_{\Theta \times \mc M} g(\theta,\mu) d \gamma(\theta,\mu) = \int_{\mc M} \int_\Theta g(\theta,\mu) \, d \lambda(\theta|\mu) \, d \pi(\mu)$.} The conditional $\Gamma$-minimax decision solves the following min-max problem for almost every realization $x^n$ of $X^n$:
\begin{equation}
	\min_{d \in {\cal D}} \; \sup_{\gamma \in \Gamma} \;
	 \int_{\Theta \times \mc M} r(d,\theta,\mu)  \, d \gamma_n(\theta,\mu|x^n), \label{eq:DM.risk.equations.1}
\end{equation}
where $\gamma_n(\theta,\mu|x^n)$ is the posterior for $(\theta,\mu)$ given $X^n = x^n$ for the prior $\gamma \in \Gamma$. This is a robust Bayes criterion for the DM's ex-post problem faced after observing $X^n$. Accordingly, $X^n$ is treated as fixed: the DM and nature know $X^n$ when choosing their actions. 

In responding to the DM's choice of $d$ in (\ref{eq:DM.risk.equations.1}), nature has to choose a set of conditional priors $\{\lambda(\theta|\mu) : \mu \in {\cal M}\}$ where each $\lambda(\cdot|\mu)$ has support contained in $\Theta_0(\mu)$. For each $\mu \in {\cal M}$, define $\theta_*(\mu;d) \in \mr{arg}\sup_{\theta \in \Theta_0({\mu})} r(d,\theta,\mu)$, assuming that the arg sup is non-empty, and let $\lambda_*(\theta|\mu,d)$ be a point mass at $\theta_*({\mu};d)$. In combination, $\pi$ and $\{\lambda_*(\theta|\mu,d) : \mu \in \mc M\}$ define a joint prior $\gamma_* = \pi \otimes \lambda_*$ over $(\theta,\mu)$. Then, by construction, the posterior risk of $d$ under the $\gamma_*$ prior equals the posterior maximum risk
\[
	\int_{\Theta \times \mc M} r(d,\theta,\mu)   \, d \gamma_{*n}(\theta,\mu|x^n) \label{eq:risk.lambdastar} 
	= \int_{ \mc M} \bigg( \sup_{\theta \in \Theta_0(\mu)}   r(d,\theta,\mu) \bigg) \, d \pi_n(\mu|x^n).
\]
Because the supremum over $\theta \in \Theta_0(\mu)$ is weakly larger than the average under any $\lambda(\theta|\mu)$, we can deduce that, provided $\gamma_* \in \Gamma$, 
\begin{equation} \label{eq:DM.risk.condgammamm.max}
 \sup_{\gamma \in \Gamma} \;
	 \int_{\Theta \times \mc M}  r(d,\theta,\mu)  \, d \gamma_n(\theta,\mu|x^n)
	=
  \int_{ \mc M} \bigg( \sup_{\theta \in \Theta_0(\mu)} r(d,\theta,\mu) \bigg)  d \pi_n(\mu|x^n).
\end{equation}
Combining (\ref{eq:DM.risk.equations}) and (\ref{eq:DM.risk.condgammamm.max}), we see that the decision rules that are considered optimal under Definition~\ref{def:efficient.robust} can be viewed as solutions to a conditional $\Gamma$-minimax problem.

As is well known, criterion (\ref{eq:DM.risk.equations.1}) can lead to different optimal decision than the worst-case Bayes risk (or unconditional $\Gamma$-minimax) criterion for the DM's ex-ante problem faced before $X^n$ is realized:
\begin{equation}\label{eq:robust.unconditional.gammamm}
 \inf_{\delta_n} \; \sup_{\gamma \in \Gamma} \; \int_{\Theta \times \mc M} \mb E_\mu \left[ r(\delta_n(X^n),\theta,\mu) \right]  d\gamma(\theta,\mu)  ,
\end{equation}
where the infimum is over all measurable $\delta_n : \mc X^n \to \mc D$; see \cite{GiacominiKitagawaRead2021} and \cite{AFMQT} for discussions. 
Thus, decision rules that are optimal under criterion (\ref{eq:robust.unconditional.gammamm}) may not align with optimal actions in the conditional $\Gamma$-minimax problem (\ref{eq:DM.risk.equations.1}). Further, optimal decisions under criterion (\ref{eq:robust.unconditional.gammamm}) are, in general, intractable (see, e.g., \cite{GiacominiKitagawaRead2021}).

Our proposal---namely, to use criterion (\ref{eq:imr}) for the ex-ante problem---circumvents both of these problems. Optimal decisions under criterion (\ref{eq:imr}) are tractable, align with optimal actions under the conditional $\Gamma$-minimax criterion (\ref{eq:DM.risk.equations.1}), and are asymptotically efficient in a frequentist sense that we formalize in Section~\ref{sec:optimality}.

Finally, we note our approach departs from the textbook Wald approach in two regards. The Wald approach would seek a minimax decision rule that solves
\[
 \inf_{\delta_n} \sup_{\mu \in \mc M} \sup_{\theta \in \Theta_0(\mu)} \E_\mu\left[ r(\delta_n(X^n), \theta, \mu) \right].
\]
Comparing with criterion (\ref{eq:imr}), we see that our approach moves the supremum over $\theta$ inside the expectation, potentially making the criterion more conservative. But as we discussed above, this conservativeness is constrained by the identified set $\Theta_0(\mu)$. Moreover, this conservativeness is offset, at least in part, by the fact that we average across a prior $\pi$ for $\mu$ rather than taking a supremum over $\mu$. These departures seem reasonable given the asymmetric nature of the identification problem. They also buy tractability, allowing the derivation of optimal decisions in a much broader class of problems than the standard Wald approach.

\section{Optimality Theory for Parametric Models}
\label{sec:optimality}

We now present the main optimality results for parametric models. We first outline the assumptions in Section~\ref{subsec:assumption}. Section~\ref{subsec:optimality.parametric} presents two main results. Theorem \ref{t1} establishes that Bayes decision rules $\delta_n^*(\,\cdot\,;\pi)$, defined in (\ref{eq:decision}), are optimal. Theorem \ref{t2} shows that any decision whose asymptotic behavior is different from $\delta_n^*(\,\cdot\,;\pi)$ is sub-optimal. In particular, we prove in Section~\ref{subsec:optimality.plugin} that plug-in decisions $\delta_n^{plug}$ are sub-optimal when the maximum risk $R(d,\mu)$ is a non-smooth function of $\mu$. As we documented in the empirical illustration in Section~\ref{subsec:treatment.empirical.numerics}, this finding can have important implications for treatment assignment under partial identification. Finally, in Section~\ref{subsec:finiteness} we provide a refinement of our analysis for settings in which the average maximum risk at the optimal decision not finite.

\subsection{Assumptions}
\label{subsec:assumption}

We first place some assumptions on the risk functions. Say $f : \mathcal M \to \mathbb R^k$ is \emph{directionally differentiable} at $\mu_0$ if there is a continuous map  $ \dot f_{\mu_0}[\,\cdot\,] : \mathbb R^K \to \mathbb R^k$ such that
\[
\lim_{n \to \infty} \frac{f(\mu_0 + t_n h_n) - f(\mu_0)}{t_n} = \dot f_{\mu_0}[h] 
\]
for all sequences $\{t_n\} \subset \mathbb R_+$ and $\{h_n\} \subset \mathbb R^K$ with $t_n \downarrow 0$ and $h_n \to h \in \mathbb R^K$. If so, we refer to  $ \dot f_{\mu_0}[\,\cdot\,]$ as the directional derivative of $f$ at $\mu_0$. Note that $ \dot f_{\mu_0}[\,\cdot\,]$ is positively homogeneous of degree one but not necessarily linear. If $\dot f_{\mu_0}[\,\cdot\,]$ is linear, then $f$ is \emph{(fully) differentiable} at $\mu_0$.

\medskip
\begin{assumption}\label{a1}
\begin{enumerate}[label={(\roman*)}, nosep]
\item\label{a1.1} $R(d,\,\cdot\,)$ is bounded and continuous on $\mc M$ for each $d \in \mc D$; 
\item\label{a1.2} For every $\mu_0 \in \mathcal M$ for which $|\underline{\mc D}_{\mu_0}| > 1$, $R(d,\,\cdot\,)$ is directionally differentiable at $\mu_0$ for each $d \in \underline{\mc D}_{\mu_0}$. 
\end{enumerate}
\end{assumption}
\medskip

We note that $R(d,\,\cdot\,)$ may be directionally but not fully differentiable in many empirically relevant cases of treatment assignment under partial identification (see Sections~\ref{sec:treatment.empirical} and~\ref{sec:treatment}). We only require directional differentiability to hold when $|\underline{\mathcal D}_{\mu_0}| > 1$. In this case, optimal actions can differ for different $\mu$ close to $\mu_0$ and the problem of distinguishing between optimal actions remains nontrivial as the sample size increases. In these scenarios, we use  directional differentiability  to derive asymptotic approximations for different decision rules.

We also place some regularity conditions on the statistical model $\mathcal P = \{P_\mu : \mu \in \mathcal M\}$. We present assumptions for a random sample to simplify exposition --- so $X^n = (X_1,\ldots,X_n)$ where each $X_i$ is an independent draw from $P_\mu$ --- though it is straightforward to extend our theory to weakly dependent data. We assume implicitly  that each $P_\mu$ admits a density $p_\mu$ with respect to a common dominating measure $\nu$. Say that  $\mathcal P$ is \emph{differentiable in quadratic mean} (DQM) at $\mu$ if there exists a vector of measurable functions $\dot \ell_\mu$ such that 
\[
 \int \left( \sqrt{p_{\mu + h}} - \sqrt{p_{\mu}} - \frac 12 h^T \dot \ell_\mu \sqrt{p_\mu} \right)^2 d\nu = o \left( \|h\|^2 \right) 
\]
as $h \to 0$. Under DQM, the Fisher information matrix is $I_\mu := \int \dot \ell_\mu \dot \ell_\mu^T \, d P_\mu$. Let $D_{KL}(p_{\mu}\|p_{\mu'}) = \int p_\mu \log(p_{\mu}/p_{\mu'}) d\nu$ if $P_\mu(p_{\mu'}(X) = 0) = 0$ and $D_{KL}(p_\mu\|p_{\mu'}) = +\infty$ if $P_\mu(p_{\mu'}(X) = 0) > 0$. Say that $\mathcal P$ is \emph{locally quadratic} for $D_{KL}$ if for all $\mu_0 \in \mathcal M$,
\[
 D_{KL}(p_\mu\|p_{\mu'}) \leq 2(\mu - \mu')^T I_{\mu_0}(\mu - \mu')
\]
holds for all $\mu,\mu'$ in a neighborhood of $\mu_0$. Following \cite{ClarkeBarron}, we say that $\mathcal P$ is \emph{sound} if weak convergence of $P_\mu$ to $P_{\mu'}$ is equivalent to convergence of $\mu$ to $\mu'$.

\medskip
\begin{assumption}\label{a2}
\begin{enumerate}[(i), nosep]
\item \label{a2.1} $\mathcal M$ is an open subset of $\mathbb R^K$;
\item \label{a2.2} $\mathcal P$ is differentiable in quadratic mean at each $\mu_0 \in \mathcal M$;
\item \label{a2.3} $I_{\mu_0}$ is finite and nonsingular at each $\mu_0 \in \mathcal M$;
\item \label{a2.4} $\mathcal P$ is locally quadratic for $D_{KL}$;
\item \label{a2.5} $\mathcal P$ is sound.
\end{enumerate}
\end{assumption}
\medskip

Assumptions~\ref{a2}\ref{a2.1}-\ref{a2.3} are similar to the conditions for parametric models in \cite{HiranoPorter2009}. DQM and local quadraticity hold under standard smoothness and integrability conditions. These conditions rule out certain ``irregular'' models, such as those with parameter-dependent support and parameters on the boundary. Assumption~\ref{a2}\ref{a2.4} holds under conditions similar to those used to establish asymptotic normality of maximum likelihood estimators. 
Soundness is a minimal identifiability condition. Following \cite{Schwartz1965}, posterior consistency is typically established in the weak topology. Soundness converts this to consistency for the parameter $\mu$. This condition trivially holds: (i) in exponential families \cite[p.~470]{ClarkeBarron}; or (ii) if $\mu$ is identified (i.e., $P_\mu = P_{\mu'}$ if and only if $\mu = \mu'$), $\mathcal M$ is precompact, and $\mu \mapsto P_\mu$ is weakly continuous on the closure of $\mathcal M$.\footnote{Alternatively, one could assume the existence of uniformly consistent tests of $H_0: \mu = \mu_0$ against $H_1: \|\mu - \mu_0\| \geq \epsilon$ \citep{Schwartz1965,vanderVaart1998}. Soundness is a weak sufficient condition for the existence of such tests \citep{ClarkeBarron}. It also allows us to control the error rate of tests along $\{P_{n,h}\}$, for which the usual classical testing condition seemed inadequate.} 

\bigskip

\paragraph{Example: Treatment Assignment.} 

Consider Assumption~1. In the empirical application, the lower bound for male youths behaved approximately like the maximum of two independent Gaussian random variables (corresponding to estimates from US and Brazilian RCTs) while the upper bound behaved approximately like a third independent Gaussian random variable (corresponding to the estimate from a Colombian RCT). We can mimic this setting with the following stylized example. Let 
\begin{equation}\label{eq:stylized_example}
	\mu = (\mu_1,\mu_2,\mu_3), \quad b_L(\mu) = (\mu_1 \vee \mu_2), \quad b_U(\mu) = \mu_3.
\end{equation}
As the bounds from the three studies are roughly the same magnitude, let $\mu_0 = (-c,-c,c)$ for $c > 0$. Then $R(0,\mu) = (\mu_3)_+$ and $R(1,\mu) = -(\mu_1 \vee \mu_2)_-$, and so $R(0,\mu_0) = R(1,\mu_0) = c$ and $\ul{\mc D}_{\mu_0} = \{0,1\}$. We deduce that $R(d,\mu)$ is bounded (provided ${\cal M} \subset \mathbb R^3$ is  bounded) and continuous in $\mu$. Here $R(0,\mu)$ is smooth at $\mu_0$ with $\dot R_{0,\mu_0}[h] = h_3$. However, $R(1,\mu)$ is only directionally differentiable at $\mu_0$ with $\dot R_{1,\mu_0}[h] = -(h_1 \vee h_2)$. Overall, $R(d,\mu)$ satisfies Assumption~\ref{a1}.

Assumption~\ref{a2} places restrictions on the family of probability distributions $\{P_\mu \, : \, \mu \in {\cal M}\}$. One formalization of the application described in Section~\ref{subsec:treatment.empirical.numerics} is to regard the published estimates, rather than the observations upon which these estimates are based, as data. So $X^n = \hat{\mu}$. Using the Normality assumption underlying our empirical illustration, $P_\mu$ is a multivariate Normal distribution with mean $\mu$ and a diagonal covariance matrix with elements $s^2_k$. The Normal family clearly satisfies Assumption~\ref{a2}. In Section~\ref{sec:semiparametric} we will relax the strict Normality assumption and cast the paper in a semiparametric framework. $\square$

\subsection{Main Results}
\label{subsec:optimality.parametric}

We first introduce some terminology. Let $\Pi$ denote the class of all priors on $\mc M$ with positive, continuous, and bounded Lebesgue density. We say that $\{\delta_n\},\{\delta_n'\} \in \mb D$ are \emph{asymptotically equivalent} if
\[
 \lim_{n \to \infty} P_{n,h}(\delta_n(X^n) = d) = \lim_{n \to \infty} P_{n,h}(\delta_n'(X^n) = d)
\]
for all $d \in \mc D$, $h \in \mb R^K$, and $\mu_0 \in \mc M$. Proposition~\ref{prop:BVM} in Appendix~\ref{appsec:BVM} shows that $\sqrt n (\bar R_n(d) - R(d,\mu_0))$ converges in distribution along $P_{n,h}$ to $\E^*[\dot R_{d,\mu_0}[Z^* + Z]|Z ]$, where $\dot R_{d,\mu_0}$ is the directional derivative of $R(d,\mu)$ at $\mu_0$, $Z \sim N(h,I_{\mu_0}^{-1})$ represents the asymptotic behavior of an efficient estimator of $\mu$, $Z^* \sim N(0,I_{\mu_0}^{-1})$ independently of $Z$ represents asymptotic uncertainty about the local parameter $h$, and $\E^*$ denotes expectation with respect to $Z^*$, which integrates over the uncertainty in the local parameter. We say there are \emph{no first-order ties} if for each $\mu_0 \in \mc M$ the minimizer of $d \mapsto \E^*[\dot R_{d,\mu_0}[Z^* + z] ]$ over $\ul{\mc D}_{\mu_0}$ is unique for almost every $z$. This condition trivially holds if $\underline{\mc D}_{\mu_0}$ is a singleton. To interpret the condition when $|\underline{\mc D}_{\mu_0}| > 1$, suppose that $R$ is differentiable at $\mu_0$, which is analogous to the case of smooth welfare contrasts studied in \cite{HiranoPorter2009}. Then $\E^*[\dot R_{d,\mu_0}[Z^* + z] ] = g_{d,\mu_0}^T z$ with $g_{d,\mu_0} =  \frac{\partial R(d,\mu_0)}{\partial \mu}$. In that case, a sufficient condition for no first-order ties is that the derivatives are different: $g_{d,\mu_0} \neq g_{d',\mu_0}$ for $d,d' \in \underline{\mc D}_{\mu_0}$. 

Our first main result establishes optimality of Bayes decisions $\{\delta_n^*(\,\cdot\,;\pi)\}$ for $\pi \in \Pi$.

\medskip
\begin{theorem}\label{t1}
Suppose that Assumptions~\ref{a1} and~\ref{a2} hold, ${\cal R}(\{\delta_n\}; \mu_0)$ is finite for at least one $\{\delta_n\} \in \mathbb D$, and there are no first-order ties. Then
\begin{enumerate}[(i), nosep]
	\item\label{t1.1} $\{\delta_n^*(\,\cdot\,;\pi)\}$ is optimal for any $\pi \in \Pi$;
	\item\label{t1.2} $\{\delta_n^*(\,\cdot\,;\pi)\}$ and 	$\{\delta_n^*(\,\cdot\,;\pi') \}$ are asymptotically equivalent for all $\pi,\pi' \in \Pi$;
	\item\label{t1.3} $\{\delta_n\} \in \mb D$ is optimal if it is asymptotically equivalent to $\{\delta_n^*(\,\cdot\,;\pi)\}$ for some $\pi \in \Pi$.
\end{enumerate} 
\end{theorem}
\medskip

According to Theorem~\ref{t1}, Bayes decisions with priors $\pi \in \Pi$ are asymptotically equivalent and optimal. All such Bayes decisions are asymptotically independent of $\pi$. Part~\ref{t1.3} implies that any decision that is asymptotically equivalent to a Bayes decision under a prior $\pi \in \Pi$ is optimal. 
Optimality of the bootstrap implementation discussed in Section~\ref{subsec:decisions.bootstrap} follows from~\ref{t1.3} under suitable regularity conditions. 

We now show that asymptotic equivalence to a Bayes decision is \emph{necessary} for optimality. Say $\{\delta_n\},\{\delta_n'\} \in \mb D$ fail to be asymptotically equivalent at $\mu_0$ if 
\[
 \lim_{n \to \infty} P_{n,h}(\delta_n(X^n) = d) \neq \lim_{n \to \infty} P_{n,h}(\delta_n'(X^n) = d)
\]
for some $h \in \mathbb R^K$ and some $d \in {\cal D}$.

\medskip
\begin{theorem}\label{t2}
Suppose that Assumptions~\ref{a1} and~\ref{a2} hold, ${\cal R}(\{\delta_n\}; \mu_0)$ is finite for at least one $\{\delta_n\} \in \mathbb D$, and there are no first-order ties. Suppose that $\{\delta_n\} \in \mb D$ is not asymptotically equivalent to $\{\delta_n^*(\,\cdot\,;\pi)\}$ for some (equivalently, all) $\pi \in \Pi$. Then at any $\mu_0$ at which asymptotic equivalence fails,
\[
 \mc R(\{\delta_n\}; \mu_0) > \mc R(\{\delta_n^*(\,\cdot\,;\pi)\}; \mu_0) \quad \mbox{for all} \; \pi \in \Pi.
\]
\end{theorem}
\medskip

In the next subsection we discuss further implications of Theorem~\ref{t2} with respect to decision rules based on plugging-in efficient estimators of $\mu$.

\subsection{Plug-in Rules}
\label{subsec:optimality.plugin}

A natural alternative to $\delta_n^*(X^n; \pi)$ is to plug in an efficient estimator $\hat \mu = \hat \mu(X^n)$ of $\mu$ into the oracle decision rule, yielding the ``plug-in'' rule $\delta^{plug}_n(X^n) = \delta^o(\hat \mu)$. \cite{Manski2021Haavelmo,Manski2021} refers to this approach as ``as-if'' optimization: the estimator $\hat \mu$ is treated ``as if'' it is the true parameter.\footnote{This approach also has connections to anticipated utility \citep{Kreps1998,CogleySargent2008}.} We show that asymptotic equivalence of the Bayes and plug-in rules can fail when the maximum risk does not depend smoothly on $\mu$; if so, Theorem~\ref{t2} implies plug-in rules are sub-optimal. We illustrate this difference within the context of the application to treatment assignment from Section~\ref{sec:treatment.empirical}.

To understand when asymptotic equivalence holds, we turn to Lemma~\ref{l1} in Appendix~\ref{ax:proofs}, which characterizes the asymptotic behavior of sequences of decision rules. Lemma~\ref{l1} implies 
\[
 \lim_{n \to \infty} P_{n,h} \left( \delta_n^*(X^n;\pi) = d \right) =
 \begin{cases} 
  \mathbb P_h \Big( {\textstyle \argmin_{i \in \underline{\mc D}_{\mu_0}} } \mathbb E^* \left[ \left. \dot R_{i,\mu_0}[Z^* + Z] \right| Z \right] = d \Big), & d \in \ul{\mc D}_{\mu_0}, \\[-4pt]
  0, & \mbox{otherwise},
  \end{cases}
\]
where $\mathbb P_h$ is the probability measure of $Z \sim N(h,I_{\mu_0}^{-1})$. One can similarly show 
\begin{equation}\label{eq:plug}
 \lim_{n \to \infty} P_{n,h} \left( \delta_n^{plug}(X^n) = d \right) =
 \begin{cases} 
  \mathbb P_h \Big( {\textstyle \argmin_{i \in \underline{\mc D}_{\mu_0}} } \dot R_{i,\mu_0}[Z] = d \Big), & d \in \ul{\mc D}_{\mu_0}, \\[-4pt]
  0, & \mbox{otherwise}.
  \end{cases}
\end{equation}
Asymptotic equivalence of the Bayes and plug-in rules holds when the right-hand side probabilities in the above two displays are equal.

Suppose $R(d,\mu)$ is fully differentiable at $\mu_0$ for all $d \in \ul{\mc D}_{\mu_0}$. Then each $\dot R_{d,\mu_0}[\,\cdot\,]$ is linear, and 
\[
 \mathbb E^* \left[ \left. \dot R_{d,\mu_0}[Z^* + Z] \right| Z \right] = \dot R_{d,\mu_0}[ \mathbb E^* \left[Z^* \right]+ Z] \equiv \dot R_{d,\mu_0}[Z] .
\]
Hence, the plug-in and Bayes rules are asymptotically equivalent, and therefore optimal by Theorem~\ref{t1}. Formally:

\begin{corollary}\label{cor:plug}
Let the conditions of Theorem~\ref{t1} hold, let the plug-in rule $\delta_n^{plug}$ satisfy (\ref{eq:plug}), and let $R(d,\mu)$ be fully differentiable at $\mu_0$ for all $d \in \ul{\mc D}_{\mu_0}$ and all $\mu_0 \in \mc M$. Then $\{\delta_n^{plug}\}$ is optimal.
\end{corollary}

Corollary~\ref{cor:plug} is consistent with Theorem~3.2 of \cite{HiranoPorter2009}, which shows plug-in rules are optimal when welfare contrasts depend smoothly on a point-identified, regularly estimable parameter.

If $R(d,\mu)$ is only directionally differentiable at $\mu_0$, then $\dot R_{d,\mu_0}[\,\cdot\,]$ is not linear and the above reasoning no longer applies. In this case,  asymptotic equivalence of the Bayes and plug-in rules cannot be guaranteed. If it fails, Theorem~\ref{t2} implies the plug-in rule is sub-optimal.

\bigskip
\paragraph{Example: Treatment Assignment.} 

We continue the calculations for the stylized example introduced in Section~\ref{subsec:assumption}. By Jensen's inequality,
\[
 \begin{aligned}
 \mb P_h \Big( {\textstyle \argmin_{d \in \{0,1\} }} \mathbb E^* \left[ \left. \dot R_{d,\mu_0}[Z^* + Z] \right| Z \right] = 1 \Big)
 & = \mb P_h \Big( \E^*[(Z_1^* + Z_1) \vee (Z_2^* + Z_2)|Z] + Z_3 \geq 0  \Big) \\
 & > \mb P_h \Big( (Z_1 \vee Z_2) + Z_3 \geq 0  \Big) \\
 & = \mb P_h \Big( {\textstyle \argmin_{d \in \{0,1\} }} \dot R_{d,\mu_0}[Z] = 1 \Big).
 \end{aligned}
\]
Hence, asymptotic equivalence fails and the plug-in rule is sub-optimal: the Bayes rule recommends treatment more aggressively than the plug-in rule. This is reflected in the empirical application, where the optimal rule recommends treatment but the plug-in rule does not. $\square$

\subsection{Finiteness of the Criterion}
\label{subsec:finiteness}

Criterion $\mc R(\{\delta_n\};\mu_0)$ in (\ref{eq:criterion_asymptotic}) is formed by integrating $\mc R(\{\delta_n\};\mu_0, h)$ with respect to Lebesgue measure on $\mathbb R^K$. This raises the possibility that $\mc R(\{\delta_n\};\mu_0) = +\infty$ for all $\{\delta_n\} \in \mb D$, in which case it does not produce a meaningful ranking. We first note this is not possible if $K = 1$: 

\medskip

\begin{proposition}\label{prop:plug}
Suppose that $K = 1$ and Assumptions \ref{a1} and \ref{a2} hold. Fix any $\mu_0 \in \mc M$ for which there are no first-order ties. Then $\mc R(\{\delta_n^*(\,\cdot\,;\pi)\}; \mu_0) < \infty$ for all $\pi \in \Pi$.
\end{proposition}

\medskip

If $K > 1$, however, we may have $\mc R(\{\delta_n\};\mu_0) = +\infty$ for all $\{\delta_n\} \in \mb D$. For instance, one could take the $K = 1$ case  and augment the parameter space with redundant parameters. To handle cases where criterion  (\ref{eq:criterion_asymptotic}) is infinite, we approximate the improper Lebesgue prior on $h$ by a sequence of proper priors. We used Lebesgue measure on $\mathbb R^K$ in criterion (\ref{eq:criterion_asymptotic}) because there is no a priori reason to view one local parameter as more likely than another. A similar effect is achieved for a $N(0,\sigma I)$ prior with large $\sigma$. However, for any finite $\sigma$ the integrated risk under this proper prior is also finite because $R(d,\cdot)$ is bounded according to Assumption~\ref{a1}(i). Let $\pi_\sigma$ denote the $N(0,\sigma I)$ density. Define 
\begin{equation} \label{eq:criterion_asymptotic_normal}
 \mc R_\sigma(\{\delta_n\};\mu_0) = \int \mc R(\{\delta_n\};\mu_0, h) \pi_\sigma(h) \, d h.
\end{equation}
Lemma~\ref{l1} implies $\mc R_\sigma(\{\delta_n\};\mu_0)$ is finite for all $\{\delta_n\} \in \mb D$. Analogously to Definition~\ref{def:efficient.robust}, we define the concept of \emph{$\sigma$-optimality}:

\begin{definition}
	\label{def:efficient.robust.sigma}
	Say $\{\delta_n\} \in \mb D$ is \emph{$\boldsymbol\sigma$-{\bf optimal}} if $\mc R_\sigma(\{\delta_n\};\mu_0) = \inf_{\{\delta_n'\} \in \mathbb D} \mc R_\sigma(\{\delta_n'\}; \mu_0) $ for all $\mu_0 \in \mc M$.
\end{definition}

As we are using the large-$\sigma$ limit of the $N(0,\sigma I)$ prior to approximate Lebesgue measure, we are really interested in the behavior of $\sigma$-optimal decisions as $\sigma \to \infty$. The following result shows  Bayes and $\sigma$-optimal decisions lead to the same choices in large samples.

\medskip

\begin{theorem}\label{t3}
Suppose that Assumptions~\ref{a1} and~\ref{a2} hold and there are no first-order ties. Let $\{\delta_{n,\sigma}^*\} \in \mb D$ be $\sigma$-optimal at $\mu_0$. Then for all $d \in \mc D$ and $h \in \mb R^K$,
\[
 \lim_{n \to \infty} P_{n,h}(\delta_n^*(X^n;\pi) = d) = \lim_{\sigma \to \infty} \lim_{n \to \infty} P_{n,h}(\delta_{n,\sigma}^* (X^n) = d).
\]
\end{theorem}

\medskip

The proof of Theorem~\ref{t3} shows that Bayes and $\sigma$-optimal decisions are \emph{identical} in large samples, for almost every realization of the data, provided $\sigma$ is sufficiently large. We omit the details here to avoid introducing additional technicalities, and instead illustrate the idea within the context of our running example.\footnote{Here is some intuition: in many Bayesian settings posterior distributions of parameters are proper even under improper priors. But integrated risk is typically only finite under a proper prior. As long as the likelihood function asymptotically dominates the prior, posteriors, and decisions derived from them, under an improper prior are very close to those obtained from a prior with a large variance.}

\bigskip
\paragraph{Example: Treatment Assignment.} 
We continue with our discussion of the stylized example from (\ref{eq:stylized_example}) that mimics the empirical application. We have $\sqrt n(\hat \mu - \mu) \to_d Z \sim N(h,\Sigma)$ along $P_{n,h}$, where $\Sigma = \mathrm{diag}(s_1,s_2,s_3)$, say, since the RCTs are independent. Lemma~\ref{l1} in Appendix~\ref{ax:proofs} implies that asymptotically, any Bayes decision behaves like 
\[
 \delta^*_\infty(z) = \mb I[ \E^*[(Z_1^* + z_1) \vee (Z_2^* + z_2)] + z_3 \geq 0 ],
\]
for almost every realization $z$ of $Z$, where $\E^*$ denotes expectation taken with respect to $Z^* \sim N(0,\Sigma)$. The expectation can be computed in closed-form (see, e.g., \cite{NKotz}) to give
\begin{align*}
 & \delta^*_\infty(z) = \mb I\left[ f(z;s_1,s_2) \geq 0 \right], \\
 & \text{where } f(z;s_1,s_2)  = z_1 \Phi\left(\frac{z_1-z_2}{\sqrt{s_1 + s_2}}\right) + z_2 \Phi\left(\frac{z_2-z_1}{\sqrt{s_1 + s_2}}\right) + \sqrt{s_1 + s_2} \phi\left(\frac{z_1-z_2}{\sqrt{s_1 + s_2}}\right) + z_3,
\end{align*}
and $\Phi$ and $\phi$ are the standard normal CDF and PDF, respectively. 
Similarly, Lemma~\ref{lem:sigma.optimal} in Appendix~\ref{ax:proofs} implies that asymptotically, the $\sigma$-optimal decision behaves like
\begin{align*}
 & \delta^*_\sigma(z) = \mb I\left[f_\sigma(z;s_1,s_2) \geq 0 \right], \\
 & \text{where } f_\sigma(z;s_1,s_2) = f\left( (I + \sigma^{-1}\Sigma)^{-1}z ; \frac{\sigma s_1}{\sigma + s_1}, \frac{\sigma s_2}{\sigma + s_2} \right).
\end{align*}
The difference between the two arises because in the $\sigma$-optimal decision the posterior mean is shrunk from $z$ to $(I + \sigma^{-1}\Sigma)^{-1}z$ and the posterior variance is shrunk from $\Sigma$ to $(\Sigma^{-1} + \sigma^{-1} I)^{-1}$.
There are no first-order ties because the set of $z$ values for which $f(z;s_1,s_2) = 0$ has measure zero. For any $z$ outside this negligible set, we have $\mathrm{sign}(f(z;s_1,s_2)) = \mathrm{sign}(f_\sigma(z;s_1,s_2))$ for $\sigma$ sufficiently large. Hence, for almost every realization $z$ of $Z$, there is a minimal value of $\sigma$ above which the Bayes and $\sigma$-optimal decisions are identical. $\square$

\section{Optimality Theory for Semiparametric Models}\label{sec:semiparametric}

This section extends our approach to semiparametric models, which is relevant in many empirical contexts. For example, the data may not follow a specific parametric model, or $X^n$ may be a vector of summary statistics with unknown finite-sample distribution, as in the empirical application in Section~\ref{sec:treatment.empirical}. We present the model in Section~\ref{subsec:model.semiparametric} then  generalize our optimality concept in Section~\ref{subsec:optimality.semiparametric}. Section~\ref{subsec:decision.semiparametric} describes a quasi-Bayesian implementation of optimal decisions. Section~\ref{subsec:theory.semiparametric} presents the main optimality results. Theorem~\ref{t1s} shows that quasi-Bayes decision rules are optimal, while Theorem~\ref{t2s} shows that any decision whose asymptotic behavior differs from quasi-Bayes rules is sub-optimal.

\subsection{Model}\label{subsec:model.semiparametric}

Let $X^n \sim P_{n,(\mu,\eta)}$ with  $\mu \in \mc M \subseteq \mb R^K$ and $\eta \in \mc H$, an infinite-dimensional space. In a GMM model, $\mu$ is a finite-dimensional parameter vector, $\eta$ is the marginal distribution of each observation $X_i \in \mc X$, and $\mc H = \{\eta \in \mb M(\mc X) : \int g(x,\mu) \,  d \eta(x) = 0$ for some $\mu \in \mc M\}$ for a vector of moment functions $g$, where $\mb M(\mc X)$ is the set of all probability measures on $\mc X$.
We again assume that, given $\mu$, the structural parameter $\theta$ takes values in a set $\Theta_0(\mu)$. Therefore, the set of payoff distributions is indexed only by the parametric component $\mu$ and the nonparametric component $\eta$ is a nuisance parameter. For instance, $\mu$ may be a vector of population moments used to construct bounds on $\theta$. The nuisance parameter $\eta$ represents other features of the distribution of $X^n$ that are irrelevant for the DM's decision problem.

\subsection{Optimality Criterion}\label{subsec:optimality.semiparametric}

Our optimality criterion (\ref{eq:criterion_asymptotic}) integrates the excess maximum risk of $\delta_n(X^n)$ relative to the oracle using Lebesgue measure on the local perturbations $h \in \mb R^K$ of $\mu_0$. This approach does not  extend to perturbations of $(\mu_0,\eta_0)$ 
due to measure-theoretic complications in infinite-dimensional spaces. 
We therefore form our optimality criterion using local perturbations of $\mu_0$ within a \emph{least favorable submodel} in which $X^n$ carries the least information about $\mu$ of all parametric submodels. The problem of parameter estimation in the least favorable submodel is asymptotically equivalent to the problem of estimating $\mu$ in the full semiparametric model. 

To simplify exposition, we present the following discussion and results within the context of a random sample --- so $X^n = (X_1,\ldots,X_n)$ where each $X_i$ is an independent draw from $P_{(\mu,\eta)}$ --- though it is straightforward to extend our theory to weakly dependent data. We say that $\mc P = \{P_{(\mu,\eta)} : \mu \in \mc M, \eta \in \mc H\}$ has a \emph{least favorable submodel} at $(\mu,\eta)$ if there exists an open neighborhood $\mc M_{(\mu,\eta)}$ of $\mu$ and a map $t \mapsto \eta_t$ from $\mc M_{(\mu,\eta)}$ into $\mc H$ such that $\{P_{\beta(t)} : t \in \mc M_{(\mu,\eta)}\}$ with $\beta(t) =(t,\eta_t)$ have densities $p_{\beta(t)}$ with respect to a common dominating measure $\nu$ that satisfy the DQM condition
\[
 \int \left( \sqrt{p_{\beta(\mu + h)}} - \sqrt{p_{\beta(\mu)}} - \frac 12 h^T \dot \ell_{(\mu,\eta)}\sqrt{p_{\beta(\mu + h)}} \right)^2 d \nu = o(\|h\|^2)
\]
as $h \to 0$, where $\dot \ell_{(\mu,\eta)} : \mc X \to \mb R^K$ is the efficient score for $\mu$ and $I_{(\mu,\eta)} := \int \dot \ell_{(\mu,\eta)}\dot \ell_{(\mu,\eta)}^T dP_{(\mu,\eta)}$ is the semiparametric information bound. In other words, there is a regular parametric submodel $\{P_{\beta(t)}: t \in \mc M_{(\mu,\eta)}\}$ whose information matrix at $t = \mu$ is the semiparametric information bound at $(\mu,\eta)$. The path $t \mapsto \eta_t$ and dominating measure $\nu$ can depend on $(\mu,\eta)$, but we suppress this to simplify notation. We note that our approach only requires the least favorable model to exist: the researcher doesn't need to derive it in order to implement optimal decisions. The least favorable model also needn't be unique, but any such model satisfying the above conditions will lead to the same optimality criterion.

Our optimality criterion is analogous to the parametric case. For each $(\mu_0,\eta_0) \in \mc M \times \mc H$, we reparametrize the least favorable submodel $\{P_{\beta(t)} : t \in \mc M_{(\mu_0,\eta_0)}\}$ using $t = \mu_0 + h/\sqrt n$ for $h \in \mathbb R^K$.  Let $P_{n,h}$ denote the distribution of $X^n$ under $P_{\beta(\mu_0 + h/\sqrt n)}$ with $\beta(\mu_0 + h/\sqrt n) = (\mu_0 + h/\sqrt n,\eta_{\mu_0 + h/\sqrt n})$ and $\E_{n,h}$ denote expectation under $P_{n,h}$. 
Define
\begin{multline}  \label{eq:criterion_asymptotic_semiparametric_h}
  \mc R(\{\delta_n\};(\mu_0,\eta_0), h) \\
   =  \limsup_{n \to \infty} \sqrt n \left( \E_{n,h}\left[R\big(\delta_n(X^n), \mu_0 + h/\sqrt n \big)\right] - \min_{d\in \mc D} R\big(d,\mu_0 + h/\sqrt n \big) \right),
\end{multline}
and let
\begin{equation} \label{eq:criterion_asymptotic_semiparametric}
 \mc R(\{\delta_n\};(\mu_0,\eta_0)) = \int \mc R(\{\delta_n\};(\mu_0,\eta_0), h) \, d h.\footnote{Note that $\mc R(\{\delta_n\};(\mu_0,\eta_0), h)$ and $\mc R(\{\delta_n\};(\mu_0,\eta_0))$ remain well defined despite the fact that $\mu_0 + h/\sqrt n$ may be outside $\mathcal M_{(\mu_0,\eta_0)}$ for small $n$.}
\end{equation}
The class  $\mathbb D$ is defined analogously to the parametric case (cf. Definition~\ref{def:D}). Recall that $\underline{\mathcal D}_{\mu_0} = \argmin_{d \in \mathcal D} R(d,\mu_0)$ denotes the set of optimal choices at $\mu_0$.

\medskip

\begin{definition} \label{def:Ds}
$\mathbb D$ consists of all sequences $\{\delta_n\}$ such that, for all $(\mu_0,\eta_0) \in \mathcal M \times \mathcal H$, 
\begin{enumerate}[label={(\roman*)}, nosep]
\item \label{Ds.1} $\lim_{n \to \infty} P_{n,h}(\delta_n(X^n) = d)$ exists for all $d \in \mathcal D$ and $h \in \mathbb R^K$;
\item \label{Ds.2} $\lim_{n \to \infty} \sqrt n \, P_{n,h} \left( \delta_n(X^n) \not \in \underline{\mathcal D}_{\mu_0} \right) = 0$
for all $h \in \mathbb R^K$.
\end{enumerate}
\end{definition}

\medskip

Finally, we say a sequence of decision rules $\{\delta_n\}$ is optimal if it minimizes the criterion $\mc R(\,\cdot\,;(\mu_0,\eta_0))$ in (\ref{eq:criterion_asymptotic_semiparametric}) over $\mb D$ for all $(\mu_0,\eta_0) \in \mc M \times \mc H$ (cf. Definition~\ref{def:efficient.robust}):

\begin{definition}
	\label{def:efficient.robust.semiparametric}
	Say $\{\delta_n\} \in \mb D$ is \textit{\textbf{optimal}} if $\mc R(\{\delta_n\};(\mu_0,\eta_0)) = \inf_{\{\delta_n'\} \in \mathbb D} \mc R(\{\delta_n'\}; (\mu_0,\eta_0)) $ for all $(\mu_0,\eta_0) \in \mc M \times \mc H$.
\end{definition}

If $\mc R(\{\delta_n\};(\mu_0,\eta_0))$ is infinite for all $\{\delta_n\} \in \mathbb D$, then a similar approach to  Section~\ref{subsec:finiteness} can be followed whereby the Lebesgue prior on $h$ is approximated by a sequence of proper priors.

\subsection{Quasi-Bayesian Implementation}\label{subsec:decision.semiparametric}

Optimal decisions are formed similarly to the parametric case, but we replace the  posterior with a \emph{quasi-posterior} formed from a limited-information criterion for $\mu$. Following \cite{DoksumLo1990}, \cite{Kim2002}, and \cite{Mueller2013}, consider a limited information $N(\hat \mu, (n \hat I)^{-1})$ quasi-likelihood for $\mu$, where $\hat \mu$ is an efficient estimator of $\mu$ and $\hat I^{-1}$ is a consistent estimator of its asymptotic variance, namely $I_{(\mu,\eta)}^{-1}$. Combining the quasi-likelihood with a prior $\pi$ on $\mc M$ yields the \emph{quasi-posterior}
\begin{equation} \label{eq:quasi_posterior}
 \pi_n(\mu|X^n) \propto e^{-\frac{1}{2}(\mu - \hat \mu)^T (n \hat I)(\mu - \hat \mu)} \pi(\mu) .
\end{equation}
The \emph{quasi-posterior maximum risk} $\bar R_n(d)$ is calculated by averaging $R(d,\mu)$ across the quasi-posterior, as in (\ref{eq:pmr}). The \emph{quasi-Bayes decision} $\delta_n^*(X^n;\pi)$ is chosen to minimize the quasi-posterior maximum risk $\bar R_n(d)$, as in (\ref{eq:decision}).

Unlike the parametric case, here the optimal decision cannot be justified on the basis of robust Bayes analysis. A formal Bayesian approach would require specifying a prior on $\mc M \times \mc H$ then forming a marginal posterior for $\mu$.\footnote{See, for instance, the Bayesian exponentially tilted empirical likelihood approach of \cite{Schennach2005} or the Bayesian GMM approaches of \cite{Shin2015} and \cite{Walker2025}. Our  approach is computationally simple and avoids the delicate issue of specifying priors in infinite-dimensional parameter spaces.} 

\bigskip

\paragraph{Example: Treatment Assignment.}  In Section~\ref{subsec:assumption} we rationalized the empirical illustration in Section~\ref{subsec:treatment.empirical.numerics} through the assumption that the vector $\hat{\mu}$ is exactly Normally distributed. The semiparametric extension allows us to relax this assumption. Treating the estimates $\hat{\mu}$ as data, the semiparametric model allows for more general distributions of the form $\hat{\mu} \sim P_{n,(\mu,\eta)}$, satisfying the moment restriction $\int (\hat{\mu} - \mu) d P_{n,(\mu,\eta)} = 0$. The density in (\ref{eq:quasi_posterior}) is identical to the one implied by the Normal model in Section~\ref{subsec:assumption}, but the interpretation changes from an exact posterior to a quasi-posterior. $\square$ 

\subsection{Optimality of Quasi-Bayes Decisions}\label{subsec:theory.semiparametric}

We first state and discuss regularity conditions, then present our optimality results. Let $\hat \lambda_{\min}$ and $\hat \lambda_{\max}$ denote the smallest and largest eigenvalues of $\hat I$. Let $\overset{P_{n,(\mu_0,\eta_0)}}{\to}$ denote convergence in probability under $\{P_{n,(\mu_0,\eta_0)}\}$. Let $\overset{P_{n,h}}{\rightsquigarrow}$ denote convergence in distribution under $\{P_{n,h}\}$.

\medskip
\begin{assumption}\label{a2s}
\begin{enumerate}[(i), nosep]
\item \label{a2s.1} $\mathcal M$ is an open subset of $\mathbb R^K$;
\item \label{a2s.2} $\mathcal P$ has a least favorable submodel at each $(\mu_0,\eta_0) \in \mc M \times \mc H$;
\item \label{a2s.3} $I_{(\mu_0,\eta_0)}$ is finite and nonsingular at each $(\mu_0,\eta_0) \in \mathcal M \times \mc H$;
\item \label{a2s.4} For each $(\mu_0,\eta_0) \in \mathcal M \times \mc H$ and $h \in \mathbb R^K$:
\begin{enumerate}[nosep]
\item \label{a2s.4.a} $\sqrt n P_{n,h}(\|\hat \mu - \mu_0\| > \epsilon) \to 0$ for each $\epsilon > 0$;
\item \label{a2s.4.b} there exists  $c \in (0,1)$ such that  $\sqrt n P_{n,h}( c \leq \hat \lambda_{\min}$, $\hat \lambda_{\max} \leq c^{-1}) \to 0$;
\end{enumerate}
\item \label{a2s.5} For each $(\mu_0,\eta_0) \in \mathcal M \times \mc H$:
\begin{enumerate}[nosep]
\item \label{a2s.5.a} $ \sqrt n ( \hat \mu - \mu_0) \overset{P_{n,h}}{\rightsquigarrow} Z$ with $Z \sim N(h,I_{(\mu_0,\eta_0)}^{-1})$ for all $h \in \mathbb R^K$;
\item \label{a2s.5.b} $\hat I \overset{P_{n,(\mu_0,\eta_0)}}{\to} I_{(\mu_0,\eta_0)}$.
\end{enumerate}
\end{enumerate}
\end{assumption}
\medskip

Assumptions~\ref{a2s}\ref{a2s.1}-\ref{a2s.3} are analogous to Assumptions~\ref{a2}\ref{a2.1}-\ref{a2.3}. We do not require versions of Assumptions~\ref{a2}\ref{a2.4} and \ref{a2}\ref{a2.5} because the quasi-likelihood here has a particular quadratic structure. But we need to ensure that the estimators $\hat \mu$ and $\hat I$ used in the quasi-likelihood are sufficiently well behaved, which is the role of Assumptions~\ref{a2s}\ref{a2s.4} and~\ref{a2s}\ref{a2s.5}. These may be verified for a wide variety of estimators $\hat \mu$ and $\hat I$ under suitable regularity conditions that cover the RCT estimators $\hat{\mu}$ of the empirical illustration in Section~\ref{subsec:treatment.empirical.numerics}.

\bigskip

\paragraph{Example: Treatment Assignment.}
Each subpopulation corresponds to an independent randomized experiment. In each subpopulation $k$ we observe a random sample of $(X_i,Y_i)$ of size $n$ drawn from a distribution $\eta_k$,\footnote{It is straightforward to extend the analysis to have different sample sizes $n_k$ in each subpopulation, provided the $n_k$ all grow at the same rate.} where $X_i$ is a binary treatment indicator and $Y_i = X_i Y_{i1} + (1-X_i) Y_{i0}$ with $Y_{i1}$ and $Y_{i0}$ denoting the treated and untreated outcomes  for individual $i$. Recall $\mu = (\mu_{k})_{k=1}^K$ and $\hat \mu = (\hat \mu_k)_{k=1}^K$, where we estimate the subpopulation-$k$ ATE $\mu_k$ by, for example, using the slope coefficient $\hat \mu_k$ in an OLS regression of $Y_i$ on $X_i$. Let $s_k$ denote the heteroskedasticity-robust standard error of $\hat \mu_k$. Then $\hat I$ is given by $(n\hat I)^{-1} = \mathrm{diag}(s_1^2,\ldots,s_K^2)$. The nuisance parameter is the joint distribution $\eta = \times_{k=1}^K \eta_k$ across the independent subpopulations. We have $P_{(\mu,\eta)} \equiv P_\eta = \eta$. We similarly drop dependence of $\mu$ in quantities such as the score and information matrix that follow.

Each $\hat \mu_k$ is semiparametrically efficient \citep{Hahn1998} with influence function
\[
 \psi_{k}(x,y) = \frac{x}{p_k}(y - \bar \mu_{1,k}) - \frac{1-x}{1-p_k}(y - \bar \mu_{0,k}),
\]
where $p_k$, $\bar \mu_{1,k}$, and $\bar \mu_{0,k}$ represent the means of $X_i$, $Y_{i1}$, and $Y_{i0}$ in subpopulation $k$ and are available from the moments of $X$, $XY$ and $(1-X)Y$ under $\eta_k$. We construct a least favorable submodel as follows. For each subpopulation $k$, fix any $\eta_{0,k}$ with $0 < p_k < 1$ and $0 < \int \psi_{k}^2 d \eta_{0,k} < \infty$. Let 
\[
 \dot \ell_{k,\eta_0}(x,y) = \frac{\psi_k(x,y)}{\int \psi_k^2 d \eta_{0,k}}.
\]
Without confusion we also let $\eta_{0,k}$ denote the density of $\eta_{0,k}$ with respect to some dominating measure $\nu$. We define a smooth parametric model $\eta_{t_k,k}$ passing through $\eta_{0,k}$ at $t_k = \mu_{0,k}$ with score $\dot \ell_{k,\eta_0}$ by
\[
 \eta_{t_k,k}(x,y) = \eta_{0,k}(x,y) \frac{v((t_k - \mu_{0,k})\dot \ell_{k,(\mu_0,\eta_0)}(x,y))}{\int v((t_k - \mu_{0,k})\dot \ell_{k,(\mu_0,\eta_0)}(x,y)) d \eta_{0,k}(x,y)},
\]
where $v(u) = 2(1+e^{-2u})^{-1}$ satisfies $v(0) = v'(0) = 1$ \cite[Example 25.16]{vanderVaart1998}. Letting $\eta_t = \times_{k=1}^K \eta_{t_k,k}$ for $t = (t_k)_{k=1}^K$, we have a smooth parametric family passing through $\eta_0$ at $t = \mu_0$ whose score $\dot \ell_{\eta_0} = (\dot \ell_{k,\eta_0})_{k=1}^K$ satisfies $\int \dot \ell_{\eta_0}\dot \ell_{\eta_0}^T d \eta_0 = \mathrm{diag}(\sigma_1^{-2},\ldots,\sigma_K^{-2}) \equiv I_{\eta_0}^{-1}$, where $\sigma_k^2 = \int \psi_k^2 d \eta_{0,k}$ is the (efficient) asymptotic variance of $\hat \mu_k$. 

By Le Cam's third lemma \cite[Example~6.7]{vanderVaart1998}, for any $h \in \mathbb R^K$ we may deduce that  $\sqrt n (\hat \mu - \mu_0)$ converges in distribution to a $N(h,I_{\eta_0}^{-1})$ random vector under $\{P_{n,h}\}$, where $P_{n,h}$ is the product measure formed by drawing $n$ copies of $(X,Y)$ under $\eta_{\mu_{0,k} + h_k/\sqrt n,k}$ for each subpopulation. This verifies Assumption~\ref{a2s}\ref{a2s.5.a}. Assumption~\ref{a2s}\ref{a2s.5.b} holds by standard consistency arguments for heteroskedasticity-robust standard errors. It is also straightforward to verify, for instance by using suitable concentration inequalities, that Assumption~\ref{a2s}\ref{a2s.4} holds.
$\square$

\bigskip

We now present analogues of Theorems~\ref{t1} and~\ref{t2} for the semiparametric case. Recall that $\Pi$ denotes the class of all priors on $\mc M$ with positive, continuous, and bounded Lebesgue density. Similar to the parametric case, we say that $\{\delta_n\},\{\delta_n'\} \in \mb D$ are \emph{asymptotically equivalent} if
\[
 \lim_{n \to \infty} P_{n,h}(\delta_n(X^n) = d) = \lim_{n \to \infty} P_{n,h}(\delta_n'(X^n) = d)
\]
for all $d \in \mc D$, $h \in \mb R^K$, and $(\mu_0,\eta_0) \in \mc M \times \mc H$.

\medskip
\begin{theorem}\label{t1s}
Suppose that Assumptions~\ref{a1} and~\ref{a2s} hold, ${\cal R}(\{\delta_n\}; (\mu_0,\eta_0))$ is finite for at least one $\{\delta_n\} \in \mathbb D$, and there are no first-order ties. Then
\begin{enumerate}[(i), nosep]
	\item\label{t1s.1} $\{\delta_n^*(\,\cdot\,;\pi)\}$ is optimal for any $\pi \in \Pi$;
	\item\label{t1s.2} $\{\delta_n^*(\,\cdot\,;\pi)\}$ and 	$\{\delta_n^*(\,\cdot\,;\pi') \}$ are asymptotically equivalent for all $\pi,\pi' \in \Pi$;
	\item\label{t1s.3} $\{\delta_n\} \in \mb D$ is optimal if it is asymptotically equivalent to $\{\delta_n^*(\,\cdot\,;\pi)\}$ for some $\pi \in \Pi$.
\end{enumerate} 
\end{theorem}
\medskip

It follows from Theorem~\ref{t1s} that quasi-Bayes decisions with priors $\pi \in \Pi$ are asymptotically equivalent and optimal. Moreover, any decision that is asymptotically equivalent to a quasi-Bayes decision under a prior $\pi \in \Pi$ is optimal. In particular, bootstrap rules will be asymptotically equivalent to quasi-Bayes rules (and hence optimal) provided the bootstrap distribution for an efficient estimator $\hat \mu$ of $\mu$ and the quasi-posterior are sufficiently close.

Similar to the parametric case, we say $\{\delta_n\},\{\delta_n'\} \in \mb D$ fail to be asymptotically equivalent at $(\mu_0,\eta_0)$ if 
\[
 \lim_{n \to \infty} P_{n,h}(\delta_n(X^n) = d) \neq \lim_{n \to \infty} P_{n,h}(\delta_n'(X^n) = d)
\]
for some $h \in \mathbb R^K$ and some $d \in {\cal D}$.

\medskip
\begin{theorem}\label{t2s}
Suppose that Assumptions~\ref{a1} and~\ref{a2s} hold, ${\cal R}(\{\delta_n\}; (\mu_0,\eta_0))$ is finite for at least one $\{\delta_n\} \in \mathbb D$, and there are no first-order ties. Suppose that $\{\delta_n\} \in \mb D$ is not asymptotically equivalent to $\{\delta_n^*(\,\cdot\,;\pi)\}$ for some (equivalently, all) $\pi \in \Pi$. Then at any $(\mu_0,\eta_0)$ at which asymptotic equivalence fails,
\[
 \mc R(\{\delta_n\};  (\mu_0,\eta_0)) > \mc R(\{\delta_n^*(\,\cdot\,;\pi)\};  (\mu_0,\eta_0)) \quad \mbox{for all} \; \pi \in \Pi.
\]
\end{theorem}
\medskip

As in the parametric case, asymptotic equivalence to $\{\delta_n^*(\,\cdot\,;\pi)\}$ for some $\pi \in \Pi$ is necessary for optimality whenever there are no first-order ties. Thus, as discussed in Section~\ref{subsec:optimality.plugin}, plug-in rules may fail to be optimal in settings where the maximum risks $R(d,\mu)$ is only directionally differentiable at $\mu_0$. This includes the empirical application from Section~\ref{sec:treatment.empirical} and the further examples to treatment assignment and optimal pricing that we discuss in the next two sections.

\section{Treatment Rules: Further Applications}\label{sec:treatment}

This section expands on our running example of treatment assignment under partial identification. We first review several empirically relevant approaches for constructing bounds on the ATE in Section~\ref{subsec:treatment.bounds}. Each of these constructions leads to bounds $b_L(\mu)$ and $b_U(\mu)$ that will in general be only directionally differentiable in a vector of reduced-form parameters $\mu$. We then discuss implementation of the optimal decision rules in these settings in Section~\ref{subsec:treatment.implementation}. As far as we are aware, ours is the first work to propose optimal decisions for these realistic empirical settings.

\subsection{Examples of Treatment Effect Bounds}
\label{subsec:treatment.bounds}

\paragraph{Intersection Bounds.}
This setting generalizes the empirical application from Section~\ref{sec:treatment.empirical}. Suppose $X^n$ consists of data from $K$ observational studies. In each study $k$ we can consistently estimate lower and upper bounds $b_{L,k}(\mu_k)$ and $b_{U,k}(\mu_k)$ on the ATE as a function of population moments $\mu_k$. We then obtain the intersection bounds 
\[
b_L(\mu) = \max_{1 \leq k \leq K} b_{L,k}(\mu_k) \,, \quad \quad b_U(\mu) = \min_{1 \leq k \leq K} b_{U,k}(\mu_k) ,
\]
with $\mu = (\mu_k)_{k=1}^K$. While the bounds $b_{L,k}$ and $b_{U,k}$ may themselves be smooth in $\mu_k$, the presence of the $\min$ and $\max$ operations makes the intersection bounds $b_L(\mu)$ and $b_U(\mu)$ only directionally differentiable in $\mu$. 

\bigskip

\paragraph{Bounds via IV-like Estimands.}
\cite{MST} present an approach for bounding the ATE and other causal effects using IV-like estimands from observational studies. Suppose treatment is determined by $D = \mb I[U \leq v(Z)]$ where $U \sim \mathrm{Uniform}(0,1)$ and $Z = (X,Z_0)$ collects control variables $X$ and instrumental variables $Z_0$. According to \cite{HV1999,HV2005}, the ATE may be expressed as a functional $\Gamma_0(m)$, where $m = (m_0,m_1)$ are the marginal treatment response (MTR) functions
\[
m_d(u,x) = \E[ Y_d | U = u, X = x] \,, \quad d \in \{0,1\},
\]
and
\[
\Gamma_0(m) = \E \left[ \int_0^1 m_1(u,X) \, du - \int_0^1 m_0(u,X) \,  du \right].
\]

\cite{MST} show the MTR functions, and hence the identified set for the ATE, can be disciplined if we know the value of certain IV-like estimands. For ease of exposition, consider a single IV estimand
\[
\beta_{IV} = \frac{\mr{Cov}(Y,Z_0)}{\mr{Cov}(D,Z_0)} \,,
\]
resulting from using $Z_0$ as an instrument for treatment status dummy $D$ in the observational study. The IV estimand may be expressed as $\beta_{IV} = \Gamma_\beta(m)$ where
\[
\Gamma_\beta(m) = \E \left[ \int_0^1 m_0(u,X) s(0,Z_0) \mb I[u > p(z_0)] \,  du + \int_0^1 m_1(u,X) s(1,Z_0) \mb I[u \leq p(z_0)]\,   du \right]
\]
with $s(d,z) = \frac{z_0 - \E[Z_0]}{\mr{Cov}(Z_0,D)}$ and where $p(z_0) = \E[D|Z_0 = z_0]$ is the propensity score. In this case, the bounds of \cite{MST} are
\[
b_L(\mu) = \inf_{m \in \mc S : \Gamma_\beta(m) = \beta_{IV}} \Gamma_0(m) ,  \quad \quad 
b_U(\mu) =  \sup_{m \in \mc S : \Gamma_\beta(m) = \beta_{IV}} \Gamma_0(m) ,
\]
where $\mc S$ is a class of functions and $\mu = (\mr{Cov}(Y,Z_0), \mr{Cov}(D, Z_0), \E[Z_0], p)$, which is finite-dimensional if $Z_0$ has finite support (e.g. binary $Z_0$). They show that $b_L(\mu)$ and $b_U(\mu)$ may be expressed as the optimal values of linear programs parameterized by $\mu$. It is known from \cite{MilgromSegal2002} (see also  \cite{Mills1956} and \cite{Williams1963} for linear programs) that the value of optimization problems may  only be directionally differentiable in parameters. 

\bigskip

\paragraph{Non-separable Panel Data Models.}
Suppose the outcome for individual $i$ at date $t$ is of the form $Y_{it} = g(X_{it}, \alpha_i, \varepsilon_{it})$ where $X_{it}$ is a vector of covariates, $\alpha_i$ is a latent individual effect, and $\varepsilon_{it}$ is a vector of disturbances, which are independent across individuals and time. Consider an intervention that changes in covariates from $x^0$ to $x^1$. The ATE associated with the intervention is 
\[
\int  \left( g(x^1,\alpha,\varepsilon) - g(x^0,\alpha,\varepsilon) \right)  d Q(\alpha,\varepsilon),
\]
where $Q$ is a distribution over $(\alpha,\varepsilon)$. When outcomes and covariates are discrete, parametric restrictions on the distribution of $\varepsilon$ and functional form restrictions on $g$ are generally insufficient to point identify the ATE without parametric restrictions on the distribution of $\alpha$. A leading example is dynamic panel data models in which $X_{it}$ collects lagged values of a discrete outcome $Y_{it}$---see, e.g., \cite{HonoreTamer2006} and \cite{Torgovitsky2019}. Building on \cite{HonoreTamer2006}, \cite{Chernozhukovetal2013} and \cite{TorgovitskyPies2019} derive bounds on the ATE without parametric assumptions on $Q$. Their bounds may be expressed as the value of optimization problems (linear programs) parameterized by a finite-dimensional vector of choice probabilities $\mu$. As in the previous example, $b_L(\mu)$ and $b_U(\mu)$ may therefore only be directionally differentiable in $\mu$.

\subsection{Implementation}
\label{subsec:treatment.implementation}

The first two examples are semiparametric and we can take $X^n$ to be a vector of estimators $\hat \mu$ of $\mu$. Assuming $\hat \mu$ is asymptotically efficient and the DM has available a consistent estimator $\hat I^{-1}$  of the asymptotic variance of $\hat \mu$, then the optimal decision can be implemented based on a $N(\hat \mu, (n \hat I)^{-1})$ quasi-posterior as in (\ref{eq:quasi_posterior}).

In the third example with discrete outcomes and covariates, the distribution $P_{n,\mu}$ of the data $((Y_{it},X_{it})_{t=1}^T)_{i=1}^n$ can be identified with a multinomial distribution parameterized by $\mu$. In this case our parametric theory applies and the optimal decision can easily be implemented using either the bootstrap or Bayesian methods.

\section{Optimal Pricing with Unobserved Heterogeneity}
\label{sec:pricing}

Our methods can also be applied to make optimal pricing decisions in models with rich unobserved heterogeneity using revealed-preference demand theory. Section~\ref{subsec:counterfactual.model} provides the model and empirical setting.  Section~\ref{subsec:counterfactual.demand} presents techniques for computing sharp bounds on functionals of counterfactual demand using linear programming. Finally, we discuss how to implement our methods in Section~\ref{subsec:counterfactual.implementation}. 

\subsection{Model}\label{subsec:counterfactual.model}

The DM observes repeated cross sections of individual demands $X^n = \big( X_{b,1},\ldots,X_{b,n} \big)_{b=1}^{B}$, where each $X_{b,i} \in \mb R^m$ is the demand of individual $i$ for $m$ goods under prices $q_b$:
\[ 
 X_{b,i} = \mr{arg}\max_{x \in \mc B_b} u_i(x) ,
\]
where $\mc B_b = \{x \in \mb R^m : x'q_b = 1\}$ is the budget set (expenditure is normalized to one). Individuals are heterogeneous in their utility functions $u_i$. We assume the demand system is rationalized by a random utility model with a probability distribution $F$ over utility functions $u$. The demand under $q_b$ of a randomly selected individual may therefore be interpreted as stochastic. The mass $p_b(s)$ of individuals whose demand is in a set $s \subset \mc B_b$ at price $q_b$ is
\begin{equation}
 p_b(s) = \int \mb I \Big[{\textstyle \argmax_{x \in \mc B_b}} u(x) \in s \Big] d F(u), \; s \in \mc{B}_b, \; b=1,\ldots,B\label{def:RUM}
\end{equation}
(see, e.g., \cite{KitamuraStoye2018}, henceforth KS18). 

The DM wishes to choose a new price vector $q_d$ for $d \in \mc D = \mc O \cup \mc C$. The price vectors $\{q_d : d \in \mc O\}$ are a subset of the observed prices $q_1,\ldots,q_B$, while each $\{q_d : d \in \mc C\}$ is a set of counterfactual price vectors. In principle, the set $\mc C$ of new price vectors could be large, representing prices rounded to nearest currency units or tax rates rounded to the nearest percentage. Let $w_d(X_d)$ represent a functional of demand (e.g., revenue or welfare) under prices $q_d$. The DM's goal is to choose $d \in \mc D$ to maximize the average of $w_d(X_d)$. For $d \in \mc O$ the average demand $\E[w_d(X_d)]$ is identified from observed choice behavior under $q_d$. However, the observed choice behavior is only sufficient to bound, but not point-identify, $\E[w_d(X_d)]$ for counterfactual prices $q_d$, $d \in \mc C$.

\subsection{Bounds on Functionals of Counterfactual Demand}\label{subsec:counterfactual.demand}

\begin{figure}[t!]
	\begin{center}
		\begin{tikzpicture}[scale=0.9]
			
			\def\ma{12} 
			\def\paa{4} 
			\def\pba{2} 
			
			\def\mb{12} 
			\def\pab{2} 
			\def\pbb{4} 
			
			\def\mb{12} 
			\def\pac{2.5} 
			\def\pbc{2.5} 

			\coordinate (inta) at ({\ma/\paa}, 0);  
			\coordinate (intb) at (0, {\ma/\pba});  
			
			\coordinate (inta2) at ({\ma/\pac}, 0);  
			\coordinate (intb2) at (0, {\ma/\pbc});  
			
			\coordinate (intc) at ({\mb/\pab}, 0);  
			\coordinate (intd) at (0, {\mb/\pbb});  
			
			\coordinate (inte) at (0, {(\ma/\pba - \mb/\pbb)/(\paa/\pba - \pab/\pbb)});
			\coordinate (intf) at ({(\ma/\pba - \mb/\pbb)/(\paa/\pba - \pab/\pbb)}, {(\ma/\pba - \mb/\pbb)/(\paa/\pba - \pab/\pbb)}); 
			
			\coordinate (intg) at (0, {(\mb/\pbc - \mb/\pba)/(\pac/\pbc - \paa/\pba)});
			\coordinate (inth) at ({(\mb/\pbc - \mb/\pbb)/(\pac/\pbc - \pab/\pbb)}, {(\mb/\pbc - \mb/\pba)/(\pac/\pbc - \paa/\pba)}); 
			
			
			
			\draw[ultra thick, color=blue] (intb) -- (inta) node[pos = 0.2, right] {$s_{11}$} node[pos = 0.9, left] {$s_{31}$} node[pos = 0.5, below] {$s_{21}\;\;\;$}; 
			
			\draw[ultra thick, dotted, color=red] (intb2) -- (inta2) node[pos = 0.15, below] {$s_{12}\;\;$} node[pos = 0.8, below] {$s_{32}\;\;$ } node[pos = 0.5, above] {$\;\;s_{22}$}; 

			\draw[ultra thick, dashed, color=black] (intc) -- (intd) node[pos = 0.2, above] {$s_{30}$} node[pos = 0.9, below] {$s_{10}$} node[pos = 0.5, left] {$s_{20}$}; 
			
			\draw[ultra thick, color=blue] (4,4.5) -- (5,4.5) node[right] {{\color{black}Observed budget $\mathcal{B}_1$}};
			\draw[ultra thick, dotted, color=red] (4,4) -- (5,4) node[right] {{\color{black}Observed budget $\mathcal{B}_2$}};			
			\draw[ultra thick, dashed, color=black] (4,3.5) -- (5,3.5) node[right] {{\color{black}Counterfactual budget $\mathcal{B}_0$}};
			
			
			
			\draw[thick, ->] (0,0) -- (6.5,0) node[right] {Good 1};
			\draw[thick, ->] (0,0) -- (0,6.5) node[above] {Good 2};
			
		\end{tikzpicture}
	\end{center}
	\caption{Choices and Budget Lines}
	\label{fig:ex.pricing}
\end{figure}
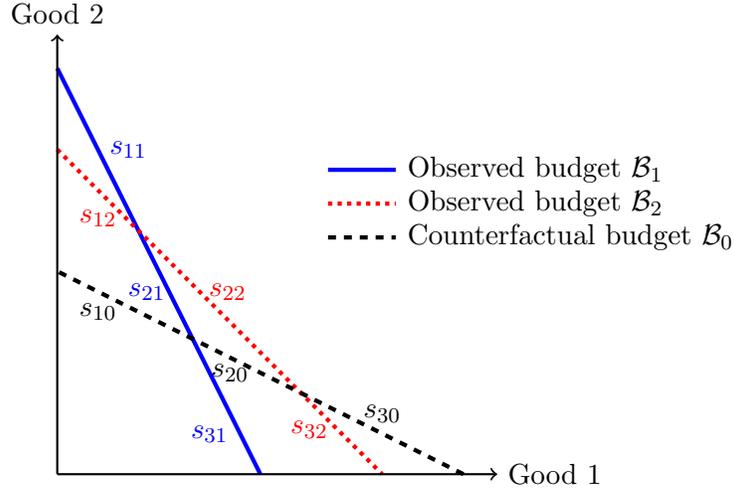

\cite{KitamuraStoye2019} present a general approach using linear programming to bound functionals of counterfactual demand. We introduce their approach with an example. Consider Figure~\ref{fig:ex.pricing}. There are two goods, and the DM has observed the demand for two price vectors $q_1$ and $q_2$, which generate the budget sets ${\cal B}_1$ and ${\cal B}_2$. The counterfactual price vector is $q_0$ which generates the counterfactual budget set $\mc B_0$. 

The budget lines in Figure~\ref{fig:ex.pricing} are divided into segments $s_{jb}$, called ``patches'' in KS18. Patch $s_{11}$ is the segment of $\mathcal{B}_1$ from the $y$-axis to the intersection of $\mathcal{B}_1$ and $\mathcal{B}_2$, patch $s_{21}$ is the segment of $\mathcal{B}_1$ between $\mathcal{B}_2$  and $\mathcal{B}_0$, and so forth.\footnote{Like KS18, we suppose for simplicity that the distribution of demand is continuous so we can disregard the ``intersection patches'' formed at the intersections of budget lines.}
There are 9 potential combinations of patches consumers may choose from among ${\cal B}_1$ and ${\cal B}_2$: $(s_{11},s_{12})$, $(s_{11},s_{22})$, $(s_{11},s_{32})$, $(s_{21},s_{12})$, $(s_{21},s_{22})$, $(s_{21},s_{32})$, $(s_{31},s_{12})$, $(s_{31},s_{22})$, $(s_{31},s_{32})$. Each combination corresponds to a consumer type. By revealed preference we know a consumer will never choose $(s_{21},s_{12})$ or $(s_{31},s_{12})$. This leaves a total of 7 rational types of consumer. Let
\[
 A = \left[ \begin{array}{ccccccc}
 1 & 1 & 1 & 0 & 0 & 0 & 0 \\
 0 & 0 & 0 & 1 & 1 & 0 & 0 \\
 0 & 0 & 0 & 0 & 0 & 1 & 1 \\
 1 & 0 & 0 & 0 & 0 & 0 & 0 \\
 0 & 1 & 0 & 1 & 0 & 1 & 0 \\
 0 & 0 & 1 & 0 & 1 & 0 & 1
 \end{array} \right].
\]
The rows of $A$ correspond to $s_{11}, s_{21}, s_{31}, s_{12}, s_{22}, s_{32}$ and the columns of $A$ correspond to the 7 rational types.  Let $p = (p_1(s_{11}),p_1(s_{21}), p_1(s_{31}), p_2(s_{12}), p_2(s_{22}), p_2(s_{32}))$ collect the corresponding choice probabilities. KS18 showed that the demand system $p$ is rationalizable if and only if
\[
 p = A f,
\]
for some $f \in \Delta^6$, the unit simplex in $\mb R^7$, representing the probabilities of the $7$ rational types. These probabilities constrain the distribution $F$ of utilities.

Now consider choice behavior on the counterfactual budget set $\mc B_0$. Each of the 7 rational types  may choose a counterfactual demand in patch $s_{10}$, $s_{20}$, or $s_{30}$, for a total of 21 potential types. By revealed preference, a consumer who chose $s_{31}$ must choose $s_{20}$ or $s_{30}$, and a consumer who chose $s_{32}$ must choose $s_{30}$. This leaves a total of 16 rational types. We may represent the system as 
\begin{equation} \label{eq:rationalize.demand.counterfactual}
 \left[ \begin{array}{c} p \\ p^* \end{array} \right]  
 = A^* f^* \,,
\end{equation}
where $p^* = (p_0(s_{10}),p_0(s_{20}),p_0(s_{30}))$ collects the counterfactual choice probabilities for patches $s_{10}$, $s_{20}$, and $s_{30}$, $f^* \in \Delta^{15}$ collects the probabilities of observing each rational type, and 
\[
 A^* = \left[ \begin{array}{cccccccccccccccc}
 1 & 1 & 1 & 1 & 1 & 1 & 1 & 1 & 1 & 0 & 0 & 0 & 0 & 0 & 0 & 0 \\
 0 & 0 & 0 & 0 & 0 & 0 & 0 & 0 & 0 & 1 & 1 & 1 & 1 & 0 & 0 & 0 \\
 0 & 0 & 0 & 0 & 0 & 0 & 0 & 0 & 0 & 0 & 0 & 0 & 0 & 1 & 1 & 1 \\
 1 & 1 & 1 & 0 & 0 & 0 & 0 & 0 & 0 & 0 & 0 & 0 & 0 & 0 & 0 & 0 \\
 0 & 0 & 0 & 1 & 1 & 1 & 0 & 0 & 0 & 1 & 1 & 1 & 0 & 1 & 1 & 0 \\
 0 & 0 & 0 & 0 & 0 & 0 & 1 & 1 & 1 & 0 & 0 & 0 & 1 & 0 & 0 & 1 \\
 1 & 0 & 0 & 1 & 0 & 0 & 1 & 0 & 0 & 1 & 0 & 0 & 0 & 0 & 0 & 0 \\
 0 & 1 & 0 & 0 & 1 & 0 & 0 & 1 & 0 & 0 & 1 & 0 & 0 & 1 & 0 & 0 \\
 0 & 0 & 1 & 0 & 0 & 1 & 0 & 0 & 1 & 0 & 0 & 1 & 1 & 0 & 1 & 1
 \end{array} \right] ,
\]
where the rows correspond to choosing $s_{11}, s_{21}, s_{31}, s_{12}, s_{22}, s_{32}, s_{10}, s_{20}, s_{30}$. Partition $A^*$ as 
\[
 A^* = \left[ \begin{array}{c} A^*_{obs} \\ A^*_{unobs} \end{array} \right],
\]
where $A^*_{obs}$ collects the first $6$ rows of $A^*$ (corresponding to the observed budget sets) while $A^*_{unobs}$ collects the rows of $A^*$ corresponding to the patches $s_{10}$, $s_{20}$, and $s_{30}$. 

Following \cite{KitamuraStoye2019}, we may deduce sharp bounds on $\E[w_0(X_0)]$ as follows. Let $(\underline w_1, \underline w_2, \underline w_3)$ and $(\overline w_1, \overline w_2, \overline w_3)$ denote row vectors which collect the smallest and largest values of $w_0(x)$ for $x$ in $s_{10}$, $s_{20}$ and $s_{30}$. Then the bounds on $\E[w_0(X_0)]$ are
\[ 
\begin{aligned}
 w_L(p) & = \min \left\{ (\underline w_1, \underline w_2, \underline w_3)^T A^*_{unobs} f^* : p = A^*_{obs} f^*, f^* \in \Delta^{15} \right\} ,\\
 w_U(p) & = \max \left\{ (\overline w_1, \overline w_2, \overline w_3)^T A^*_{unobs} f^* : p = A^*_{obs} f^*, f^* \in \Delta^{15} \right\}.
 \end{aligned} 
\]

More generally, the preceding argument applies with a collection of $B$ observed budget sets and multiple goods. In that case, representation (\ref{eq:rationalize.demand.counterfactual}) holds for observed choice probabilities $p$ of patches on the $B$ observed budget sets and counterfactual choice probabilities $p^*$ of patches on the counterfactual budget set $\mc B_d$ generated by $q_d$ for $d \in \mc C$. Suppose there are $J$ patches across $\mc B_d$ and $T$ rational types. Then letting row vectors $\underline w_d$ and $\overline w_d$ collect the smallest and largest values of $w_d(x)$ for $x$ in each of the $J$ patches, sharp bounds on $\E[w_d(X_d)]$ are
\begin{equation}\label{eq:linear.program.demand}
 \begin{aligned}
 w_{d,L}(p) & = \min \left\{ \underline w_d^T A^*_{unobs} f^* : p = A^*_{obs} f^*, f^* \in \Delta^{T-1} \right\}, \\
 w_{d,U}(p) & = \max \left\{ \overline w_d^T A^*_{unobs} f^* : p = A^*_{obs} f^*, f^* \in \Delta^{T-1} \right\},
 \end{aligned} 
\end{equation}
where we have partitioned the $A^*$ matrix analogously to the simple 3 budget example.

\subsection{Implementation}\label{subsec:counterfactual.implementation}

As in \cite{BergemannSchlag2011}, we assume the DM has a preference for robustness. That is, the DM wishes to choose $d$ to minimize maximum risk, where the maximum is taking over the identified set of counterfactual demand responses.

The DM's problem maps into our framework as follows. For each $d \in \mc O$, the value $\E[w_d(X_d)]$ is identified from observed choice behavior under $q_d$. The remaining reduced-form parameters are the patch probabilities $p$. Thus, $\mu = (p,(\E[w_d(X_d)])_{d \in \mc O})$. Here $\theta = F$ and $\Theta_0(\mu)$ is all $F$ that rationalize the patch probabilities $p$ consistent with revealed preference. The vector $Y = (w_d(X_d))_{d \in \mc C}$ collects functionals of demand under the counterfactual prices. Each $ \theta \equiv F \in \Theta_0(\mu)$ induces a distribution $G_\theta$ for $Y$. Although we suppressed it in the previous subsections, we now write $\E_\theta[w_d(X_d)]$ for $d \in \mc C$ to denote that the average counterfactual demand functional depends on the structural parameter $\theta$.

For $d \in \mc O$, the DM incurs risk 
\[
 r(d,\theta,\mu) = -\E[w_d(X_d)],
\]
which is point identified from $\mu \equiv (p,(\E[w_d(X_d)])_{d \in \mc O})$. For $d \in \mc C$, the DM incurs risk
\[
 r(d,\theta,\mu) = -\E_\theta[w_d(X_d)],
\]
which is set-identified and may be bounded as described in the previous subsection. The maximum risk is
\[
 R(d,\mu) = \begin{cases}
 -\E[w_d(X_d)] & d \in \mc O, \\
 -w_{d,L}(p) & d \in \mc C,
 \end{cases}
\]
where $w_{d,L}(p)$ was defined in (\ref{eq:linear.program.demand}).

This is a semiparametric model and our implementation follows the steps described in Section~\ref{subsec:decision.semiparametric}. Partition $\mu = (\mu_1,\ldots,\mu_B)$ where $\mu_b$ collects the parameters corresponding to budget set $\mc B_b$. As we have assumed that each of the $B$ observed budgets is sampled in a repeated cross section, we can estimate each $\mu_b$ by just-identified GMM. As we observe repeated cross sections under $B$ different price vectors, we can simply take $\eta = \times_{b=1}^B \eta_b$, where $\eta_b$ is the distribution of $X_{b,i}$. We can then form a quasi-posterior $\pi_n(\mu|X^n)$ based on a limited-information Gaussian quasi-likelihood as in~(\ref{eq:quasi_posterior}). For $d \in \mc O$, the expected value $\E[w_d(X_d)]$ is an element of $\mu$ and so $\bar R_n(d)$ is simply its quasi-posterior mean:
\[
 \bar R_n(d) =  -\int \E[w_d(X_d)] \,  d \pi_n(\mu)\,.
\]
For $d \in \mc C$, the posterior maximum risk is
\[
 \bar R_n(d) = -\int w_{d,L}(p) \,  d \pi_n(\mu)\,.
\]
This may be computed by sampling $p$ from the quasi-posterior, solving the linear program~(\ref{eq:linear.program.demand}) defining $w_{d,L}(p)$ for each draw of $p$, then taking the average across draws.

The optimal decision  minimizes $\bar R_n(d)$ for $d \in \mathcal D = \mc O \cup \mc C$. As the value of a linear program is typically  directionally differentiable, the asymptotic distribution of the posterior mean $\bar R_n(d)$ for $d \in \mc C$ may be different from that of the corresponding plug-in values $-w_{d,L}(\hat p)$. In this case, the optimal decision $\delta_n^*$ may outperform the plug-in rule.

\section{Conclusion}
\label{sec:conclusion}

We derived optimal statistical decision rules for discrete choice problems when payoffs depend on a set-identified parameter $\theta$ and the decision maker can use a point-identified parameter $\mu$ to deduce restrictions on $\theta$. Our notion of optimality combines a minimax approach to handle the ambiguity from partial identification of $\theta$ given $\mu$ with average risk minimization for $\mu$. In many empirically relevant applications, the maximum risk depends non-smoothly on $\mu$, making plug-in rules sub-optimal. We provided detailed applications to optimal treatment choice under partial identification and optimal pricing with rich unobserved heterogeneity. Our asymptotic approach is well suited for empirical settings in which the derivation of finite-sample optimal rules is intractable. While continuous decisions fall outside the scope of our theory and analysis, it would be interesting to study asymptotic efficiency in the continuous case as well.

{
\bibliography{ref_discrete}
}

\clearpage

\renewcommand{\thepage}{A.\arabic{page}}
\setcounter{page}{1}

\begin{appendix}

\section*{Online Appendix}

\section{Preliminaries}
\label{ax:preliminaries}

We first introduce relevant concepts on limit experiments and asymptotic representations. We refer the reader to \cite{Vandervaart1991,vanderVaart1998} for  comprehensive treatments. 

An \emph{experiment} is a measurable space equipped with a family of probability measures. Let $\mc E_{n} = (\mc X^n,\mc B_n , P_{n,h} : h \in \mb R^K)$ where $\mc B_n$ denotes the Borel $\sigma$-algebra on $\mc X^n$. Each $\mc E_n$ corresponds to observing a random sample $X_1,\ldots,X_n$ from $P_{\mu_0 + h/\sqrt n}$ with parameter $h \in \mathbb R^K$. Also let $\mc E = (\mb R^K, \mc B_K, N(h, I_{\mu_0}^{-1}) : h \in \mb R^K)$ where $\mc B_K$ denotes the Borel $\sigma$-algebra on $\mb R^K$. This experiment corresponds to observing a single $Z \sim N(h, I_{\mu_0}^{-1})$ with unknown $h$.  We suppress dependence of $\mc E_n$, $\mc E$, $P_{n,h}$, and $N_h$ on $\mu_0$ to simplify notation.

The sequence of experiments $\{\mc E_{n}\}$ is said to converge to the \emph{limit experiment} $\mc E$ if for all finite subsets $H$ of $\mathbb R^K$ and every $h_0 \in \mathbb R^K$, we have
\[
 \left( \frac{ d P_{n,h}}{ d P_{n,h_0}}(X^n) \right)_{h \in H} \overset{P_{n,h_0}}{\rightsquigarrow}
 \;\; \left( \frac{ d N(h,I_{\mu_0}^{-1})}{ d N(h_0,I_{\mu_0}^{-1})}(Z) \right)_{h \in H},
\]
where $\overset{P_{n,h}}{\rightsquigarrow}$ denotes convergence in distribution along $\{P_{n,h}\}$. This convergence holds under Assumption~\ref{a2}\ref{a2.1}-\ref{a2.3} \cite[Example~7.15 and Corollary~9.5]{vanderVaart1998}.

Because $\mc E$ is dominated,\footnote{In particular, each $N(h,I_{\mu_0}^{-1})$ distribution has density with respect to Lebesgue measure.} we can invoke an asymptotic representation theorem of \cite{Vandervaart1991}. For each $\{\delta_n\} \in \mb D$ there exists a \emph{matching decision} $\delta_\infty(Z,U)$ given by a measurable map $\delta_\infty : \mb R^K \times [0,1]\to \mc D$, the observation $Z \sim N(h,I_{\mu_0}^{-1})$, and an independent randomization $U \sim \mathrm{Uniform}[0,1]$, with the property that for each $h \in \mb R^K$,
\[
 \lim_{n \to \infty} P_{n,h}(\delta_n(X^n) = d) = \mathbb P_h (\delta_\infty(Z,U) = d), \quad d \in \mathcal D,
\]
where $\mathbb P_h$ is the product of the $N(h,I_{\mu_0}^{-1})$ and $U[0,1]$ distributions. Although $\delta_\infty(Z,U)$ depends on the randomization $U$ in general, under our regularity conditions this randomization is unnecessary and any optimal decision depends only on $Z$.

\section{Proof of Main Results}
\label{ax:proofs}

\subsection{Proofs for Section~\ref{sec:optimality}}

We first state and prove a lemma which is used to prove Theorems~\ref{t1} and \ref{t2}. The first part shows that if $\pi \in \Pi$, then $\{\delta_n^*(\,\cdot\,;\pi)\} \in \mathbb D$ and its matching decision is independent of the prior $\pi$ and nonrandomized. Thus, we drop dependence of the matching decision on $\pi$ and $U$ and simply write $\delta_\infty^*(Z)$. In the second part, we derive expressions for criteria (\ref{eq:criterion_asymptotic_h}) and  (\ref{eq:criterion_asymptotic}) in terms of the limit experiment. Finally, in the third part, we show that our optimality criterion  (\ref{eq:criterion_asymptotic}) is minimized for matching decisions corresponding to $\{\delta_n^*(\,\cdot\,;\pi)\}$ for $\pi \in \Pi$. 

Recall that $\dot R_{d,\mu_0}$ denotes the directional derivative of $R(d,\mu)$ at $\mu_0$, $Z \sim N(h,I_{\mu_0}^{-1})$ represents the asymptotic behavior of an efficient estimator of $\mu$, $Z^* \sim N(0,I_{\mu_0}^{-1})$ independently of $Z$ represents asymptotic uncertainty about the local parameter $h$, and $\E^*$ denotes expectation with respect to $Z^*$.

\medskip
	
\begin{lemma}\label{l1}
Suppose that Assumptions \ref{a1} and \ref{a2} hold. Fix any $\mu_0 \in \mc M$ for which there are no first-order ties.
\begin{enumerate}[label={(\roman*)}, nosep]
\item \label{l1.1} If $\pi \in \Pi$, then $\{\delta_n^*(\,\cdot\,;\pi)\} \in \mb D$ and the matching decision of $\{\delta_n^*(\,\cdot\,;\pi)\}$ is $\delta_\infty^*(Z)$ with
\[
 \delta_\infty^*(z) = \argmin_{d \in \underline{\mc D}_{\mu_0}}  \mathbb E^* \left[ \dot R_{d,\mu_0}[Z^* + z] \right] 
\]
for almost every $z \in \mb R^K$.

\item \label{l1.2} For any $\{\delta_n\} \in \mathbb D$, 
\begin{align}
 \mc R(\{\delta_n\};\mu_0, h) 
 & = \sum_{d \in \underline{\mc D}_{\mu_0}} \left( \dot R_{d,\mu_0}[h] - \min_{i \in \underline{\mc D}_{\mu_0}} \dot R_{i,\mu_0}[h] \right) \mathbb P_h( \delta_\infty(Z,U) = d) ,  \notag\\
 \mc R(\{\delta_n\};\mu_0) 
 & = \int \Bigg( \sum_{d \in \underline{\mc D}_{\mu_0}} \left( \dot R_{d,\mu_0}[h] - \min_{i \in \underline{\mc D}_{\mu_0}} \dot R_{i,\mu_0}[h] \right)  \mathbb P_h( \delta_\infty(Z,U) = d) \Bigg) d h, \label{eq:l1.0}
\end{align}
where $\delta_\infty(Z,U)$ is the matching decision of $\{\delta_n\}$.
\item \label{l1.3} The matching decision $\delta_\infty^*(Z)$ in (i) minimizes the right-hand side of (\ref{eq:l1.0}) over all matching decisions $\delta_\infty(Z,U)$ of sequences $\{\delta_n\} \in \mathbb D$.
\end{enumerate} 
\end{lemma}

\begin{proof}[Proof of Lemma \ref{l1}]
\textbf{Part \ref{l1.1}:} Let $A_n$ denote the event $\delta_n^*(X^n;\pi) \in \underline{\mc D}_{\mu_0}$. By Lemma~\ref{lem:mistake_prob_bayes}, we have $\lim_{n \to \infty} \sqrt n \, P_{n,h}(A_n^c) = 0$ for all $h \in \mathbb R^K$. Hence, condition~\ref{D.2} in Definition~\ref{def:D} holds. In particular, if $\underline{\mc D}_{\mu_0} \subsetneq \mc D$, then 
\begin{equation}\label{eq:l1.1}
 \lim_{n \to \infty} P_{n,h}(\delta_n^*(X^n;\pi) = d) = 0 \quad \mbox{all $h \in \mathbb R^K$ and $d \not \in \underline{\mc D}_{\mu_0}$}.
\end{equation}
Hence, if $\underline{\mc D}_{\mu_0} = \{d\}$ we have $\lim_{n \to \infty} P_{n,h}(\delta_n^*(X^n;\pi) = d) = 1$. 

It remains to characterize $\lim_{n \to \infty} P_{n,h}(\delta_n^*(X^n;\pi) = d)$ for $d \in \underline{\mc D}_{\mu_0}$ when $|\underline{\mc D}_{\mu_0}| \geq 2$. We require a rule to break ties when $\argmin_{d \in \mc D} \bar R_n(d)$ is not a singleton. We shall take the smallest index $d$ among the set of minimizers, though it will be seen below that the matching decision doesn't depend on the tie-breaking rule. Thus, any (possibly randomized) tie-breaking rule would lead to the same matching decision. Under this tie-breaking rule, we have
\[ 
 \mathbb I[\delta_n^*(X^n;\pi) = d]  =  \mathbb I \left[ \bar R_n(d) < \min_{i \in \underline{\mc D}_{\mu_0}: i  < d} \bar R_n(i) , \;\mbox{and}\; \bar R_n(d) \leq \min_{i \in \underline{\mc D}_{\mu_0}: i \geq d} \bar R_n(i) \right] \;\; \mbox{on $A_n$},
\]
where the minimum over an empty set is $+\infty$. Since $|\Pr(A) - \Pr(B)| \leq 2 \Pr(C^c)$ if $A = B$ holds on $C$, we obtain
\begin{multline*}
 \left| P_{n,h}(\delta_n^*(X^n;\pi) = d) - P_{n,h} \left( \bar R_n(d) < \min_{i \in \underline{\mc D}_{\mu_0}: i  < d} \bar R_n(i) , \; \mbox{and}\; \bar R_n(d) \leq \min_{i \in \underline{\mc D}_{\mu_0}: i \geq d} \bar R_n(i) \right) \right| \\
 \leq 2 P_{n,h}(A_n^c) \to 0.
\end{multline*}
As $R(d,\mu_0) = R(d',\mu_0)$ for all $d,d' \in \underline{\mc D}_{\mu_0}$, we have
\begin{multline*}
 P_{n,h} \left( \bar R_n(d) < \min_{i \in \underline{\mc D}_{\mu_0}: i  < d} \bar R_n(i) ,\; \mbox{and}\; \bar R_n(d) \leq \min_{i \in \underline{\mc D}_{\mu_0}: i \geq d} \bar R_n(i) \right) \\
 = P_{n,h} \bigg( \int \sqrt n (R(d,\mu) - R(d,\mu_0)) \, d \pi_n(\mu) < \min_{i \in \underline{\mc D}_{\mu_0}: i  < d} \int \sqrt n (R(i,\mu) - R(i,\mu_0))  \, d \pi_n(\mu), \\
  \mbox{and}\;  \int \sqrt n (R(d,\mu) - R(d,\mu_0)) \, d \pi_n(\mu) \leq \min_{i \in \underline{\mc D}_{\mu_0}: i  \geq d} \int \sqrt n (R(i,\mu) - R(i,\mu_0))  \, d \pi_n(\mu)\bigg).
\end{multline*}

Now by Proposition~\ref{prop:BVM}, we have
\[
 \bigg( \int \sqrt n (R(d,\mu) - R(d,\mu_0)) \, d \pi_n(\mu) \bigg)_{d \in \underline{\mc D}_{\mu_0}} \overset{P_{n,h}}{\rightsquigarrow} \;\; \bigg( \E^* \big[ \dot R_{d,\mu_0} \left[ Z^* + Z \big] \big| Z  \right] \bigg)_{d \in \underline{\mc D}_{\mu_0}}.
\]
Define $U_d = \{(u_i)_{i \in \underline{\mathcal D}_{\mu_0}}  : u_d < \min_{i \in \underline{\mc D}_{\mu_0}: i  < d} u_i$, and $u_d \leq \min_{i \in \underline{\mc D}_{\mu_0}: i \leq d} u_i\} \subset \mathbb R^{|\underline{\mathcal D}_{\mu_0}|}$ for $d \in \underline{\mathcal D}_{\mu_0}$. These sets form a partition of $\mathbb R^{|\underline{\mathcal D}_{\mu_0}|}$. Evidently, $\mr{int}(U_d) \supseteq \{(u_i)_{i \in \underline{\mathcal D}_{\mu_0}}  : u_d < \min_{i \in \underline{\mc D}_{\mu_0}: i  \neq d} u_i \}$. Thus, boundary points of $U_d$ are points at which $\argmin_{i \in \underline{\mc D}_{\mu_0}} u_i$ is not a singleton. Since $\argmin_{d \in \underline D_{\mu_0}} \E^*[\dot R_{d,\mu_0}[Z^* + z]]$ is a singleton for almost every $z$, it follows that
\[
 \mathbb{P}_h \left( \left( \E^* \left[ \left. \dot R_{d,\mu_0} \left[  Z^* + Z \right] \right| Z \right]  \right)_{d \in \underline{\mc D}_{\mu_0}} \in \partial U_d \right) = 0.
\]
Hence, by the Portmanteau theorem and the preceding four displays, for all $d \in \underline{\mc D}_{\mu_0}$ we have
\begin{multline}\label{eq:l1.2}
 \lim_{n \to \infty} P_{n,h}(\delta_n^*(X^n;\pi) = d)
 = \mathbb P_h \Big( \Big( \mathbb E^*[ \dot R_{d,\mu_0}[Z^* + Z] |Z] < \min_{i \in \underline{\mc D}_{\mu_0}: i  < d}  \mathbb E^*[ \dot R_{i,\mu_0}[Z^* + Z]|Z] \Big), \; \mbox{and } \\
  \Big( \mathbb E^*[ \dot R_{d,\mu_0}[Z^* + Z]|Z] \leq \min_{i \in \underline{\mc D}_{\mu_0}: i  \geq d} \mathbb E^*[ \dot R_{i,\mu_0}[Z^* + Z]|Z] \Big) \Big) \,.
\end{multline}
Thus, we have shown that $\{\delta_n^*(\,\cdot\,;\pi)\} \in \mb D$. 

We can now invoke Theorem 3.1 of \cite{Vandervaart1991} to guarantee existence of the matching form $\delta^*_\infty(Z,U;\pi)$ of $\{\delta_n^*(\,\cdot\,;\pi)\}$. By (\ref{eq:l1.1}), we see that $\delta^*_\infty(z,u;\pi)$ must take values in $\underline{\mc D}_{\mu_0}$ for almost every $(z,u)$. Moreover, as $\argmin_{d \in \underline{\mc D}_{\mu_0}} \E^*[\dot R_{d,\mu_0}[Z^* + z]]$ is a singleton for almost every $z$, we may restate (\ref{eq:l1.2}) as 
\[
 \lim_{n \to \infty} P_{n,h}(\delta_n^*(X^n;\pi) = d) = \mathbb P_h \Big( d =  \argmin_{i \in \underline{\mc D}_{\mu_0}}  \mathbb E^*[ \dot R_{i,\mu_0}[Z^* + Z]|Z]  \Big) .
\]
Hence, we may take $\delta^*_\infty(z,u;\pi) = \argmin_{i \in \underline{\mc D}_{\mu_0}}  \mathbb E^*[ \dot R_{i,\mu_0}[Z^* + z]]$ for almost every $(z,u)$. As $\delta^*_\infty(z,u;\pi)$ doesn't depend on $u$ or $\pi$, we simply write $\delta^*_\infty(z)$.

\medskip

\noindent
\textbf{Part \ref{l1.2}:} By Assumption \ref{a1}\ref{a1.1}, $\min_{d \in \mc D} R(d,\mu_0 + h/\sqrt n) = \min_{d \in \underline{\mc D}_{\mu_0}}R(d,\mu_0 + h/\sqrt n)$ holds for all $n$ sufficiently large. Hence, for all $n$ sufficiently large,
\begin{multline*}
 \sqrt n \left( \E_{n,h}\left[R\left(\delta_n(X^n), \mu_0 + h/\sqrt n \right)\right] - \min_{i \in \mc D} R\left(i,\mu_0 + h/\sqrt n \right) \right) \\
 = \sum_{d \in \underline{\mc D}_{\mu_0}} P_{n,h} \left( \delta_n(X^n) = d \right) \sqrt n  \Big( R(d,\mu_0 + h/\sqrt n) - \min_{i \in \underline{\mc D}_{\mu_0}} R(i,\mu_0 + h/\sqrt n) \Big) \\
 \quad \quad + \sum_{d \not \in \underline{\mc D}_{\mu_0}} \sqrt n \, P_{n,h} \left( \delta_n(X^n) = d \right)  \Big( R(d,\mu_0 + h/\sqrt n) - \min_{i \in \underline{\mc D}_{\mu_0}} R(i,\mu_0 + h/\sqrt n) \Big) ,
\end{multline*}
where the second sum is zero if $\mc D = \underline{\mc D}_{\mu_0}$; otherwise, it converges to zero as $n \to \infty$ by Definition~\ref{def:D}\ref{D.2} and Assumption \ref{a1}\ref{a1.1}. For the first sum, Theorem~3.1 of \cite{Vandervaart1991} implies
\[
  \lim_{n \to \infty} P_{n,h}( \delta_n(X^n) = d ) = \mb P_h(\delta_\infty(Z,U) = d), \quad d \in \underline{\mc D}_{\mu_0},
\]
where $\delta_\infty(Z,U)$ is the matching decision of $\{\delta_n\}$. As $R(d,\mu_0) = R(d',\mu_0)$ for all $d,d' \in \underline{\mc D}_{\mu_0}$, we have
\[
 \begin{aligned}
 & \sqrt n  \Big( R(d,\mu_0 + h/\sqrt n) -  \min_{i \in \underline{\mc D}_{\mu_0}} R(i,\mu_0 + h/\sqrt n) \Big) \\
 & = \sqrt n \left( R(d,\mu_0 + h/\sqrt n) - R(d,\mu_0)\right) - \sqrt n \left(  \min_{i \in \underline{\mc D}_{\mu_0}} R(i,\mu_0 + h/\sqrt n) -  \min_{i \in \underline{\mc D}_{\mu_0}} R(i,\mu_0) \right).\\
 & = \dot R_{d,\mu_0}[h] - \min_{i \in \underline{\mc D}_{\mu_0}} \dot R_{i,\mu_0}[h],
 \end{aligned}
\]
where the final line follows from the fact that if $\{f_i\}_{i \in \mc I}$ is a finite set of functions that are directionally differentiable at $\mu_0$ with $f_i(\mu_0) = f_{i'}(\mu_0)$ for all $i,i' \in \mc I$, then $\min_{i \in \mc I} f_i(\mu)$ has directional derivative $\min_{i \in \mc I} \dot f_{i,\mu_0}[h]$ at $\mu_0$.

\medskip

\noindent
\textbf{Part \ref{l1.3}:}  Let $g_{d,\mu_0}[h] = \dot R_{d,\mu_0}[h] - \min_{i \in \underline{\mc D}_{\mu_0}} \dot R_{i,\mu_0}[h]$. By part~\ref{l1.2}, we have
\begin{align*}
 \mc R(\{\delta_n\};\mu_0) & = \int \Bigg( \sum_{d \in \underline{\mc D}_{\mu_0}} g_{d,\mu_0}[h]  \mathbb P_h( \delta_\infty(Z,U) = d) \Bigg) d h \\
 & \propto \sum_{d \in \underline{\mc D}_{\mu_0}}  \int \int \int_0^1 \mathbb I[\delta_\infty(z,u) = d] e^{-\frac{1}{2} (z - h)^T I_{\mu_0} (z - h) }\, d u \, d z \left( g_{d,\mu_0}[h] \right)  \, d h \\
 & = \int_0^1 \int \sum_{d \in \underline{\mc D}_{\mu_0}}  \mathbb I[\delta_\infty(z,u) = d] \left( \int  g_{d,\mu_0}[h] e^{-\frac{1}{2} (z - h)^T I_{\mu_0} (z - h) } \, d h \right) d z \, d u.
\end{align*}
Changing the order of integration in the final line above is justified by Tonelli's theorem, which only requires non-negativity of the integrand. Define
\begin{equation} \label{eq:I_argmin}
 \mc I(z) = \argmin_{i \in \underline{\mc D}_{\mu_0}} \int \dot R_{i,\mu_0}[h] e^{-\frac{1}{2} (z - h)^T I_{\mu_0} (z - h) }  \,  d h  \equiv \argmin_{i \in \underline{\mc D}_{\mu_0}} \E^*\left[\dot R_{i,\mu_0}[Z^* + z] \right].
\end{equation}
Evidently, any $\delta_\infty$ with $\delta_\infty(z,u) \in \mc I(z)$ for almost every $(z,u)$ minimizes the criterion. Since there are no first-order ties at $\mu_0$, the set $\mc I(z)$ is a singleton for almost every $z$ and the criterion is minimized by $\delta_\infty^*(z)$ derived in part \ref{l1.1}.
\end{proof}

\

\begin{proof}[Proof of Theorem \ref{t1}]
By Lemma~\ref{l1}, for any $\{\delta_n\} \in \mathbb D$, we have
\begin{multline*}
 \mc R(\{\delta_n\};\mu_0)
 = \int \Bigg( \sum_{d \in \underline{\mc D}_{\mu_0}} \left( \dot R_{d,\mu_0}[h] - \min_{i \in \underline{\mc D}_{\mu_0}} \dot R_{i,\mu_0}[h] \right)  \mathbb P_h( \delta_\infty(Z,U) = d) \Bigg) d h \\
  \geq \int \Bigg( \sum_{d \in \underline{\mc D}_{\mu_0}} \left( \dot R_{d,\mu_0}[h] - \min_{i \in \underline{\mc D}_{\mu_0}} \dot R_{i,\mu_0}[h] \right)  \mathbb P_h( \delta_\infty^*(Z) = d) \Bigg) d h = \mc R(\{\delta_n^*(\,\cdot\,;\pi)\};\mu_0).
\end{multline*}
Result~\ref{t1.1} now follows by taking the infimum over $\{\delta_n\} \in \mb D$ and noting that $\{\delta_n^*(\,\cdot\,;\pi)\} \in \mathbb D$ by Lemma~\ref{l1}\ref{l1.1}.

Result~\ref{t1.2} follows from the fact that the matching forms $\delta_\infty^*$ of $\{\delta_n^*(\,\cdot\,;\pi)\}$ derived in Lemma \ref{l1}\ref{l1.1} do not depend on the prior $\pi \in \Pi$. 

Finally, if $\{\delta_n\} \in \mathbb D$ is asymptotically equivalent to $\{\delta_n^*(\,\cdot\,;\pi)\}$, then $\{\delta_n\}$ must also have matching form $\delta_\infty^*(Z)$. Hence, by similar arguments to the above, we have $\mc R(\{\delta_n'\};\mu_0) \geq \mc R(\{\delta_n\};\mu_0)$ for any $\{\delta_n'\} \in \mathbb D$. Taking the infimum over $\{\delta_n'\} \in \mathbb D$ yields result~\ref{t1.3}.
\end{proof}

\

\begin{proof}[Proof of Theorem \ref{t2}]
First note that asymptotic equivalence cannot fail if $|\underline{\mc D}_{\mu_0}| = 1$, since if $\underline{\mc D}_{\mu_0} = \{d\}$, say, then $\lim_{n \to \infty} P_{n,h}(\delta_n(X^n) = d) = 1$ must hold for any $\{\delta_n\} \in \mb D$.

Now suppose $|\underline{\mc D}_{\mu_0}| \geq 2$. Let $g_{d,\mu_0}[h] = \dot R_{d,\mu_0}[h] - \min_{i \in \underline{\mc D}_{\mu_0}} \dot R_{i,\mu_0}[h]$. By Lemma \ref{l1}\ref{l1.2}-\ref{l1.3}, it suffices to show
\[
 \int \Bigg( \sum_{d \in \underline{\mc D}_{\mu_0}} g_{d,\mu_0}[h] \mathbb P_h( \delta_\infty(Z,U) = d) \Bigg) d h 
 > \int \Bigg( \sum_{d \in \underline{\mc D}_{\mu_0}} g_{d,\mu_0}[h]  \mathbb P_h( \delta_\infty^*(Z) = d) \Bigg) d h,
\]
where $\delta_\infty(Z,U)$ is the matching form of $\{\delta_n\}$ and $\delta_\infty^*(Z)$ is the matching form of $\{\delta_n^*(\,\cdot\,;\pi)\}$ from Lemma~\ref{l1}\ref{l1.1}. Since both integrands are non-negative, by Tonelli's theorem we may restate this inequality as
\begin{multline*}
 \int_0^1  \int \sum_{d \in \underline{\mc D}_{\mu_0}} \mathbb I [ \delta_\infty(z,u) = d ]   \left( \int g_{d,\mu_0}[h] \, e^{-\frac{1}{2}(z-h)^T I_{\mu_0}(z-h)} \, d h  \right) d z \, du \\
 > \int_0^1 \int \sum_{d \in \underline{\mc D}_{\mu_0}} \mathbb I [ \delta_\infty^*(z) = d ] \left( \int g_{d,\mu_0}[h] \, e^{-\frac{1}{2}(z-h)^T I_{\mu_0}(z-h)} \, d h  \right) d z \, du.
\end{multline*}

As $\{\delta_n\}$ and $\{\delta_n^*(\,\cdot\,;\pi)\}$ are not asymptotically equivalent, we have
\[
 \lim_{n \to \infty} P_{n,h_0}(\delta_n(X^n) = d) \neq \lim_{n \to \infty} P_{n,h_0}(\delta_n^*(X^n;\pi) = d)
\]
for some $h_0 \in \mathbb R^K$ and some $d \in {\cal D}$. As $\{\delta_n\}$ and $\{\delta_n^*(\,\cdot\,;\pi)\}$ are both elements of $\mathbb D$, we can use Theorem~3.1 of \cite{Vandervaart1991} and restate this in terms of their matching forms $\delta_\infty(Z,U)$ and $\delta^*_\infty(Z)$:
\[
 \mathbb P_{h_0}(\delta_\infty(Z,U) = d) \neq \mathbb P_{h_0}(\delta_\infty^*(Z) = d).
\]
Hence, it must be the case that $\delta_\infty(z,u)$ and $\delta_\infty^*(z)$ disagree on a set $S$ of positive Lebesgue measure. As $\delta^*_\infty(z) \in \mc I(z)$ for almost every $z$, and $\mc I(z)$ is a singleton for almost every $z$ as there are no first-order ties, we must have that $\delta_\infty(z,u) \not \in \mathcal I(z)$ on a set $S'$ of positive Lebesgue measure. Now by definition of $\mc I(z)$, we also have that
\begin{multline*}
 \sum_{d \in \underline{\mc D}_{\mu_0}} \mathbb I [ \delta_\infty(z,u) = d ] \left( \int g_{d,\mu_0}[h] \, e^{-\frac{1}{2}(z-h)^T I_{\mu_0}(z-h)} \, d h  \right)  \\
 \geq \sum_{d \in \underline{\mc D}_{\mu_0}} \mathbb I [ \delta_\infty^*(z) = d ] \left( \int g_{d,\mu_0}[h] \, e^{-\frac{1}{2}(z-h)^T I_{\mu_0}(z-h)} \, d h  \right) ,
\end{multline*}
holds for almost every $(z,u)$, with strict inequality for $(z,u)$ in a set of positive measure. Integrating with respect to $(z,u)$ yields the strict inequality above.
\end{proof}

\

\begin{proof}[Proof of Corollary~\ref{cor:plug}]
Immediate from Theorem~\ref{t1} and the argument given in the text preceding the statement of this corollary.
\end{proof}

\

\begin{proof}[Proof of Proposition~\ref{prop:plug}]
Let $g_{d,\mu_0}[h] = \dot R_{d,\mu_0}[h] - \min_{i \in \underline{\mc D}_{\mu_0}} \dot R_{i,\mu_0}[h]$. Note $g_{d,\mu_0}[h] \geq 0$. By Lemma~\ref{l1}, we have
\[
 \mc R(\{\delta_n^*(\,\cdot\,;\pi)\};\mu_0) \leq \sum_{d \in \underline{\mc D}_{\mu_0}}  \int g_{d,\mu_0}[h] \, \mathbb P_h\left(   {\textstyle \argmin_{i \in \underline{\mc D}_{\mu_0}} } \dot R_{i,\mu_0}[Z] = d \right)  d h,
\]
where interchanging summation and integration is justified by Tonelli's theorem. Since $h \in \mb R$, we split the integral into an integral over $[0,\infty)$ and an integral over $(-\infty,0)$. Suppose $h \geq 0$. 
For each $d \in \ul{\mc D}_{\mu_0}$, note $g_{d,\mu_0}[h] = h g_{d,\mu_0}[1]$  by positive homogeneity. Suppose $g_{d,\mu_0}[1] > 0$ for some $d \in \ul{\mc D}_{\mu_0}$, else the integral is zero. Then $d = \argmin_{i \in \underline{\mc D}_{\mu_0}} \dot R_{i,\mu_0}[z]$ implies $g_{d,\mu_0}[z] = 0$. And since $g_{d,\mu_0}[1] > 0$ and $g_{d,\mu_0}[\,\cdot\,]$ is positively homogeneous, $g_{d,\mu_0}[z] = 0$ in turn implies $z \leq 0$. Hence,
\[
 \mathbb P_h( {\textstyle \argmin_{i \in \underline{\mc D}_{\mu_0}} } \dot R_{i,\mu_0}[Z] = d)  \leq \mathbb P_h( Z \leq 0) = \Phi(-h/\sigma),
\]
where $\sigma = {I_{\mu_0}^{-1/2}} > 0$, and $\Phi$ is the CDF of the $N(0, 1)$ distribution. Then
\[
 \int_0^\infty  g_{d,\mu_0}[h] \, \mathbb P_h\left(   {\textstyle \argmin_{i \in \underline{\mc D}_{\mu_0}} } \dot R_{i,\mu_0}[Z] = d \right)  d h 
 \leq g_d[1] \int_0^\infty h \, \Phi(-h/\sigma) \,  d h  < \infty.
\]
One may similarly show the integral over $(-\infty,0)$ is finite.
\end{proof}

\medskip

We first present a preliminary result before turning to the proof of Theorem~\ref{t3}.

\begin{lemma}\label{lem:sigma.optimal}
Suppose that Assumption~\ref{a1} and~\ref{a2} hold. Then $\{\delta_n\} \in \mb D$ is $\sigma$-optimal at $\mu_0$ if its matching decision is $\delta^*_\sigma(Z,U)$ with 
\[
 \delta^*_\sigma(z,u) \in \argmin_{d \in \ul{\mc D}_{\mu_0}} \E^*_\sigma \left[ \dot R_{d,\mu_0}[Z^*_\sigma + (I + (\sigma I_{\mu_0})^{-1})^{-1}z] \right]
\]
for almost every $(z,u) \in \mb R^K \times [0,1]$, where $\E^*_\sigma$ denotes expectation with respect to $Z^*_\sigma \sim N( 0, (I_{\mu_0} + \sigma^{-1} I)^{-1})$.
\end{lemma}

\begin{proof}[Proof of Lemma~\ref{lem:sigma.optimal}]
By Lemma~\ref{l1}\ref{l1.2}, for any $\{\delta_n\} \in \mb D$ we have
\[
 \mc R_\sigma(\{\delta_n\};\mu_0) 
 \propto \int \sum_{d \in \underline{\mc D}_{\mu_0}} g_{d,\mu_0}[h] \, \mathbb P_h( \delta_\infty(Z,U) = d) e^{-\frac{1}{2\sigma} \|h\|^2} \, d h ,
\]
where $g_{d,\mu_0}[h] = \dot R_{d,\mu_0}[h] - \min_{i \in \underline{\mc D}_{\mu_0}} \dot R_{i,\mu_0}[h]$ and $\delta_\infty(Z,U)$ is the matching decision of $\{\delta_n\}$. By Tonelli's theorem, we have 
\[
 \mc R_\sigma(\{\delta_n\};\mu_0) \propto \int_0^1 \int \sum_{d \in \underline{\mc D}_{\mu_0}} \mb I[\delta_\infty(z,u) = d] \left( \int  g_{d,\mu_0}[h] e^{-\frac{1}{2} (z - h)^T I_{\mu_0} (z - h) - \frac{1}{2 \sigma} \|h\|^2} \, d h \right) dz \, du .
\]
Therefore, $\mc R_\sigma(\{\delta_n\};\mu_0)$ is minimized when 
\[
 \delta_\infty(z,u) \in \argmin_{i \in \underline{\mc D}_{\mu_0}} \int \dot R_{i,\mu_0}[h] e^{-\frac{1}{2} (z - h)^T I_{\mu_0} (z - h) - \frac{1}{2 \sigma} \|h\|^2 }  \,  d h 
\]
for almost every $(z,u)$. The result follows by completing the square. 
\end{proof}

\

\begin{proof}[Proof of Theorem~\ref{t3}]
Lemma~\ref{l1}\ref{l1.1} implies 
\[
 \lim_{n \to \infty} P_{n,h} \left( \delta_n^*(X^n;\pi) = d \right) =
 \mathbb P_h ( \delta^*(Z) = d ),
\]
for all $d \in \mc D$ and $h \in \mb R^K$, where 
\[
 \delta^*(z) = \mathcal I(z) := {\textstyle \argmin_{i \in \underline{\mc D}_{\mu_0}} } \mathbb E^* \left[ \dot R_{i,\mu_0}[Z^* + z]  \right]
\]
for almost every $z$, since there are no first-order ties. Similarly, Lemma~\ref{lem:sigma.optimal} implies
\[
 \lim_{n \to \infty} P_{n,h} \left( \delta_{n,\sigma}^*(X^n) = d \right) =
 \mathbb P_h ( \delta^*_\sigma(Z,U) = d ),
\]
for all $d \in \mc D$ and $h \in \mb R^K$, where
\[
 \delta^*_\sigma(z,u) \in \mc I_\sigma(z) := {\textstyle \argmin_{i \in \underline{\mc D}_{\mu_0}} } \mathbb E^*_\sigma \left[ \dot R_{i,\mu_0}[Z^*_\sigma + (I + (\sigma I_{\mu_0})^{-1})^{-1}z]  \right]
\]
for almost every $(z,u) \in \mb R^K \times [0,1]$. 

For almost every $z$, we have by dominated convergence that 
\[
 \lim_{\sigma \to \infty} \mathbb E^*_\sigma \left[ \dot R_{i,\mu_0}[Z^*_\sigma + (I + (\sigma I_{\mu_0})^{-1})^{-1}z]  \right] = \E^* \left[ \dot R_{i,\mu_0}[Z^* + z]  \right].
\]
Hence, $\lim_{\sigma \to \infty} \mc I_\sigma(z) = \mc I(z)$ for almost every $z$, and so $\lim_{\sigma \to \infty} \delta^*_\sigma(z,u) = \delta^*(z)$ for almost every $(z,u)$. Hence, by dominated convergence and Theorem 3.1 of \cite{Vandervaart1991}, 
\begin{multline*}
 \lim_{n \to \infty} P_{n,h}(\delta_n^*(X^n;\pi) = d) = \mathbb P_h(\delta^*(Z) = d) \\
 = \lim_{\sigma \to \infty} \mathbb P_h(\delta^*_\sigma(Z,U) = d) = \lim_{\sigma \to \infty} \lim_{n \to \infty} P_{n,h} \left( \delta_{n,\sigma}^*(X^n) = d \right),
\end{multline*}
as required.
\end{proof}

\subsection{Proofs for Section~\ref{sec:semiparametric}}

Once we fix $(\mu_0,\eta_0)$, the least favorable model $\{P_{\beta(t)} : t \in \mc M_{(\mu_0,\eta_0)}\}$ is a parametric model. By Assumption~\ref{a2s}\ref{a2s.1}-\ref{a2s.3}  and Example~7.15 and Corollary~9.5 of \cite{vanderVaart1998}, the sequence of experiments $\{\mc E_n\}$, with $\mc E_n = (\mc X^n, \mc B_n, P_{n,h} : h \in \mathbb R^K)$ and $P_{n,h}$ denoting the distribution of $X^n$ under $P_{\beta(\mu_0 + h/\sqrt n)}$, converges to a limit experiment $\mc E = (\mb R^K, \mc B_K, N(h,I_{(\mu_0,\eta_0)}^{-1}): h \in \mb R^K)$. Because $\mc E$ is dominated, the asymptotic representation theorem of \cite{Vandervaart1991} implies that each $\{\delta_n\} \in \mb D$ is matched by a $\delta_\infty(Z,U)$ given by a measurable map $\delta_\infty : \mb R^K \times [0,1]\to \mc D$, the observation $Z \sim N(h,I_{(\mu_0,\eta_0)}^{-1})$, and an independent randomization $U \sim \mathrm{Uniform}[0,1]$, with the property that for each $h \in \mb R^K$,
\[
 \lim_{n \to \infty} P_{n,h}(\delta_n(X^n) = d) = \mathbb P_h (\delta_\infty(Z,U) = d), \quad d \in \mathcal D,
\]
where $\mathbb P_h$ is the product of the $N(h,I_{(\mu_0,\eta_0)}^{-1})$ and $U[0,1]$ distributions

Once these modifications are made, the proofs of Theorems~\ref{t1s} and~\ref{t2s} follow identical arguments to the proofs of Theorems~\ref{t1} and~\ref{t2}, using Lemma~\ref{lem:mistake_prob_quasi_bayes} in place of Lemma~\ref{lem:mistake_prob_bayes} and Proposition~\ref{prop:BVM.semiparametric} in place of Proposition~\ref{prop:BVM}.

\section{Asymptotic Distribution of the Posterior Mean of Directionally Differentiable Functions}
\label{appsec:BVM}

In this section we derive the asymptotic distribution of the posterior mean of directionally differentiable functions.  Let $f : \mathcal M \to \mathbb R^k$ be a directionally differentiable function of $\mu$. The first main result is Proposition~\ref{prop:BVM}, which presents the asymptotic distribution of the centered and scaled posterior mean
\[
 \bar f_n := \int \sqrt n (f(\mu) - f(\mu_0)) d \pi_n (\mu)
\]
along sequences of local parameters $P_{n,h}$. In Proposition~\ref{prop:BVM.semiparametric} we then extend this result to semiparametric models. Both results are of independent interest. The most closely related work is \cite{KitagawaOleaPayneVelez2020}, which characterizes the asymptotic behavior of the posterior distribution of directionally differentiable functions. We instead characterize the (frequentist) asymptotic distribution of its posterior mean. For both results, we assume the following:

\medskip
\begin{assumption}\label{af}
\begin{enumerate}[label={(\roman*)}, nosep]
\item\label{af.1} $f$ is integrable with respect to the prior $\pi$ for $\mu$; 
\item\label{af.2} $f$ is directionally differentiable at $\mu_0$. 
\end{enumerate}
\end{assumption}
\medskip

\subsection{Parametric Models}

We first consider parametric models as described in Section~\ref{sec:optimality}. We present some regularity conditions and preliminary lemmas before giving the main result.

Recall the notation from Section~\ref{sec:optimality}. Define $Z_{n,\mu_0} = \frac{1}{\sqrt n} \sum_{i=1}^n I_{\mu_0}^{-1} \dot \ell_{\mu_0}(X_i)$. Let $\tilde \pi_n$ denote the posterior induced by $\pi_n(\mu)$ under a change of variables from $\mu$ to $h = \sqrt n (\mu - \mu_0) - Z_{n,\mu_0}$. Let $P^*$ denote the distribution of $Z^* \sim N(0,I_{\mu_0}^{-1})$ and let $\|\cdot\|_{TV}$ denote total variation distance.

\begin{lemma}\label{lem:derivative}
Suppose that Assumptions~\ref{a2}\ref{a2.1}-\ref{a2.3} and \ref{a2.5} hold, Assumption~\ref{af}\ref{af.2} holds, and $\pi \in \Pi$. Then there exists a sequence $\{B_n\} \subset \mathbb R_+$ with $B_n \uparrow +\infty$, $B_n = o(\sqrt n)$, such that for all $n$ sufficiently large,
\[
 \sup_{h : \|h\| \leq 2B_n} \left\| \sqrt n \left( f(\mu_0 + h/\sqrt n) - f(\mu_0) \right) - \dot f_{\mu_0}[h] \right\| \leq \frac{1}{B_n},
\]
and
\[
 P_{n,\mu_0} \left( \left\| \tilde \pi_n - P^* \right\|_{TV} > \frac{1}{B_n^2} \right) \leq \frac{1}{B_n}.
\]
\end{lemma}

\begin{proof}[Proof of Lemma~\ref{lem:derivative}]
By Assumption~\ref{af}\ref{af.2} and Proposition~3.3 of \cite{Shapiro1990}, for each $a \in \mathbb N$ there exists $n_{a,f} \in \mathbb N$ such that $n \geq n_{a,f}$ implies
\[
 \sup_{h : \|h\| \leq 2a} \left\| \sqrt n \left( f(\mu_0 + h/\sqrt n) - f(\mu_0) \right) - \dot f_{\mu_0}[h] \right\| \leq \frac 1a. 
\]

Under Assumptions~\ref{a2}\ref{a2.1}-\ref{a2.3} and since $\pi \in \Pi$, we may invoke Theorem~10.1 of \cite{vanderVaart1998}, using Assumption~\ref{a2}\ref{a2.5} and Proposition~6.2 of \cite{ClarkeBarron} to deduce existence of tests with the desired properties. Further, we may use invariance of total variation under location changes to deduce that for all $a \in \mathbb N$ there exists $n_{a,\pi} \in \mathbb N$ such that $n \geq n_{a,\pi}$ implies
\[
 P_{n,\mu_0} \left(  \left\| \tilde \pi_n - P^* \right\|_{TV} > \frac{1}{a^2} \right) \leq \frac 1a.
\]
The result follows by setting $B_n = \sup\{a \in \mathbb N : n \geq (n_{a,f} \vee n_{a,\pi})\}$ for $n \geq (n_{1,f} \vee n_{1,\pi})$, then letting $B_n$ increase slower if necessary so that $B_n = o(\sqrt n)$.
\end{proof}

\medskip

Let $\overset{P_{n,\mu_0}}{\to}$ denote convergence in probability under $\{P_{n,\mu_0}\}$. Let $\mathbb E^*$ denote expectation with respect to $Z^* \sim N(0,I_{\mu_0}^{-1})$ independent of $Z_{n,\mu_0}$.

\begin{lemma}\label{lem:BVM.linearize}
Suppose Assumptions~\ref{a2} and~\ref{af} hold and $\pi \in \Pi$. Then
\[
 \left\| \bar f_n - \E^* \left[ \left.  \dot f_{\mu_0} \left[ Z^* + Z_{n,\mu_0} \right] \right| Z_{n,\mu_0}  \right] \right\|
 \overset{P_{n,\mu_0}}{\to} 0
\]
as $n \to \infty$.
\end{lemma}

\begin{proof}[Proof of Lemma~\ref{lem:BVM.linearize}]
Let $\{B_n\}$ be as in Lemma~\ref{lem:derivative}. By a change of variables, we have
\begin{multline*}
 \bar f_n
 = \int_{h : \|h\| \leq B_n} \sqrt n \left( f(\mu_0 + n^{-1/2}(Z_{n,\mu_0} + h)) - f(\mu_0) \right) d \tilde \pi_n(h) \\
 + \int_{\mu : \|\sqrt n(\mu - \mu_0) - Z_{n,\mu_0}\| > B_n} \sqrt n \left( f(\mu) - f(\mu_0) \right) d \pi_n(\mu) =: \mc J_{1,n} + \mc J_{2,n}.
\end{multline*}
Let $A_n$ denote the event $\|Z_{n,\mu_0}\| \leq B_n$ and $\left\| \tilde \pi_n - P^* \right\|_{TV} \leq B_n^{-2}$. Assumption~\ref{a2}\ref{a2.3} and Lemma~\ref{lem:derivative} imply $\lim_{n \to \infty} P_{n,\mu_0}(A_n^c) = 0$. Again by Lemma~\ref{lem:derivative}, on $A_n$ we have
\[
 \Bigg\| 
 \mc J_{1,n} -
 \int_{h : \|h\| \leq B_n} \dot f_{\mu_0} \left[ Z_{n,\mu_0} + h \right] d \tilde \pi_n(h) 
 \Bigg\| \leq \frac{1}{B_n} \to 0.
\]
Note that $\dot f_{\mu_0}[\,\cdot\,]$ is continuous under Assumption~\ref{af}\ref{af.2} \cite[Proposition~3.1]{Shapiro1990} and hence that 
\begin{equation} 
	C_f := \sup_{u:\|u\| = 1} \| \dot f_{\mu_0}[u]\| < \infty. \label{eq:ap.def.Cf}
\end{equation}
 Moreover, $\dot f_{\mu_0}[\,\cdot\,]$ is positively homogeneous of degree one \cite[p.~478]{Shapiro1990}. It therefore follows by Lemma~\ref{lem:derivative} that on $A_n$ we have
\begin{multline*}
  \left\| 
 \int_{h : \|h\| \leq B_n} \dot f_{\mu_0} \left[ Z_{n,\mu_0} + h \right] d \tilde \pi_n(h) -
 \int_{h : \|h\| \leq B_n} \dot f_{\mu_0} \left[ Z_{n,\mu_0} + h \right] d P^*(h)
 \right\| \\
 \leq 2 B_n C_f \| \tilde \pi_n - P^*\|_{TV} 
 \leq \frac{2 C_f}{B_n} \to 0.
\end{multline*}
Now, as $\|Z_{n,\mu_0}\| \leq B_n$ holds on $A_n$, we have by the triangle and Chebyshev inequalities that
\[
 \begin{aligned}
 \left\| \int_{h : \|h\| > B_n} \dot f_{\mu_0} \left[ Z_{n,\mu_0} + h \right] d P^*(h) \right\| 
 & \leq C_f \int_{h : \|h\| > B_n} \left( \|Z_{n,\mu_0}\| + \|h\| \right) d P^*(h) \\
 & \leq C_f \left( B_n P^*(\|Z^*\| > B_n) + \int_{h:\|h\| > B_n} \|h\| d P^*(h) \right) \\
 & \leq \frac{2 C_f \mathrm{tr}(I_{\mu_0}^{-1})}{B_n} \to 0
 \end{aligned}
\]
by Assumption~\ref{a2}\ref{a2.3}. Combining the preceding three displays, we see that 
\[
 \bigg\|  \mc J_{1,n}
 - \E^* \left[ \left.  \dot f_{\mu_0} \left[ Z^* + Z_{n,\mu_0} \right] \right| Z_{n,\mu_0} \right] \bigg\| \to 0 \quad \mbox{on $A_n$.}
\]
It follows that $\|\mc J_{1,n} - \E^* [  \dot f_{\mu_0} [ Z^* + Z_{n,\mu_0} ] | Z_{n,\mu_0} ] \| \overset{P_{n,\mu_0}}{\to} 0$ because $P_{n,\mu_0}(A_n^c) \to 0$.

It remains to show that $\|\mathcal J_{2,n}\|\overset{P_{n,\mu_0}}{\to} 0$. First note that by Assumptions~\ref{a2}\ref{a2.1}-\ref{a2.3} we may invoke Lemma~10.3 of \cite{vanderVaart1998}, using Assumption~\ref{a2}\ref{a2.5} and Proposition~6.2 of \cite{ClarkeBarron} to deduce existence of uniformly consistent tests as required in Theorem~10.1 of \cite{vanderVaart1998}. Hence, there exists a sequence of tests $\{\phi_n\}$ and a constant $c > 0$ such that for every sufficiently large $n$ and every $\mu \in \mathcal M$ with $\|\mu - \mu_0\| \geq B_n/(2\sqrt n)$, we have $\E_{n,\mu_0}[\phi_n(X^n)] \to 0$ and $\E_{n,\mu}[1-\phi_n(X^n)] \leq e^{-cn(\|\mu - \mu_0\|^2 \wedge 1)}$. Without loss of generality we may take the tests to be nonrandomized \cite[p.~148]{vanderVaart1998}. Then
\[
 \mathcal J_{2,n} = \phi_n(X^n) \mathcal J_{2,n} + (1 - \phi_n(X^n)) \mathcal J_{2,n},
\]
where $P_{n,\mu_0}( \left\| \phi_n(X^n) \mathcal J_{2,n} \right\| > 0) \leq P_{n,\mu_0}(\phi_n(X^n) = 1) \to 0$. The remaining term is
\[
  (1 - \phi_n(X^n)) \frac{\int_{\mu  : B_n < \|\sqrt n (\mu - \mu_0) - Z_{n,\mu_0}\|}\sqrt n \left( f(\mu) - f(\mu_0) \right) \prod_{i=1}^n p_\mu(X_i)/p_{\mu_0}(X_i) d \pi(\mu)}{\mathcal J_n},
\]
where $\mathcal J_n = \int  \prod_{i=1}^n p_\mu(X_i) / p_{\mu_0}(X_i) d \pi(\mu)$. By Lemma~6.26 of \cite{GhosalVanderVaart} and Lemma~\ref{lem:prior}, there exists $C > 0$ such that for any $C_n > 0$, the inequality
\[
 P_{n,\mu_0}\left( \mathcal J_n < C n^{-K/2} e^{-2C_n} \right) \leq \frac{1}{C_n}
\]
holds for all $n$ sufficiently large. Below we shall specify a sequence $\{C_n\} \subset \mathbb R_+$ with $C_n \uparrow \infty$. Letting $A'_n$ denote the event upon which $\|Z_{n,\mu_0}\| \leq \frac 12 B_n$ and $\mathcal J_n \geq Cn^{-K/2} e^{-2C_n}$ both hold, we have by the above display and  Assumption~\ref{a2}\ref{a2.3} that $P_{n,\mu_0}(A_n^{\prime c}) \to 0$. It follows by the union bound, Markov's inequality, and Fubini's theorem that for any $\epsilon > 0$,
\begin{multline*}
 P_{n,\mu_0} \left( \left\| (1 - \phi_n(X^n)) \mathcal J_{2,n} \right\| > \epsilon \right) \\
 \leq \frac{\int_{\mu : \|\sqrt n (\mu - \mu_0)\| > \frac 12 B_n} \left\| \sqrt n \left( f(\mu) - f(\mu_0) \right) \right\| e^{-cn(\|\mu - \mu_0\|^2 \wedge 1)} d \pi(\mu)}{\epsilon Cn^{-K/2} e^{-2C_n}} + P_{n,\mu_0}(A_n^{\prime c}).
\end{multline*}
Split the numerator into integrals over $N_{1,n} = \{\mu  : \frac 12 B_n < \|\sqrt n (\mu - \mu_0) \| \leq \epsilon' \sqrt n\}$ and $N_{2,n} = \{\mu : \|\mu - \mu_0\| > \epsilon' \}$ for $\epsilon' \in (0,1)$ sufficiently small that $\pi(\mu) \leq C_2$ on $\{\mu : \|\mu - \mu_0\| \leq \epsilon'\}$. 
Finally, Proposition~3.3 of \cite{Shapiro1990} implies $\sup_{h : \|h\| \leq 1} \|f(\mu_0 + th) - f(\mu_0) - \dot f_{\mu_0}[th]\| = o(t)$ as $t \downarrow 0$. Hence, we may choose $\bar t > 0$ such that $\|f(\mu_0 + th) - f(\mu_0) - \dot f_{\mu_0}[th]\| \leq t$ holds for all $t \leq \bar t$ and all $\|h\| \leq 1$. Let $\epsilon' = \min\{\epsilon', \bar t\}$. It then follows for any $\mu \in N_{1,n}$ that, by setting $t = \|\mu - \mu_0\|$ and $h = \frac{\mu - \mu_0}{\|\mu - \mu_0\|}$, we have $\|f(\mu) - f(\mu_0)\| \leq \|\dot f_{\mu_0}[\mu - \mu_0]\| + \|\mu - \mu_0\|$. Hence, by (\ref{eq:ap.def.Cf}), we have that $\|f(\mu) - f(\mu_0)\| \leq (1+C_f)\|\mu - \mu_0\|$ holds on $N_{1,n}$.
Then
\begin{align*}
 & \int_{N_{1,n}} \| \sqrt n \left( f(\mu) - f(\mu_0) \right) \| e^{-cn(\|\mu - \mu_0\|^2 \wedge 1)} d \pi(\mu) \\
 & \quad \quad \leq C_2 (1 + C_f) \int_{N_{1,n}} \|\sqrt n(\mu - \mu_0)\| e^{-c\|\sqrt n (\mu - \mu_0)\|^2} d \mu \\
 & \quad \quad \leq C_2 (1 + C_f) n^{-K/2} \int_{h : \|h\| > \frac 12 B_n} \|h\| e^{-c\|h\|^2} d h ,
\end{align*}
where the final line is by a change of variables from $\mu$ to $h = \sqrt n (\mu - \mu_0)$. The final integral vanishes as $n \to \infty$ by dominated convergence. Further, by Assumption~\ref{af}\ref{af.1}, we have
\[
 \int_{N_{2,n}} \| \sqrt n \left( f(\mu) - f(\mu_0) \right) \| e^{-cn(\|\mu - \mu_0\|^2 \wedge 1)} d \pi(\mu)
 \leq 
 \sqrt n e^{-nc (\epsilon')^2} \int \|f(\mu) - f(\mu_0)\| d \pi(\mu).
\]
Choosing $\{C_n\} \subset \mathbb R_+$ with $C_n \uparrow \infty$ sufficiently slowly that
\[
 e^{2 C_n} \int_{h : \|h\| > \frac 12 B_n} \|h\| e^{-c\|h\|^2} d h \to 0,
\quad \quad
\mbox{and}
\quad \quad
 n^{(K+1)/2} e^{2C_n - nc(\epsilon')^2} \to 0,
\]
it follows from the preceding four displays that $P_{n,\mu_0} \left( \left\| (1 - \phi_n(X^n)) \mathcal J_{2,n} \right\| > \epsilon \right) \to 0$.
\end{proof}

\medskip

For the next result, recall that $Z \sim N(h,I_{\mu_0}^{-1})$, $\mathbb E^*$ denotes expectation with respect to $Z^* \sim N(0,I_{\mu_0}^{-1})$ independent of $Z$, and $\overset{P_{n,h}}{\rightsquigarrow}$ denotes convergence in distribution along $\{P_{n,h}\}$.

\begin{proposition}\label{prop:BVM}
Suppose that Assumptions~\ref{a2} and~\ref{af} hold and $\pi \in \Pi$. Then for every $h \in \mathbb R^K$,
\[
 \bar f_n \overset{P_{n,h}}{\rightsquigarrow} \E^* \left[ \left. \dot f_{\mu_0} \left[ Z^* + Z \right] \right| Z \right].
\]
\end{proposition}

\begin{proof}[Proof of Proposition~\ref{prop:BVM}]
Assumption~\ref{a2}\ref{a2.1}-\ref{a2.3} imply $\mathcal P = \{P_\mu : \mu \in \mathcal M\}$ is locally asymptotically normal \cite[Example~7.15]{vanderVaart1998} and so $\{P_{n,\mu_0}\}$ and $\{P_{n,h}\}$ are mutually contiguous \cite[Example~6.5]{vanderVaart1998}. Hence, with $\overset{P_{n,h}}{\to}$ denoting convergence in probability under $\{P_{n,h}\}$, we have by Lemma~\ref{lem:BVM.linearize} that 
\[
 \left\| \bar f_n - \E^* \left[ \left. \dot f_{\mu_0} \left[ Z^* + Z_{n,\mu_0} \right] \right| Z_{n,\mu_0} \right] \right\|
 \overset{P_{n,h}}{\to} 0.
\]
Moreover, under Assumption~\ref{a2}\ref{a2.1}-\ref{a2.3} we have by Le Cam's third lemma \cite[Example~6.7]{vanderVaart1998} that 
\[
 Z_{n,\mu_0} \overset{P_{n,h}}{\rightsquigarrow} Z \sim  N(h,I_{\mu_0}^{-1}).
\]
The result follows by the continuous mapping theorem, noting  $z \mapsto \E^* [ \dot f_{\mu_0} [ Z^* + z ] ] $ is everywhere continuous by Assumption~\ref{af}\ref{af.2} and Proposition~3.1 of \cite{Shapiro1990}.
\end{proof}

\subsection{Semiparametric Models}

We now extend Proposition~\ref{prop:BVM} to semiparametric models. In this case,
\[
 \bar f_n = \int \sqrt n (f(\mu) - f(\mu_0)) d \pi_n (\mu)
\]
with $\pi_n$ denoting the quasi-posterior in (\ref{eq:quasi_posterior}) based on a Gaussian quasi-likelihood. Before giving the main result we first state and prove several lemmas. The first is a Bernstein--von Mises result for the quasi-posterior. Let $Z_{n,(\mu_0,\eta_0)} = \sqrt n(\hat \mu - \mu_0)$. 
Let $\tilde \pi_{n}(h) \propto e^{-\frac 12 h^T \hat I h} \pi(\mu_0 + n^{-1/2}(h + Z_{n,(\mu_0,\eta_0)})$ denote the quasi-posterior for $h$ under a change of variables from $\mu$ to $h = \sqrt n (\mu - \mu_0) - Z_{n,(\mu_0,\eta_0)}$. Let $P^*$ denote the distribution of $Z^* \sim N(0,I_{(\mu_0,\eta_0)}^{-1})$.

\begin{lemma}\label{lem:BVM.semiparametric.prelim}
Suppose that Assumption~\ref{a2s}\ref{a2s.1},\ref{a2s.3},\ref{a2s.5} holds and $\pi \in \Pi$. Then
\[
 \| \tilde \pi_n - P^* \|_{TV} \overset{P_{n,(\mu_0,\eta_0)}}{\to} 0.
\]
\end{lemma}

\begin{proof}[Proof of Lemma~\ref{lem:BVM.semiparametric.prelim}]
Fix any Borel set $B \subset \mb R^K$. We have
\[
 \tilde \pi_n(B) = \frac{\int_{h : \mu_0 + n^{-1/2}(h + Z_{n,(\mu_0,\eta_0)}) \in \mc M} \mb I[h \in B] e^{-\frac 12 h^T \hat I h} \pi(\mu_0 + n^{-1/2} (h + Z_{n,(\mu_0,\eta_0)}) dh}{\int_{h : \mu_0 + n^{-1/2}(h + Z_{n,(\mu_0,\eta_0)}) \in \mc M} e^{-\frac 12 h^T \hat I h} \pi(\mu_0 + n^{-1/2} (h + Z_{n,(\mu_0,\eta_0)}) dh}.
\]
Let $\{M_n\} \subset \mb R_+$ be a sequence with $M_n \uparrow +\infty$, $M_n = o(n^{1/2})$. Let $A_n$ denote the event upon which $\|Z_{n,(\mu_0,\eta_0)}\| \leq M_n$ and $c \leq \hat \lambda_{\min},\hat \lambda_{\max} \leq c^{-1}$ hold, where $\hat \lambda_{\min}$ and $\hat \lambda_{\max}$ are the smallest and largest eigenvalues of $\hat I$, and $c \in (0,1)$ is chosen such that the minimum and maximum eigenvalues of $I_{(\mu_0,\eta_0)}$ are contained in $(c,c^{-1})$ (such a $c$ exists by Assumption~\ref{a2s}\ref{a2s.3}). We have $P_{n,(\mu_0,\eta_0)}(A_n) \to 1$ by Assumptions \ref{a2s}\ref{a2s.3} and~\ref{a2s}\ref{a2s.5}. Moreover, by Assumption~\ref{a2s}\ref{a2s.1},
\[
 \{\mu_0 + n^{-1/2}(h + Z_{n,(\mu_0,\eta_0)}) : \|h\| \leq M_n \} \subset \{\mu_0 + n^{-1/2}h : \|h\| \leq 2 M_n \} \subset \mc M
\]
holds on $A_n$ for all $n$ sufficiently large. It follows that for $n$ sufficiently large, we have
\[
 \sup_{\|h\| \leq M_n} | \pi(\mu_0 + n^{-1/2} (h + Z_{n,(\mu_0,\eta_0)})) - \pi(\mu_0)| \leq \sup_{\|h\| \leq 2M_n} | \pi(\mu_0 + n^{-1/2} h) - \pi(\mu_0)| =: \varepsilon_n \pi(\mu_0)
\]
on $A_n$, where $\varepsilon_n \downarrow 0$ because $\pi \in \Pi$. The density $\pi(\mu)$ is uniformly bounded above by some $\ol \pi < +\infty$ because $\pi \in \Pi$. Hence, on $A_n$, for all $n$ sufficiently large (independent of $B$),
\[
 \begin{aligned}
 \tilde \pi_n(B)
 & \leq \frac{\int_{h : \mu_0 + n^{-1/2}(h + Z_{n,(\mu_0,\eta_0)}) \in \mc M} \mb I[h \in B] e^{-\frac 12 h^T \hat I h} \pi(\mu_0 + n^{-1/2} (h + Z_{n,(\mu_0,\eta_0)}) dh}{(1-\varepsilon_n) \pi(\mu_0)\int_{h : \|h\| \leq M_n} e^{-\frac 12 h^T \hat I h}  dh} \\
 & \leq \frac{(1+\varepsilon_n)\int_{h : \|h\| \leq M_n} \mb I[h \in B] e^{-\frac 12 h^T \hat I h} \ dh}{(1-\varepsilon_n) \int_{h : \|h\| \leq M_n} e^{-\frac 12 h^T \hat I h}  dh} + \frac{ \ol \pi \int_{h : \|h\| > M_n} e^{-\frac 12 h^T \hat I h} dh}{(1-\varepsilon_n) \pi(\mu_0) \int_{h : \|h\| \leq M_n} e^{-\frac 12 h^T \hat I h}  dh} \\
 & \leq \frac{(1+\varepsilon_n)\int \mb I[h \in B] e^{-\frac 12 h^T \hat I h} \ dh}{(1-\varepsilon_n) \int_{h : \|h\| \leq M_n} e^{-\frac 12 h^T \hat I h}  dh} + \frac{ \ol \pi \int_{h : \|h\| > M_n} e^{-\frac c2 \|h\|^2} dh}{(1-\varepsilon_n) \pi(\mu_0) \int_{h : \|h\| \leq M_n} e^{-\frac{1}{2c} \|h\|^2}  dh},
 \end{aligned}
\]
where the second term on the right-hand side, say $t_{1,n}$, converges to zero by dominated convergence. 
Moreover, 
\[
 \delta_{1,n} := \frac{\int e^{-\frac 12 h^T \hat I h}  dh}{\int_{h : \|h\| \leq M_n} e^{-\frac 12 h^T \hat I h}  dh} - 1 
 = \frac{\int_{h : \|h\| > M_n} e^{-\frac 12 h^T \hat I h}  dh}{\int_{h : \|h\| \leq M_n} e^{-\frac 12 h^T \hat I h}  dh}
 \leq \frac{\int_{h : \|h\| > M_n} e^{-\frac c2 \|h\|^2}  dh}{\int_{h : \|h\| \leq M_n} e^{-\frac{1}{2c} \|h\|^2}  dh} \to 0,
\]
on $A_n$, again by dominated convergence. 
Hence, 
\begin{multline} \label{eq:BVM.semiparametric.prelim.1}
 \tilde \pi_n(B) \leq \frac{(1+\varepsilon_n)}{(1-\varepsilon_n)}  \frac{\int e^{-\frac 12 h^T \hat I h}  dh}{\int_{h : \|h\| \leq M_n} e^{-\frac 12 h^T \hat I h}  dh} \frac{\int \mathbb I[h \in B] e^{-\frac 12 h^T \hat I h} \ dh}{\int e^{-\frac 12 h^T \hat I h}  dh} + t_{1,n} \\
 \leq \frac{(1+\varepsilon_n)(1+\delta_{1,n})}{1-\varepsilon_n} \hat P(B) + t_{1,n},
\end{multline}
holds on $A_n$ for $n$ sufficiently large (independent of $B$), where $\hat P$ denotes the distribution of $Z^* \sim N(0, \hat I^{-1})$, and $\varepsilon_n$, $\delta_{1,n}$ and $t_{1,n}$ are all independent of $B$ and converge (either deterministically or in probability) to zero.

Similarly, for all $n$ sufficiently large (independently of $B$), on $A_n$, we have
\[
 \begin{aligned}
 \tilde \pi_n(B)
 & \geq \frac{\int_{h : \|h\| \leq M_n} \mb I[h \in B] e^{-\frac 12 h^T \hat I h} \pi(\mu_0 + n^{-1/2} (h + Z_{n,(\mu_0,\eta_0)}) dh}{\int_{h : \mu_0 + n^{-1/2}(h + Z_{n,(\mu_0,\eta_0)}) \in \mc M} e^{-\frac 12 h^T \hat I h} \pi(\mu_0 + n^{-1/2} (h + Z_{n,(\mu_0,\eta_0)}) dh} \\
 & \geq \frac{(1-\varepsilon_n) \pi(\mu_0) \int_{h : \|h\| \leq M_n} \mb I[h \in B] e^{-\frac 12 h^T \hat I h}  dh}{\int_{h : \mu_0 + n^{-1/2}(h + Z_{n,(\mu_0,\eta_0)}) \in \mc M} e^{-\frac 12 h^T \hat I h} \pi(\mu_0 + n^{-1/2} (h + Z_{n,(\mu_0,\eta_0)}) dh} \\
 & \geq \frac{(1-\varepsilon_n) \pi(\mu_0) \int_{h : \|h\| \leq M_n} \mb I[h \in B] e^{-\frac 12 h^T \hat I h}  dh}{(1+\delta_{2,n})\int_{h : \|h\| \leq M_n} e^{-\frac 12 h^T \hat I h} \pi(\mu_0 + n^{-1/2} (h + Z_{n,(\mu_0,\eta_0)}) dh} ,
 \end{aligned}
\]
where
\begin{multline*}
 \delta_{2,n} := \left| \frac{\int_{h : \mu_0 + n^{-1/2}(h + Z_{n,(\mu_0,\eta_0)}) \in \mc M} e^{-\frac 12 h^T \hat I h} \pi(\mu_0 + n^{-1/2} (h + Z_{n,(\mu_0,\eta_0)}) dh}{\int_{h : \|h\| \leq M_n} e^{- \frac 12 h^T \hat I h} \pi(\mu_0 + n^{-1/2} (h + Z_{n,(\mu_0,\eta_0)}) dh} - 1 \right| \\
 \leq \frac{ \ol \pi \int_{h : \|h\| > M_n} e^{-\frac 12 h^T \hat I h} d h }{(1-\varepsilon_n) \pi(\mu_0) \int_{h : \|h\| \leq M_n} e^{-\frac 12 h^T \hat I h} dh} 
 \leq \frac{ \ol \pi \int_{h : \|h\| > M_n} e^{-\frac c2 \|h\|^2} d h }{(1-\varepsilon_n) \pi(\mu_0) \int_{h : \|h\| \leq M_n} e^{-\frac{1}{2c} \|h\|^2} dh} \to 0.
\end{multline*}
Hence, on $A_n$, for all $n$ sufficiently large (independent of $B$),
\[
 \begin{aligned}
 \tilde \pi_n(B)
 & \geq \frac{(1-\varepsilon_n)}{(1+\delta_{2,n})(1+\varepsilon_n) }  \frac{\int e^{-\frac 12 h^T \hat I h} dh}{\int_{h : \|h\| \leq M_n} e^{-\frac 12 h^T \hat I h} dh} \frac{ \int_{h : \|h\| \leq M_n} \mb I[h \in B] e^{-\frac 12 h^T \hat I h}  dh}{\int e^{-\frac 12 h^T \hat I h} dh}\\
 & \geq \frac{(1-\varepsilon_n)(1-\delta_{1,n})}{(1+\delta_{2,n})(1+\varepsilon_n)} \frac{ \int_{h : \|h\| \leq M_n} \mb I[h \in B] e^{-\frac 12 h^T \hat I h}  dh}{ \int e^{-\frac 12 h^T \hat I h} dh} \\
 & \geq \frac{(1-\varepsilon_n)(1-\delta_{1,n})}{(1+\delta_{2,n})(1+\varepsilon_n)} \hat P(B) - \frac{(1-\varepsilon_n)(1-\delta_{1,n})}{(1+\delta_{2,n})(1+\varepsilon_n)} \frac{ \int_{h : \|h\| > M_n} e^{-\frac c2 \|h\|^2}  dh}{ \int e^{-\frac{1}{2c} \|h\|^2} dh},
 \end{aligned}
\]
where the second term on the right-hand side, say $t_{2,n}$, converges to zero by dominated convergence. Hence,
\begin{equation} \label{eq:BVM.semiparametric.prelim.2}
 \tilde \pi_n(B) \geq \frac{(1-\varepsilon_n)(1-\delta_{1,n})}{(1+\delta_{2,n})(1+\varepsilon_n)} \hat P(B) - t_{2,n}
\end{equation}
holds on $A_n$ for $n$ sufficiently large (independent of $B$), where $\varepsilon_n$, $\delta_{1,n}$, $\delta_{2,n}$, and $t_{2,n}$ are all independent of $B$ and converge (either deterministically or in probability) to zero. 

Combining (\ref{eq:BVM.semiparametric.prelim.1}) and (\ref{eq:BVM.semiparametric.prelim.2}) and using the fact that $P_{n,(\mu_0,\eta_0)}(A_n) \to 1$, we may deduce
\[
 \| \tilde \pi_n - \hat P\|_{TV} \equiv \sup_B \left| \tilde \pi_n(B) - \hat P(B)\right| \overset{P_{n,(\mu_0,\eta_0)}}{\to} 0,
\]
where the supremum is over all Borel sets. Finally, it is easy to deduce (e.g., by using Pinsker's inequality) that the total variation distance between the $N(0,\Sigma_1)$ and $N(0,\Sigma_2)$ distributions converges to zero as $\Sigma_1 \to \Sigma_2$ for $\Sigma_2$ positive definite. It therefore follows by Assumptions~\ref{a2s}\ref{a2s.3} and~\ref{a2s}\ref{a2s.5.b} that $\| \hat P - P^*\|_{TV}  \overset{P_{n,(\mu_0,\eta_0)}}{\to} 0$.
\end{proof}

\

\begin{lemma}\label{lem:derivative.semiparametric}
Suppose that Assumptions~\ref{a2s}\ref{a2s.1},\ref{a2s.3},\ref{a2s.5} and~\ref{af}\ref{af.2} hold and $\pi \in \Pi$. Then there exists a sequence $\{B_n\} \subset \mathbb R_+$ with $B_n \uparrow +\infty$, $B_n = o(\sqrt n)$, such that for all $n$ sufficiently large,
\[
 \sup_{h : \|h\| \leq 2B_n} \left\| \sqrt n \left( f(\mu_0 + h/\sqrt n) - f(\mu_0) \right) - \dot f_{\mu_0}[h] \right\| \leq \frac{1}{B_n},
\]
and
\[
 P_{n,(\mu_0,\eta_0)} \left( \left\| \tilde \pi_n - P^* \right\|_{TV} > \frac{1}{B_n^2} \right) \leq \frac{1}{B_n}.
\]
\end{lemma}

\begin{proof}[Proof of Lemma~\ref{lem:derivative.semiparametric}]
The result follows by similar arguments to the proof of Lemma~\ref{lem:derivative}, using Lemma~\ref{lem:BVM.semiparametric.prelim} in place of the (parametric) Bernstein--von Mises theorem.
\end{proof}

\

\noindent
For the next result we invoke Assumption~\ref{a2s}\ref{a2s.5.a}, but we only require $\|\sqrt n(\hat \mu - \mu_0)\|$ to be bounded in probability under $\{P_{n,(\mu_0,\eta_0)}\}$. Let $\mathbb E^*$ denote expectation with respect to $Z^* \sim N(0,I_{(\mu_0,\eta_0)}^{-1})$ independent of $Z_{n,(\mu_0,\eta_0)}$.

\begin{lemma}\label{lem:BVM.linearize.semiparametric}
Suppose Assumptions~\ref{a2s}\ref{a2s.1},\ref{a2s.3},\ref{a2s.5} and~\ref{af} hold and $\pi \in \Pi$. Then
\[
 \left\| \bar f_n - \E^* \left[ \left.  \dot f_{\mu_0} \left[ Z^* + Z_{n,(\mu_0,\eta_0)} \right] \right| Z_{n,(\mu_0,\eta_0)}  \right] \right\|
 \overset{P_{n,(\mu_0,\eta_0)}}{\to} 0
\]
as $n \to \infty$.
\end{lemma}

\begin{proof}[Proof of Lemma~\ref{lem:BVM.linearize.semiparametric}]
Let $\{B_n\}$ be as in Lemma~\ref{lem:derivative.semiparametric}. By a change of variables, we have
\begin{multline*}
 \bar f_n
 = \int_{h : \|h\| \leq B_n} \sqrt n \left( f(\mu_0 + n^{-1/2}(Z_{n,(\mu_0,\eta_0)} + h)) - f(\mu_0) \right) d \tilde \pi_n(h) \\
 + \int_{\mu : \|\sqrt n(\mu - \hat \mu)\| > B_n} \sqrt n \left( f(\mu) - f(\mu_0) \right) d \pi_n(\mu) =: \mc J_{1,n} + \mc J_{2,n}.
\end{multline*}
Let $A_n$ denote the event $\|Z_{n,(\mu_0,\eta_0)}\| \leq B_n$, $\left\| \tilde \pi_n - P^* \right\|_{TV} \leq B_n^{-2}$, and $c \leq \hat \lambda_{\min}$, $\hat \lambda_{\max} \leq c^{-1}$ for some $c \in (0,1)$. Assumptions~\ref{a2s}\ref{a2s.3} and~\ref{a2s}\ref{a2s.5} and Lemma~\ref{lem:derivative.semiparametric} imply $\lim_{n \to \infty} P_{n,(\mu_0,\eta_0)}(A_n^c) = 0$. Again by Lemma~\ref{lem:derivative.semiparametric}, on $A_n$ we have
\[
 \Bigg\| 
 \mc J_{1,n} -
 \int_{h : \|h\| \leq B_n} \dot f_{\mu_0} \left[ Z_{n,(\mu_0,\eta_0)} + h \right] d \tilde \pi_n(h) 
 \Bigg\| \leq \frac{1}{B_n} \to 0,
\]
where, as in the proof of Lemma~\ref{lem:BVM.linearize}, we may deduce that
\[
 \left\| 
 \int_{h : \|h\| \leq B_n} \dot f_{\mu_0} \left[ Z_{n,(\mu_0,\eta_0)} + h \right] d \tilde \pi_n(h) -
 \int_{h : \|h\| \leq B_n} \dot f_{\mu_0} \left[ Z_{n,(\mu_0,\eta_0)} + h \right] d P^*(h)
 \right\| 
 \leq \frac{2 C_f}{B_n} \to 0,
\]
on $A_n$, and
\[
 \begin{aligned}
 \left\| \int_{h : \|h\| > B_n} \dot f_{\mu_0} \left[ Z_{n,(\mu_0,\eta_0)} + h \right] d P^*(h) \right\| 
 \leq \frac{2 C_f \mathrm{tr}(I_{(\mu_0,\eta_0)}^{-1})}{B_n} \to 0,
 \end{aligned}
\]
on $A_n$, where the latter convergence holds in view of Assumption~\ref{a2s}\ref{a2s.3}. It follows that 
\[
 \|\mc J_{1,n} - \E^* [  \dot f_{\mu_0} [ Z^* + Z_{n,(\mu_0,\eta_0)} ] | Z_{n,(\mu_0,\eta_0)} ] \| \overset{P_{n,(\mu_0,\eta_0)}}{\to} 0.
\]

It remains to show that $\|\mathcal J_{2,n}\|\overset{P_{n,(\mu_0,\eta_0)}}{\to} 0$. To this end, write 
\begin{multline*}
 \mc J_{2,n} = \int_{\mu : B_n < \|\sqrt n(\mu - \hat \mu)\| \leq \epsilon \sqrt n} \sqrt n \left( f(\mu) - f(\mu_0) \right) d \pi_n(\mu) \\
 + \int_{\mu : \|\sqrt n(\mu - \hat \mu)\| > \epsilon \sqrt n} \sqrt n \left( f(\mu) - f(\mu_0) \right) d \pi_n(\mu) =: \mc J_{2,n,a} + \mc J_{2,n,b},
\end{multline*}
where, as in the proof of Lemma~\ref{lem:BVM.linearize}, we choose $\epsilon > 0$ sufficiently small that $\|f(\mu) - f(\mu_0)\| \leq (1 + C_f) \|\mu - \mu_0\|$ holds for $\mu$ with $\|\mu - \mu_0\| \leq 2\epsilon$. 
Then on $A_n$, we have
\[
 \|\mu - \mu_0\| \leq \| \mu - \hat \mu\| + \| \hat \mu - \mu_0\|  \leq \| \mu - \hat \mu\| + \frac{B_n}{\sqrt n},
\]
so $\{ \mu : \|\mu - \hat \mu\| \leq \epsilon \} \subseteq \{\mu : \|\mu - \mu_0\| \leq 2 \epsilon\}$ for all $n$ sufficiently large, in which case
\[
 \|\mc I_{2,n,a}\| \leq (1 + C_f) \Bigg( \int_{\mu : B_n < \|\sqrt n (\mu - \hat \mu)\| \leq \epsilon \sqrt n } \|\sqrt n (\mu - \hat \mu)\| d \pi_n(\mu) + B_n \pi_n (\{\mu : B_n < \sqrt n (\mu - \hat \mu) \}) \Bigg) .
\]
Taking $\epsilon$ smaller if necessary, we have $\{\mu : \|\mu - \mu_0\| \leq 2 \epsilon\} \subset \mc M$ in view of Assumption~\ref{a2s}\ref{a2s.1}. Then by dominated convergence and the fact that $c \leq \hat \lambda_{\min}$, $\hat \lambda_{\max} \leq c^{-1}$ holds on $A_n$, we have
\[
 \int_{\mu : B_n < \|\sqrt n (\mu - \hat \mu)\| \leq \epsilon \sqrt n } \|\sqrt n (\mu - \hat \mu)\| d \pi_n(\mu) 
 \leq 
 \frac{\int_{h :\|h \| > B_n } \|h\| e^{-\frac{c}{2}\|h\|^2} \pi(\mu_0 + n^{-1/2} (h + Z_{n,(\mu_0,\eta_0)})  dh}{\int e^{-\frac{1}{2c} \|h\|^2} \pi(\mu_0 + n^{-1/2} (h + Z_{n,(\mu_0,\eta_0)})  dh} 
\]
and 
\[
 B_n \pi_n (\{\mu : B_n < \sqrt n (\mu - \hat \mu) \}) 
 \leq B_n \frac{\int_{h :\|h \| > B_n }  e^{-\frac{c}{2}\|h\|^2} \pi(\mu_0 + n^{-1/2} (h + Z_{n,(\mu_0,\eta_0)})  dh}{\int e^{-\frac{1}{2c} \|h\|^2} \pi(\mu_0 + n^{-1/2} (h + Z_{n,(\mu_0,\eta_0)})  dh} .
\]
It then follows by similar arguments to the proof of Lemma~\ref{lem:BVM.semiparametric.prelim} that $\|\mc J_{2,n,a}\|\overset{P_{n,(\mu_0,\eta_0)}}{\to} 0$.

Finally, for $\mc J_{2,n,b}$, on $A_n$ we have
\[
 \|\mc J_{2,n,b}\| \leq \frac{\sqrt n e^{-\frac{cn\epsilon^2}{2}} \int f(\mu) - f(\mu_0) \pi(\mu) d \mu}{ n^{-K/2} \int e^{-\frac{1}{2c} \|h\|^2} \pi(\mu_0 + n^{-1/2} (h + Z_{n,(\mu_0,\eta_0)})  dh}.
\]
The integral in the numerator is finite by Assumption~\ref{af}\ref{af.1} and the integral in the denominator may be shown to be bounded below by a positive constant on $A_n$ by similar arguments to the proof of Lemma~\ref{lem:BVM.semiparametric.prelim}, from which it follows that $\|\mc J_{2,n,b}\|\overset{P_{n,(\mu_0,\eta_0)}}{\to} 0$.
\end{proof}

\medskip

For the next result, let $Z \sim N(h,I_{\mu_0}^{-1})$ and let $\mathbb E^*$ denote expectation with respect to $Z^* \sim N(0,I_{\mu_0}^{-1})$ independent of $Z$.

\begin{proposition}\label{prop:BVM.semiparametric}
Suppose that Assumptions~\ref{a2s} and~\ref{af} hold and $\pi \in \Pi$. Then for every $h \in \mathbb R^K$,
\[
 \bar f_n \overset{P_{n,h}}{\rightsquigarrow} \E^* \left[ \left. \dot f_{\mu_0} \left[ Z^* + Z \right] \right| Z \right].
\]
\end{proposition}

\begin{proof}[Proof of Proposition~\ref{prop:BVM.semiparametric}]
Assumption~\ref{a2s}\ref{a2s.1}-\ref{a2s.3} imply $\mathcal P = \{P_{\beta(t)} : t \in \mathcal M_{(\mu_0,\eta_0)}\}$ is locally asymptotically normal \cite[Example~7.15]{vanderVaart1998} and so $\{P_{n,(\mu_0,\eta_0)}\}$ and $\{P_{n,h}\}$ are mutually contiguous \cite[Example~6.5]{vanderVaart1998}. Hence, with $\overset{P_{n,h}}{\to}$ denoting convergence in probability under $\{P_{n,h}\}$, we have by Lemma~\ref{lem:BVM.linearize.semiparametric} that 
\[
 \left\| \bar f_n - \E^* \left[ \left. \dot f_{\mu_0} \left[ Z^* + Z_{n,(\mu_0,\eta_0)} \right] \right| Z_{n,(\mu_0,\eta_0)} \right] \right\|
 \overset{P_{n,h}}{\to} 0.
\]
Moreover, under Assumption~\ref{a2s}\ref{a2s.5.a} we have
\[
 Z_{n,(\mu_0,\eta_0)} \overset{P_{n,h}}{\rightsquigarrow} Z \sim  N(h,I_{(\mu_0,\eta_0)}^{-1}).
\]
The result follows by the continuous mapping theorem, noting  $z \mapsto \E^* [ \dot f_{\mu_0} [ Z^* + z ] ] $ is everywhere continuous by Assumption~\ref{af}\ref{af.2} and Proposition~3.1 of \cite{Shapiro1990}.
\end{proof}

\section{Bounds on Mistake Probabilities}
\label{appsec:mistake}

Definition~\ref{def:D}\ref{D.2} requires mistake probabilities to vanish faster than $n^{-1/2}$.
The main results of this section are Lemma~\ref{lem:mistake_prob_bayes}, which shows that mistake probabilities for Bayes decisions vanish at rate $n^{-1}$, and Lemma~\ref{lem:mistake_prob_quasi_bayes}, which presents a result for quasi-Bayes decisions.

\subsection{Parametric Models}

We first present some preliminary lemmas before proving Lemma~\ref{lem:mistake_prob_bayes}.

\begin{lemma}\label{lem:R_inequality_bayes}
Suppose that Assumption~\ref{a1}\ref{a1.1} holds. Then for any $\mu_0 \in \mathcal M$ there exists $\epsilon,\epsilon' > 0$ such that 
\begin{equation}\label{eq:pi_n_inequality}
 \pi_n(\{ \mu : \|\mu  - \mu_0\| \geq \epsilon\}) \leq 2\epsilon'
\end{equation}
implies $\delta_n^*(X^n;\pi) \in \underline{\mc D}_{\mu_0}$.
\end{lemma}

\begin{proof}[Proof of Lemma~\ref{lem:R_inequality_bayes}]
If $\underline{\mc D}_{\mu_0} = \mc D$ then the result is trivial. Now suppose $\underline{\mc D}_{\mu_0} \subsetneq \mc D$.
By Assumption~\ref{a1}\ref{a1.1}, we may choose  $\epsilon,\varepsilon > 0$ such that
\[
 \min_{i \not \in \underline{\mc D}_{\mu_0}} R(i,\mu) - \max_{i \in \underline{\mc D}_{\mu_0}} R(i,\mu) \geq \varepsilon, \quad \quad \mu \in N : = \{\mu  : \|\mu - \mu_0\| < \epsilon\}.
\]
We have 
\[
 \begin{aligned}
 \max_{i \in \underline{\mc D}_{\mu_0}} \bar R_n(i)
 & \leq  \max_{i \in \underline{\mc D}_{\mu_0}} \int_N R(i,\mu) \, d\pi_n(\mu) + \max_{i \in \underline{\mc D}_{\mu_0}} \int_{N^c} R(i,\mu) \, d\pi_n(\mu) \\
 & \leq  \int_N \max_{i \in \underline{\mc D}_{\mu_0}} R(i,\mu) \, d\pi_n(\mu) + \pi_n(N^c) \|R\|_\infty ,
 \end{aligned}
\]
where $\|R\|_\infty := \sup_{(d,\mu) \in \mc D \times \mc M} |R(d,\mu)| < \infty$ by Assumption~\ref{a1}\ref{a1.1}, and
\[
 \begin{aligned}
 \min_{i \not \in \underline{\mc D}_{\mu_0}} \bar R_n(i)
 & \geq  \min_{i \not \in \underline{\mc D}_{\mu_0}} \int_N R(i,\mu) \, d\pi_n(\mu) + \min_{i \not \in \underline{\mc D}_{\mu_0}} \int_{N^c} R(i,\mu) \, d\pi_n(\mu) \\
 & \geq  \int_N \min_{i \not \in \underline{\mc D}_{\mu_0}} R(i,\mu) \, d\pi_n(\mu) - \pi_n(N^c) \|R\|_\infty .
 \end{aligned}
\]
Hence,
\[
 \begin{aligned}
 \min_{i \not \in \underline{\mc D}_{\mu_0}} \bar R_n(i) - \max_{i \in \underline{\mc D}_{\mu_0}} \bar R_n(i)
 & \geq  \int_N \left( \min_{i \not \in \underline{\mc D}_{\mu_0}} R(i,\mu) - \max_{i \in \underline{\mc D}_{\mu_0}} R(i,\mu) \right) \, d\pi_n(\mu) - 2\pi_n(N^c) \|R\|_\infty \\
 & \geq \pi_n(N) \varepsilon - 2 \pi_n(N^c) \|R\|_\infty \\
 & = \varepsilon - \pi_n(N^c) ( \varepsilon + 2 \|R\|_\infty).
 \end{aligned}
\]
If $\pi_n(N^c) \leq \frac{\varepsilon}{2(\varepsilon + 2 \|R\|_\infty)} =: 2\epsilon'$ holds, then
\[
 \min_{i \not \in \underline{\mc D}_{\mu_0}} \bar R_n(i) - \max_{i \in \underline{\mc D}_{\mu_0}} \bar R_n(i) \geq \varepsilon - \frac{\varepsilon}{2} = \frac{\varepsilon}{2},
\]
which implies 
\[
 \max_{i \in \underline{\mc D}_{\mu_0}} \bar R_n(i) \leq \min_{i \not \in \underline{\mc D}_{\mu_0}} \bar R_n(i) - \frac{\varepsilon}{2} < \min_{i \not \in \underline{\mc D}_{\mu_0}} \bar R_n(i) ,
\]
and hence that $\delta_n^*(X^n;\pi) \in \underline{\mc D}_{\mu_0}$.
\end{proof}

\begin{lemma}\label{lem:prob_inequality_bayes}
Suppose that Assumption~\ref{a1}\ref{a1.1} holds. Take any $\mu_0 \in \mathcal M$ and $h \in \mathbb R^K$, and let $\epsilon,\epsilon'$ be as in (\ref{eq:pi_n_inequality}) of  Lemma~\ref{lem:R_inequality_bayes}. Suppose that there exist tests $\{\phi_n\}$ with the property that $\E_{n,h}[\phi_n(X^n)] \leq C'e^{-nC}$ and $\sup_{\mu:\|\mu - \mu_0\| \geq \epsilon} \E_{n,\mu}[1-\phi_n(X^n)] \leq C'e^{-nC}$ for $C,C' > 0$. Then
\[
 P_{n,h}\left(\delta_n^*(X^n;\pi) \not \in \underline{\mc D}_{\mu_0} \right) \leq 
 \frac{C'e^{-nC}}{\epsilon'}
 + \frac{C'e^{-nC/2}}{\epsilon' \pi(\{\mu : D_{KL}(p_{\mu_0 + h/\sqrt n}\| p_\mu) < n^{-1}\})} + \frac{4}{nC}.
\]
\end{lemma}

\begin{proof}[Proof of Lemma~\ref{lem:prob_inequality_bayes}]
In view of Lemma~\ref{lem:R_inequality_bayes}, we have 
\[
 P_{n,h}\left(\delta_n^*(X^n;\pi) \not \in \underline{\mc D}_{\mu_0} \right) \leq P_{n,h}\left(\pi_n(\{\mu : \|\mu - \mu_0\| \geq \epsilon\}) > 2\epsilon' \right).
\]
We proceed by similar arguments to the consistency theorem of \cite{Schwartz1965} but centering along $P_{n,h}$. Write
\begin{multline*}
 P_{n,h}\left( \pi_n(\{\mu : \|\mu - \mu_0\| \geq \epsilon\}) > 2\epsilon' \right)
 \leq
 P_{n,h}\left( \phi_n(X^n) > \epsilon' \right) \\ + 
 P_{n,h}\left( \frac{(1 - \phi_n(X^n)) \int_{N^c} \prod_{i=1}^n p_\mu(X_i)/p_{\mu_0+h/\sqrt n}(X_i) \, d \pi(\mu)}{\mathcal J_n}  > \epsilon' \right),
\end{multline*}
where $\mathcal J_n = \int \prod_{i=1}^n p_\mu(X_i)/p_{\mu_0 + h/\sqrt n}(X_i) \, d \pi(\mu)$ and $N^c = \{\mu : \|\mu - \mu_0\| > \epsilon\}$. The first probability on the right-hand side is bounded by $C'e^{-nC}/\epsilon'$ by Markov's inequality. For the second, Lemma~6.26 of \cite{GhosalVanderVaart} implies that for any $B > 0$ and $\varepsilon \geq n^{-1}$,\footnote{The result is stated requiring $B > 1$, but inspection of the proof shows any $B > 0$ is valid: restricting to $B > 1$ is simply so the probability bound of $1/B$ is not vacuous.}
\[
 P_{n,h}\left( \mathcal J_n < \pi(\{\mu : D_{KL}(p_{\mu_0+h/\sqrt n}\| p_\mu) < \varepsilon\}) e^{-2nB \varepsilon} \right) \leq \frac 1B.
\]
Hence we may deduce
\begin{multline*}
 P_{n,h}\left( \frac{(1 - \phi_n'(X^n)) \int_{N^c} \prod_{i=1}^n p_\mu(X_i)/p_{\mu_0 + h/\sqrt n}(X_i) \, d \pi(\mu)}{\mathcal J_n}  > \epsilon' \right) \\
 \leq \frac{\E_{n,h} \left[ (1 - \phi_n'(X^n)) \int_{N^c} \prod_{i=1}^n p_\mu(X_i)/p_{\mu_0 + h/\sqrt n}(X_i) \, d \pi(\mu) \right]}{\epsilon' \pi(\{\mu : D_{KL}(p_{\mu_0 + h/\sqrt n}\| p_\mu) < \varepsilon\}) e^{-2nB \varepsilon}} + \frac 1B.
\end{multline*}
The numerator in the first term on the right-hand side may be bounded by $C'e^{-Cn}$ using Fubini's theorem (see, e.g., the proof of Lemma~6.17 of \cite{GhosalVanderVaart}). Setting $B = Cn/4$ and $\varepsilon = n^{-1}$, the right-hand side of the above display is bounded by
\[
 \frac{C'e^{-nC/2}}{\epsilon' \pi(\{\mu : D_{KL}(p_{\mu_0 + h/\sqrt n}\| p_\mu) < n^{-1}\})} + \frac{4}{nC},
\]
as required.
\end{proof}

\begin{lemma}\label{lem:prior}
Suppose that Assumption~\ref{a2}\ref{a2.1},\ref{a2.3}-\ref{a2.4} hold and $\pi \in \Pi$. Then for $\mu_0 \in \mathcal M$ and $h \in \mathbb R^K$, 
\[
 \liminf_{n \to \infty} n^{K/2} \pi(\{\mu : D_{KL}(p_{\mu_0 + h/\sqrt n}\| p_\mu) < n^{-1}\}) > 0.
\]
\end{lemma}

\begin{proof}[Proof of Lemma~\ref{lem:prior}]
Assumption~\ref{a2}\ref{a2.4} implies $D_{KL}(p_{\mu_0+h/\sqrt n}\|p_\mu) \leq B\|\mu - \mu_0 - h/\sqrt n\|^2$ on a neighborhood of $\mu_0$ for all sufficiently large $n$, with $B = 2 \lambda_{\max}(I_{\mu_0}) < \infty$ by Assumption~\ref{a2}\ref{a2.3}. Hence, for all $n$ sufficiently large, $\|\mu - \mu_0 - h/\sqrt n\|^2 < (Bn)^{-1}$ implies $D_{KL}(p_{\mu_0+h/\sqrt n}\|p_\mu) < n^{-1}$. Assumption~\ref{a2}\ref{a2.1} and the positivity and  continuity of $\pi$ imply that $\pi(\{ \mu : \|\mu - \mu_0 - h/\sqrt n\|^2 < (Bn)^{-1} \})$ is bounded below by a multiple of $n^{-K/2}$ for all $n$ sufficiently large, and hence that $\pi(\{\mu : D_{KL}(p_{\mu_0 + h/\sqrt n}\| p_\mu) < n^{-1}\})$, is bounded below by a multiple of $n^{-K/2}$ for all $n$ sufficiently large.
\end{proof}

\begin{lemma}\label{lem:mistake_prob_bayes}
Suppose that Assumptions~\ref{a1}\ref{a1.1} and~\ref{a2} hold and $\pi \in \Pi$. Then there exists $C > 0$ such that
\[
 P_{n,h}\left(\delta_n^*(X^n;\pi) \not \in \underline{\mc D}_{\mu_0} \right) \leq \frac{C}{n}.
\]
\end{lemma}

\begin{proof}[Proof of Lemma~\ref{lem:mistake_prob_bayes}]
We will prove the result by applying Lemma~\ref{lem:prob_inequality_bayes}. We first establish existence of tests as required by Lemma~\ref{lem:prob_inequality_bayes}. We proceed as in Propositions~6.1 and~6.2 of \cite{ClarkeBarron}. Let $d_G$ denote the metric they construct to metrize weak convergence. Let $\hat P_n \equiv \hat P_n(X^n)$ denote the empirical measure. As shown in the proof of Proposition~6.2 of \cite{ClarkeBarron}, for any $\varepsilon > 0$ there exists an integer $k$ depending only on $\varepsilon$ such that
\begin{equation}\label{eq:test}
 P_{n,\mu} \left( d_G(\hat P_n, P_\mu) \geq \varepsilon \right) \leq 2ke^{-n\varepsilon^2}, \quad \quad \mu \in \mathcal M.
\end{equation}
By Assumption~\ref{a2}\ref{a2.5}, there exists $\epsilon' > 0$ such that $\|\mu - \mu_0\| \geq \epsilon$ implies $d_G(P_\mu ,P_{\mu_0}) \geq \epsilon'$. Consider the test $\phi_n(X^n) = \mathbb I[d_G(\hat P_n,P_{\mu_0}) > \epsilon'/2]$. Again using Assumption~\ref{a2}\ref{a2.5}, take $n$ large enough that $d_G(P_{\mu_0 + h/\sqrt n},P_{\mu_0}) < \epsilon'/4$. Then by (\ref{eq:test}), we have 
\[
 \E_{n,h}[\phi_n(X^n)] \leq P_{n,h} \left( d_G(\hat P_n, P_{\mu_0 + h/\sqrt n}) \geq \epsilon'/4 \right) \leq C' e^{-nC}
\]
for suitable $C,C'$. We may similarly deduce as in the proof of Proposition~6.1 of \cite{ClarkeBarron} that for any $\mu$ with $\|\mu - \mu_0\| \geq \epsilon$ we have $\E_{n,\mu}[1-\phi_n(X^n)] \leq C' e^{-nC}$ for suitable $C,C'$ not depending on $\mu$. Thus, we have verified existence of a sequence of tests with the desired properties.

By Lemma~\ref{lem:prob_inequality_bayes}, it suffices to show $e^{nc} \pi(\{\mu : D_{KL}(p_{\mu_0 + h/\sqrt n}\| p_\mu) < n^{-1}\}) \to + \infty$ for any $c > 0$. This follows immediately from Lemma~\ref{lem:prior}.
\end{proof}

\subsection{Semiparametric Models}

\begin{lemma}\label{lem:mistake_prob_quasi_bayes}
Suppose that Assumptions~\ref{a1}\ref{a1.1} and~\ref{a2s} hold and $\pi \in \Pi$. Then for any $(\mu_0,\eta_0) \in \mc M \times \mc H$,
\[
 \sqrt n \, P_{n,h}\left(\delta_n^*(X^n;\pi) \not \in \underline{\mc D}_{\mu_0} \right) \to 0
\]
for all $h \in \mb R^K$. 
\end{lemma}

\begin{proof}[Proof of Lemma~\ref{lem:mistake_prob_quasi_bayes}]
In view of Lemma~\ref{lem:R_inequality_bayes}, we have 
\[
 P_{n,h}\left(\delta_n^*(X^n;\pi) \not \in \underline{\mc D}_{\mu_0} \right) \leq P_{n,h}\left(\pi_n(\{\mu : \|\mu - \mu_0\| \geq \epsilon\}) > 2\epsilon' \right)
\]
for some $\epsilon, \epsilon' > 0$. By Assumption~\ref{a2s}\ref{a2s.1}, choose $C > 0$ such that $\{\mu : \|\mu - \mu_0\| \leq C\} \subset \mc M$. Taking $\epsilon$ smaller if needed, we may assume $\epsilon \leq C$. Recall that $\hat \lambda_{\min}$ and $\hat \lambda_{\max}$ denote the smallest and largest eigenvalues of $\hat I$. Let $A_n$ denote the event upon which $\|\hat \mu - \mu_0\| \leq \epsilon/2$ and $c < \hat \lambda_{\min} \leq \hat \lambda_{\max} < c^{-1}$ both hold, where the constant $c \in (0,1)$ is from Assumption~\ref{a2s}\ref{a2s.4.b}. Note $\pi$ has a positive, continuous, and bounded Lebesgue density $\pi(\mu)$ on $\mc M$, hence $\overline \pi := \sup_{\mu \in \mc M}\pi(\mu) < \infty$ and $\underline \pi := \inf_{\mu : \|\mu - \mu_0\| \leq C} \pi(\mu) > 0$. It follows that on $A_n$, we have
\[
 \begin{aligned}
 \pi_n(\{\mu : \|\mu - \mu_0\| \geq \epsilon\})
 & \leq \pi_n(\{\mu : \|\mu - \hat \mu\| \geq \epsilon/2\}) \\
 & \leq \frac{\int_{\mu:\|\mu - \hat \mu\| \geq \epsilon/2} e^{- \frac 12 (\mu - \hat \mu)^T(n \hat I)(\mu - \hat \mu)} \, d \pi(\mu) }{\int_{\mu:\|\mu - \hat \mu\| \leq C - \epsilon/2} e^{- \frac 12 (\mu - \hat \mu)^T(n \hat I)(\mu - \hat \mu)} \, d \pi(\mu) }  \\
 & \leq \frac{\overline \pi \int_{h:\|h\| \geq \sqrt n \epsilon/2} e^{- \frac{c}{2} \|h\|^2} \, d h }{ \underline \pi \int_{h:\|h\| \leq \sqrt n (C - \epsilon/2)} e^{-  \frac{1}{2c} \|h\|^2} \, d h }  \to 0,
 \end{aligned}
\]
where the first inequality is by the reverse triangle inequality, the final inequality is by a change of variables from $\mu$ to $h = \sqrt n (\mu - \hat \mu)$, and convergence to zero is by the dominated convergence theorem. We therefore have that $\pi_n(\{\mu : \|\mu - \mu_0\| \geq \epsilon\}) < 2\epsilon'$ holds on $A_n$ for all $n$ sufficiently large, in which case
\[
 P_{n,h}\left(\pi_n(\{\mu : \|\mu - \mu_0\| \geq \epsilon\}) > 2\epsilon' \right) 
 \leq P_{n,h}(A_n^c) .
\]
Finally, $\sqrt n P_{n,h}(A_n^c) \to 0$ by Assumption~\ref{a2s}\ref{a2s.4}.
\end{proof}

\end{appendix}

\end{document}